\documentclass[superscriptaddress,twocolumn,prx,aps,preprintnumbers,notitlepage,longbibliography,floatfix]{revtex4-1}

\usepackage[latin9]{inputenc}
\usepackage{amsmath}
\usepackage{amssymb}
\usepackage{graphicx}
\usepackage{esint}
\usepackage[scr=boondoxo, scrscaled=1.05]{mathalfa}

\usepackage{censor}

\usepackage{hyperref}
\allowdisplaybreaks
\usepackage{ulem}
\makeatletter

\pdfpageheight\paperheight
\pdfpagewidth\paperwidth

\pdfoutput=1

\usepackage{epsfig}
\usepackage{graphics}
\graphicspath{{figures/}}

\usepackage{mciteplus}
\mciteErrorOnUnknownfalse

\makeatother
\usepackage{color}

\begin{document}
\StopCensoring
\preprint{\censor{FERMILAB-PUB-21-458-AE}}

\title{Angular correlations of causally-coherent primordial  quantum perturbations}

\censor{
\author{Craig  Hogan} 
\affiliation{University of Chicago, 5640 South Ellis Ave., Chicago, IL 60637}
\affiliation{Fermi National Accelerator Laboratory, Batavia, IL 60510}
\author{Stephan S. Meyer}
\affiliation{University of Chicago, 5640 South Ellis Ave., Chicago, IL 60637}
}
\begin{abstract}
 
{ We consider the hypothesis  that nonlocal, omnidirectional, causally-coherent quantum entanglement of  inflationary horizons
may account for some well-known measured anomalies of Cosmic Microwave Background (CMB) anisotropy on large angular scales. It is shown that causal coherence can lead to less cosmic variance in the large-angle power spectrum ${C}_\ell$ of primordial curvature perturbations on spherical horizons than predicted by the standard model of locality in  effective field theory, and to new symmetries of the angular correlation function ${C}(\Theta)$. 
Causal considerations are used to construct an  approximate analytic model for  ${C}(\Theta)$ on angular scales larger than a few degrees.}
Allowing for uncertainties from the unmeasured intrinsic dipole and from Galactic foreground subtraction,
{causally-coherent constraints} are shown to  be consistent  with measured  CMB correlations on large angular scales.   Reduced cosmic variance will enable powerful tests of the hypothesis  with better  foreground subtraction  and higher fidelity  measurements on large angular scales. 
\end{abstract}

\maketitle


\section{Introduction}

\subsection{Primordial structure on large angular scales}

\begin{figure*}
  \centering
  \includegraphics[width=0.49\textwidth]{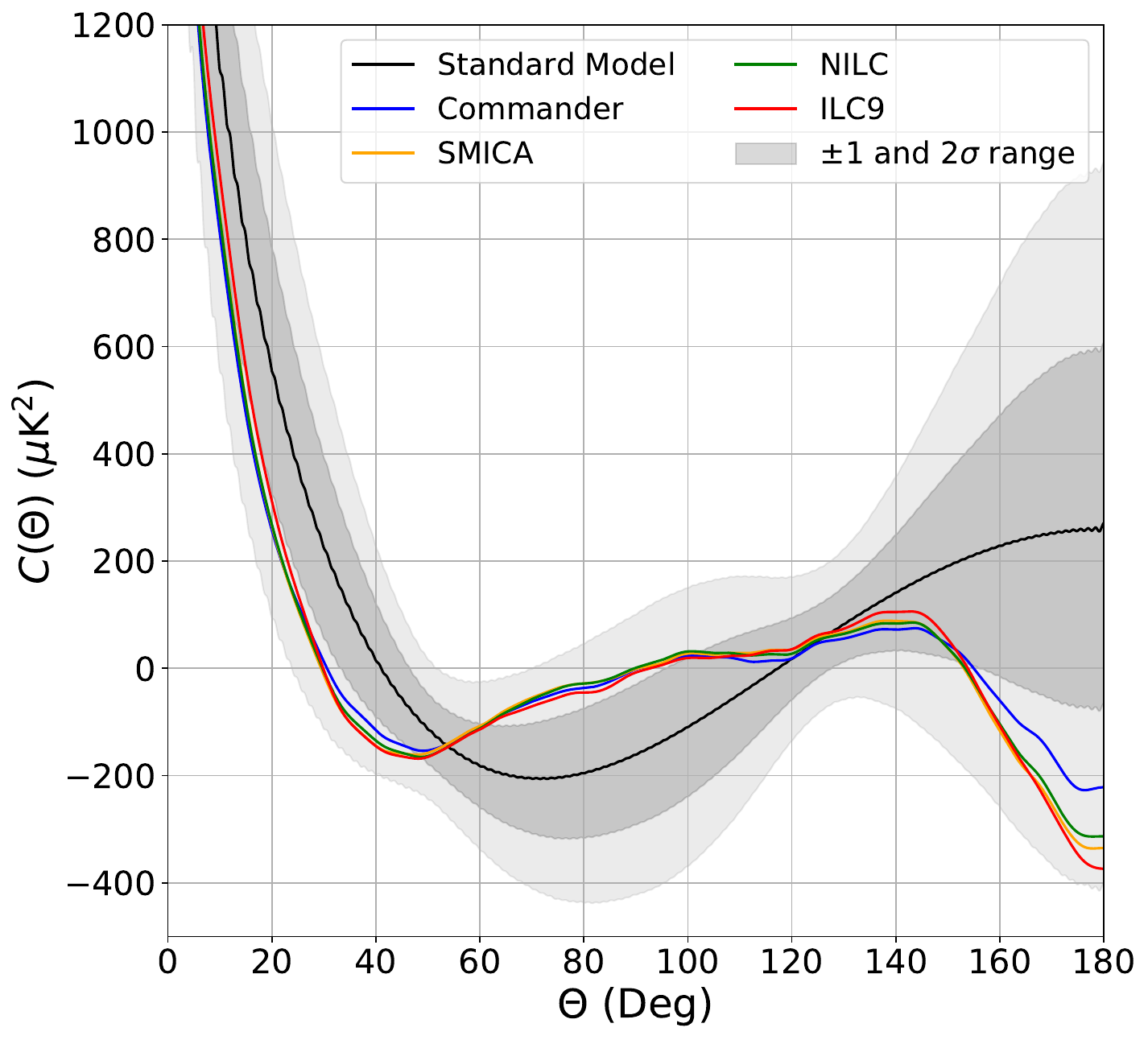}
  \includegraphics[width=0.49\textwidth]{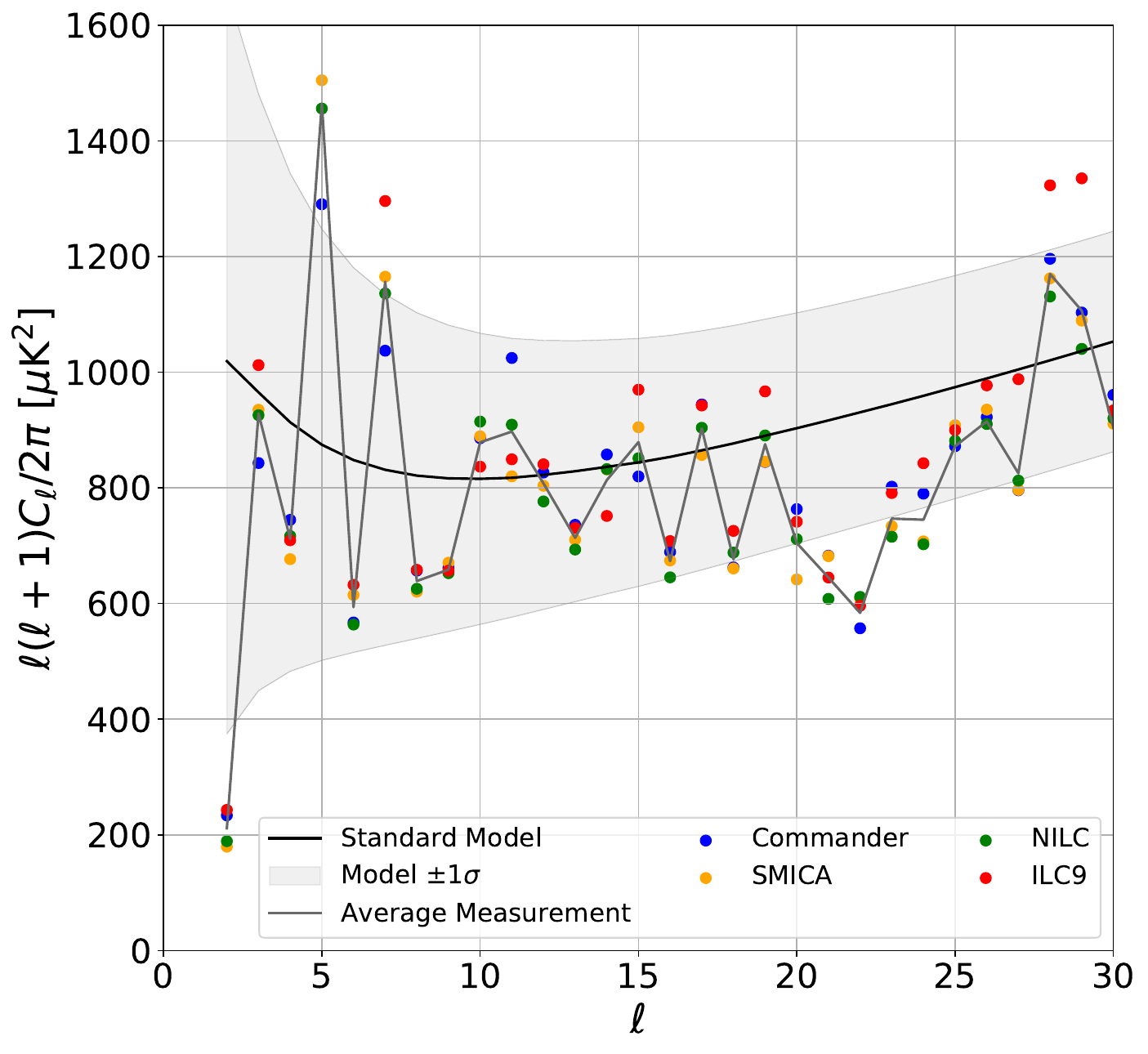}
  \caption{Angular correlation function $C(\Theta)$ and  angular power spectrum $C_\ell$ of CMB temperature anisostropy, measured with 
 the  {\sl WMAP}  and {\sl Planck} satellites. Different colors show results for smoothed, unmasked maps with four different foreground subtraction schemes: ILC9, NILC, Commander, and SMICA.  Functions are defined  in Sec. (\ref{definitions}).
The left plot shows the measured angular correlation functions of smoothed maps, together with the expectation and  range of  standard inflationary realizations, shown by the shaded bands.  
The right plot shows measurements of the angular power spectrum $C_\ell$ up to $\ell= 20$, a range generally excluded from fits to cosmological models.   
 The model curve for $C_\ell$ shows the expectation for realizations in the standard inflation scenario, and  the  band shows the ``cosmic variance'' of the predicted distribution.   These angular and spectral representations of {\sl Planck} and {\sl WMAP} data at large angles contain
similar  information about the pattern of anisotropy, but reveal different physical symmetries.
 For $\ell\lesssim 20$, or $\Theta\gtrsim 9^\circ$, the temperature anisotropy is dominated by the Sachs-Wolfe effect\cite{1967ApJ...147...73S,PhysRevD.22.1882}, so the pattern of temperature  is close to the original pattern of the primordial  curvature on our horizon.  
  \label{data}}
\end{figure*}

\begin{figure*}
\begin{centering}
 \includegraphics[width=0.7\textwidth]{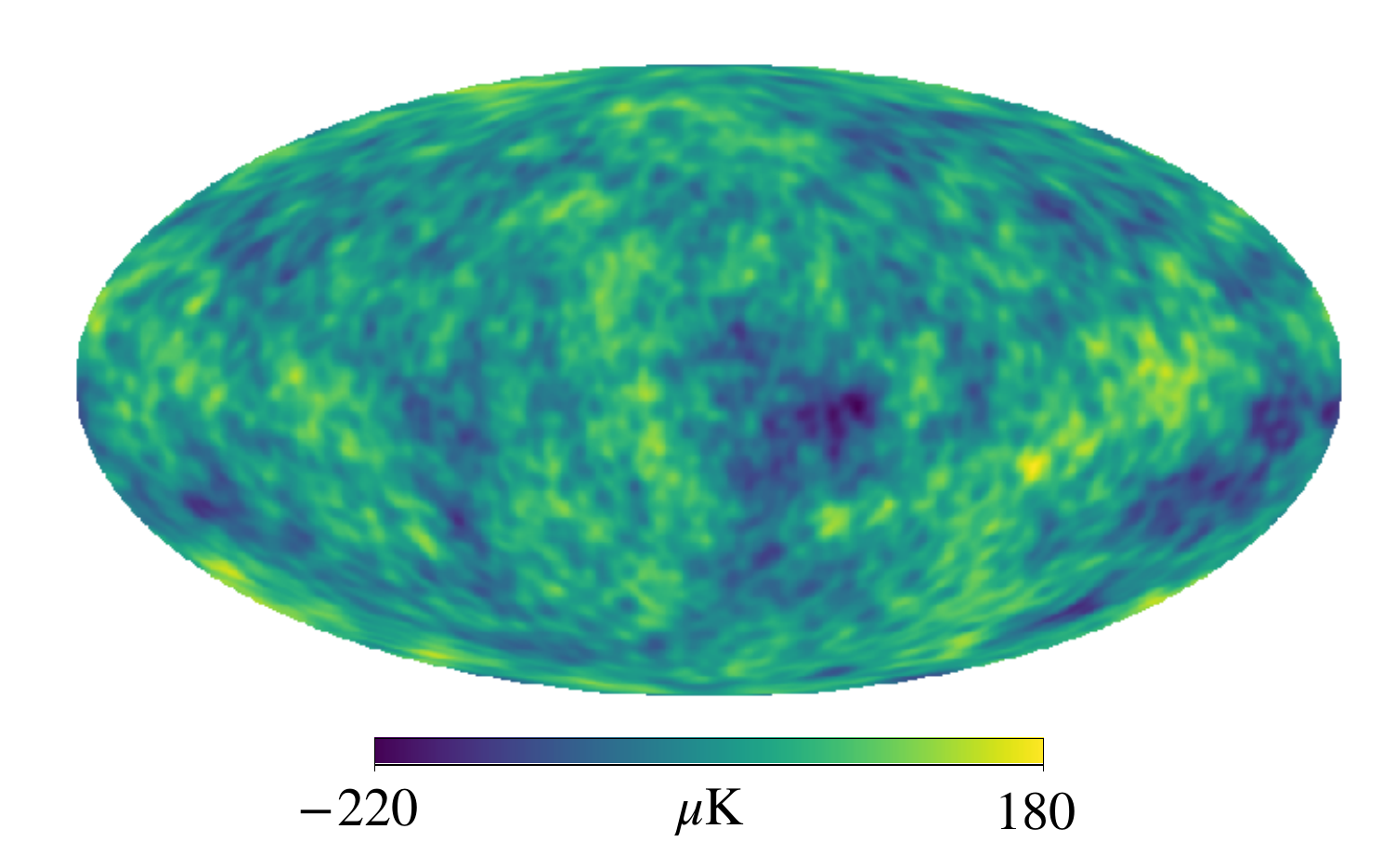}
\par\end{centering}
\protect\caption{ Foreground-subtracted smoothed average of the four different maps shows the  pattern of temperature anisotropy on the sky at  $\Theta\gtrsim 5^\circ$ and $\ell\lesssim 30$.   At this resolution and contrast scale, the map  closely approximates the actual pattern of primordial potential perturbations on our spherical horizon, apart from an unmeasured intrinsic dipole. The scale is shown in microKelvin. \label{Averagemap}}
\end{figure*}

In the standard cosmological model\cite{Weinberg:2008zzc,Baumann:2009ds},
cosmic structure originates from perturbations  generated by quantum fluctuations during an early period of cosmic inflation.  
According to this model,   the large-angle anisotropy of the cosmic microwave background (CMB)\cite{2013ApJS..208...20B,2013ApJS..208...19H,Ade:2015xua,Ade:2015lrj,Array:2015xqh,Aylor:2017haa,Akrami:2018vks,Aghanim:2018eyx,Akrami:2018odb}  preserves  intact the detailed pattern of structure created at the earliest times.
Smoothed  at an angular resolution of $\Theta\sim 5^\circ$, or filtered with  spherical harmonic numbers below about
$\ell \sim  30$, 
the temperature anisotropy  is essentially a map of  primordial curvature on our spherical horizon\cite{1967ApJ...147...73S,PhysRevD.22.1882}: on these scales,   the sky    preserves a high-fidelity,  coherent image of curvature perturbations imprinted  by quantum fluctuations during inflation (Figs. \ref{data}, \ref{Averagemap}).

The  standard inflationary picture does not  make a specific prediction for angular correlations,  only for  statistical properties of an ensemble of possible realizations.
Stochastic waves of random 3D  fluctuations  create a variety of  different patterns, with a ``cosmic variance''  between different realized skies.  The main prediction of the standard quantum model of inflation, a nearly-scale-invariant 3D power spectrum of scalar curvature perturbations,
 is best tested at  $\ell \gtrsim 30$ , because on the  largest  angular scales, the observable universe does not include a large enough volume to measure the universal power spectrum of  3D modes accurately.

It is well known that even allowing for cosmic variance,  the standard scenario does not  actually agree very well with the  measured CMB pattern on the largest angular scales;  that is, the pattern of the observed sky is highly  atypical of the ensemble of realizations.
Starting with the first measurements of primordial anisotropy with {\sl COBE}\cite{1992ApJ...396L..13W,1994ApJ...436..423B,Hinshaw_1996}, 
and continuing with improved maps from {\sl WMAP} and
{\sl Planck}\cite{WMAPanomalies,Ade:2015hxq,Akrami:2019bkn},
a variety of large-scale  ``anomalies'' have led some authors to question  whether the standard picture is  correct or complete\cite{deOliveira-Costa:2003utu,2017MNRAS.472.2410A,2012MNRAS.419.3378A,2015MNRAS.449.3458C,Schwarz:2015cma}.

The most conspicuous  anomaly is a surprisingly small angular correlation function $C(\Theta)$ on large angular scales, which corresponds to an
angular spectrum $C_\ell$ with a remarkably small quadrupole ($\ell=2$) moment and a seemingly conspiratorially-cancelling arrangement of higher multipoles.  
The largest-angle correlation at $\Theta\gtrsim 135^\circ$ also displays a  significant  anticorrelation, 
which  appears   in the angular spectrum  as  a  prevalence of odd-parity over even-parity fluctuations over a wide range of angular scales up to $\ell\sim 30$.    In the absence of a compelling physical explanation, these  anomalies can be (and  generally are) interpreted   as statistical flukes of our particular realization, with no deep physical significance. Thus, in spite of the pristine primordial provenance  of the actual pattern, the largest angular scales are generally omitted in model fits to cosmological data.

 On the other hand, there are hints that  large angle correlations may actually have a deeper significance.
One remarkable example is a nearly exactly vanishing angular two-point  correlation  function at right angles\cite{Hagimoto_2020}:
\begin{equation}\label{90}
C(90^\circ) \approx 0.
\end{equation}
The nearly-exact null at this special angle hints that it is due to an exact symmetry.   If so,  it would be evidence that primordial large-angle two-dimensional correlations in general are governed by  precisely defined fundamental principles,  rather than by chance.

Any universal exact symmetry of angular correlation  requires
departures from the standard quantum model of inflation. In this paper, we consider the hypothesis that the symmetry in Eq. (\ref{90}), as well as other anomalies, are due to constraints from a new physical effect: the nonlocal causal coherence of quantum gravity.
Although our analysis of inflation, causality, symmetry and correlation constraints lies entirely in the classical realm, this hypothesis requires radical modification of some standard assumptions about the  quantum system that generates inflationary perturbations.

\subsection{Causal coherence of quantum geometry}
As discussed in the Appendix,  our proposed departure from the standard quantum inflation model arises from different assumptions about the emergence of locality and causality from a quantum system.  
{ Measured correlations of primordial curvature perturbations in our picture are causal, but they are not local}.

In the standard quantum model of inflation, based on quantized amplitudes of plane wave modes with definite classical wave vectors $\vec k$, primordial structure is laid down  by coherent quantum states that live on infinite comoving spacelike  hypersurfaces.
  The thesis of this paper is instead that {\it   geometrical states of  potential  are coherent on causal horizons}. The inflationary horizon of a world line is a null surface that intersects  surfaces of constant time on compact 2-spheres, not planes.  
  

In the small-angle regime, less than a few degrees,  where 
the curvature of spherical surfaces is unimportant,  coherent plane waves are a good approximation.
Thus, the two models  agree about correlations at $\ell\gtrsim 30$, where cosmological predictions have been precisely  tested.
The two models  disagree in the regime where the curvature of spherical null surfaces is important.
Differences become significant on large angular scales, at $\ell\lesssim 10$, and on scales larger than the  horizon.

A complete quantum treatment of  correlations at large angles and distances would require a  holographic theory of quantum gravity, locality and emergence. 
There is no consensus on a comprehensive theory of quantum gravity, but as discussed in the Appendix, there are  indications in some formal holographic or entropic approaches\cite{Banks2011,Banks:2011qf,Banks:2015iya,Banks:2018ypk,Banks:2020dus,Banks:2021jwj} that quantum geometry entangles quantum states of coherent causal diamonds, with far fewer independent degrees of freedom in the infrared than the quantized  field theory applied to standard quantum inflation. 

Our approach in this paper is to guess some  symmetries that might govern  a coherent holographic theory, and test whether they fit the data better than the standard picture.
We seek direct evidence for  symmetries of angular correlations that could arise from causal coherence.  We analyze how classical geometrical causal relationships among causal diamonds constrain angular correlations, and compare these constraints with data. 
Evidence for these symmetries in primordial perturbations could provide clues to how space-time and locality emerge in holographic quantum gravity.

\subsection{Causal constraints on large-angle correlations}

In standard quantum inflation, the initial state of a   local scalar field vacuum is specified  at some initial time, on an infinite spacelike hypersurface in an unperturbed background.  Each mode eventually collapses coherently into a definite, classical scalar curvature perturbation $\Delta(\vec k)$ on a spacelike surface, at a time that depends on its wavelength. 
The final local value of the curvature at any point is determined by a sum of mode amplitudes at that point. The modal contributions are acausally correlated at spatial separations far greater than the horizon distance when a mode collapses, indeed infinitely far for separations normal to the wave vector.  These correlations are baked into  the random phases assigned to modes in the initial state.

In quantum mechanics, acausal spacelike correlations can happen, but only in systems where  the state is prepared causally\cite{RevModPhys.71.S288,2009FoPh...39..677B,Zych_2019}.  In those systems, preparation and measurement of a state by a timelike observer creates  nonlocal correlations inside a causal diamond.  
For the quantum mechanics of a coupled matter-geometry inflation system to be consistent with causality around a point, reduction of quantum states should occur on invariant null cones, so that   events on a light cone always correlate with local measurements at  its apex.
For this to be possible, {\it the  local classical invariant curvature perturbation $\Delta(\vec x)$  must be determined by  information  within the past null cone of $\vec x$.}

This causal constraint on  bulk information  is a natural property of holographic space-times
\cite{Banks2011,Banks:2011qf,Banks:2015iya,Banks:2018ypk,Banks:2020dus,Banks:2021jwj}.
Quantum states of scalar fields in effective field theory, used in standard quantum inflation, do not have this property.  They are coherent on infinite spacelike planes, not spherical null  surfaces, and
 scalar curvature mode amplitudes  freeze out independently on  surfaces of constant cosmic time. As discussed  in the Appendix, applications of this effective-field framework also lead to difficulties in other situations, such as well-known paradoxes for information flow in Hawking  evaporation of black holes.

We have argued previously\cite{Hogan:2002xs,Hogan:2003mq,10.2307/27857721,PhysRevD.99.063531,Hogan_2020} that causal constraints in  holographic space-times  lead to constraints  on large angular scale correlations  in the cosmological system.
Here, we investigate the consequences of the specific conjecture that
the angular  spectrum ${C}_\ell$ and correlation function $C(\Theta)$ of primordial  curvature perturbations on  spherical surfaces  around any comoving world line 
 approximate  universal  functions at large angles,   governed by the requirement that 
 {\it correlations of invariant scalar curvature among different world lines  arise entirely from causal entanglement within their horizons.}
Again, this 
 causal constraint does not apply in standard quantum inflation.

Causal constraints on correlated bulk information of overlapping horizons for different observers translates  into  correlations of curvature on spherical surfaces in the angular domain.   We show below that {\it causality requires that the  angular  correlation function ${C}(\Theta)$ of primordial potential on spheres exactly vanishes not only at $90^\circ$, but over a wide range of antihemispherical angles}:
 \begin{equation}\label{antizero}
{C}(90^\circ\le\Theta\le 135^\circ) = 0,
\end{equation}
{\it and is negative in the antipodal region},
 \begin{equation}\label{beyond135}
{C}(135^\circ<\Theta<180^\circ) < 0.
\end{equation}
Exact zeros can only be true for all observers if the correlation function  approximates a universal form, with much less cosmic variance than the standard picture on large angular scales.

Holographic symmetries are hidden in measured maps of the sky, but it may be possible to measure their signatures  on large angular scales.
Eqs. (\ref{antizero}) and (\ref{beyond135}) are apparently not true of the real sky: there is a small positive correlation at $\Theta>90^\circ$, as seen in Fig. (\ref{data}). {\it We will argue that  the true symmetry has been hidden from direct measurement by the subtraction of  a primordial dipole component,   
and by residual error in the subtraction of Galactic emission.}
For  a quantitative comparison with data over a wider range of angular scales and angular harmonics, we construct an analytic model to approximate the  2D  correlation function,   based on  causal relationships among intersecting horizons of different world lines. 
The parameters of the  model are uniquely determined by imposing  causal constraints, which include a prediction for the unmeasured dipole. 
We also demonstrate a  way to compare the data with causally-coherent constraints  and standard predictions  without a specific holographic model. 
We show that the causally-coherent scenario agrees much better than the standard picture with maps of
CMB  temperature anisotropy, on large angular scales $\Theta>5^\circ$ where the measured temperature anisotropy preserves
the   primordial scalar curvature pattern.

\begin{figure*}[hbt]
\begin{centering}
\includegraphics[width=.9\linewidth]{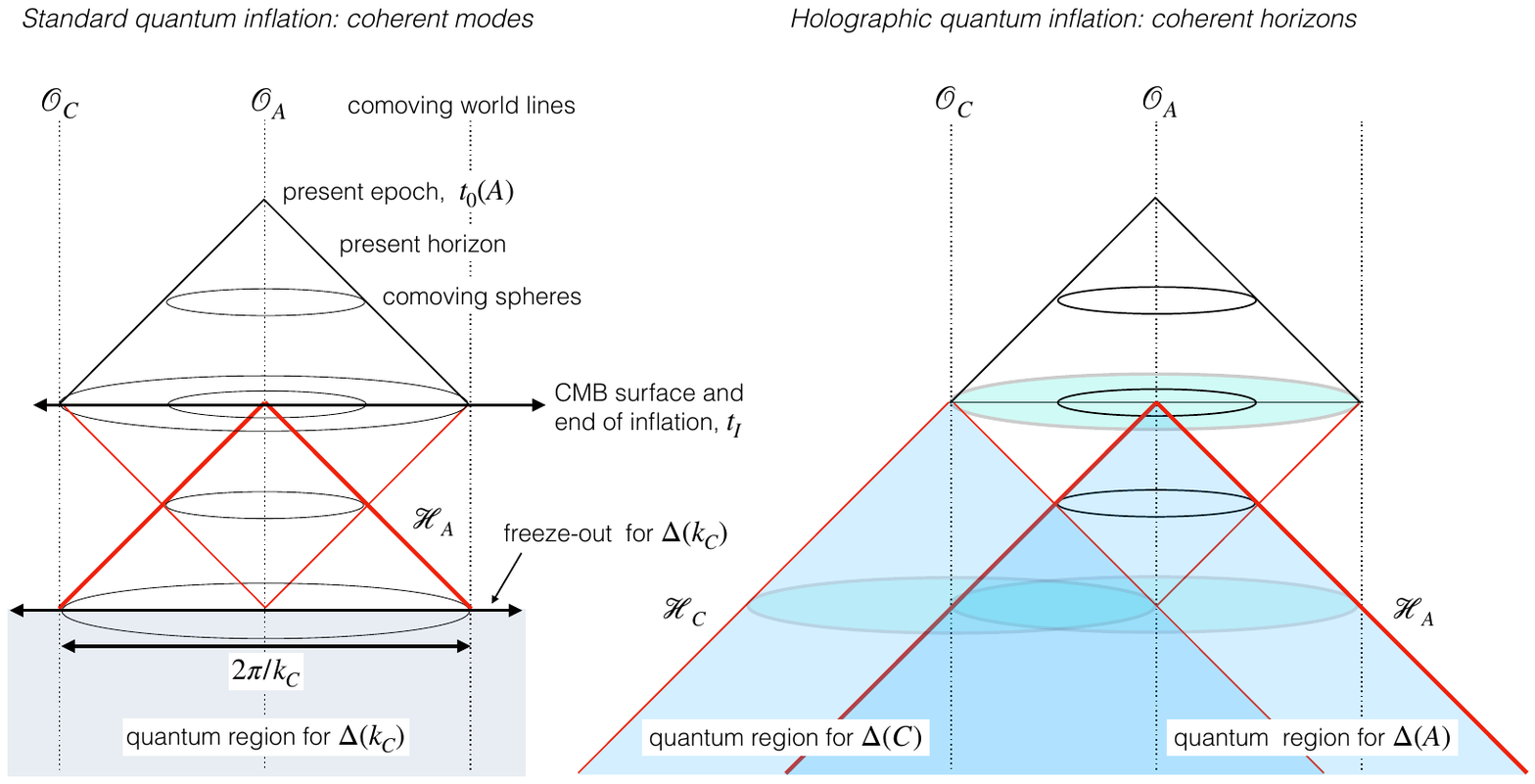}
\par\end{centering}
\protect\caption{Conformal causal structure of an inflationary universe.  Vertical lines represent the  world lines of comoving observers $\cal O$.
Horizontal planes represent spatial hypersurfaces of constant cosmic time $t$ or  conformal time $\eta$, including one at
the end of inflation, $t_I$.   The inflationary horizon $\cal H_A$ is a null surface that  forms the outer boundary of  causal diamonds of $A$ that end before $t_I$.  As shown at the left, in standard quantum inflation, the amplitude $\Delta(\vec k_C)$ of a scalar-field plane wave  mode freezes out acausally everywhere at the time  when its wavelength matches the $AC$ horizon radius,  on a 3D spacelike hypersurface. 
Thus, values of $\Delta(x)$ at arbitrary  spacelike separation in  planes normal to $\vec k$, and at longitudinal separations $>> k^{-1}$, are fixed acausally (see also Fig. \ref{standardcausal}).
The effect of  coherent  horizons is shown on the right: values of $\Delta$ are entangled among all directions on the compact spherical bounding surfaces of causal diamonds in ${\cal H}$.  Values of $\Delta$ at locations $A$ and $C$  at the end of inflation are entangled with the causal quantum interiors of their horizons, as shown by shaded light cones.  
  Directional properties of perturbations are entangled between $A$ and $C$   where their causal diamonds overlap.  \label{conformals}}
\end{figure*}

\section{Causal  Correlations }

\subsection{Causal structure of inflationary  space-time }
Our framework assumes the causal structure of a standard unperturbed classical inflationary universe.
The same nearly-scale-invariant power spectrum is predicted for $\ell\gtrsim 30$, where holographic angular correlations  are unimportant,  and where fits to cosmological parameters are made.  (As discussed in the Appendix, since holographic perturbations depend differently on background parameters,  the observational constraints on the  physical inflaton potential differ from  those of the standard picture\cite{PhysRevD.99.063531}.)

An unperturbed inflationary 
universe has a 
Friedmann-Lema\^itre-Robertson-Walker metric, with space-time interval 
\begin{equation}\label{FLRW}
ds^2 = a^2(t) [c^2d\eta^2- d\Sigma^2],
\end{equation}
where  $t$ denotes proper cosmic time for any comoving observer, $d\eta\equiv dt/ a(t)$ denotes a conformal time interval, and  $a(t)$ denotes the cosmic scale factor, determined by the equations of motion. These are summarized in the Appendix.
The  spatial 3-metric in comoving coordinates is
\begin{equation}\label{flatspace}
d\Sigma^2 = dr^2 + r^2 d\Omega^2,
\end{equation}
where the angular interval in standard polar notation is $d\Omega^2 = d\theta^2 + \sin^2 \theta d\phi^2$.
Future and past light cones from an event are defined by a null path,
\begin{equation}\label{null}
d\Sigma = \pm cd\eta.
\end{equation}
Causal diagrams for an inflationary metric are  shown in Figure   (\ref{conformals}).
A causal diamond  for an observer ${\cal O}$   with boundary at $t$ corresponds   to an interval with  equal conformal time before and after $t$.
The end of inflation $t_I$ is taken to be the time   
 when the  expansion changes from accelerating, $\ddot a>0$, to decelerating, $\ddot a<0$.



{
\subsection{Causality and locality in quantum gravity}

 Quantum mechanics is inherently ``nonlocal'': it does not automatically include any notion of space or time, so a model for the quantum system that creates primordial perturbations  also needs to  include some  assumptions about the  emergence  of position in space and time, or locality, from a quantum system.

As discussed  in
in the Appendix, the standard approach to quantizing $\Delta$ during inflation adopts assumptions about quantum locality that are built into 
standard effective field theory.  The scalar $\Delta$ is represented by a sum of  field modes, which are standing waves in comoving, conformal coordinates.   Each mode has a fixed space-time structure, a sum of plane waves  propagating in opposite directions from and to spacelike infinity. The quantum wave function for the amplitude of each mode is that of a quantum harmonic oscillator in its ground state.
Coupling of $\Delta$  to other quantized fields is determined by linearized gravity.
Because  momenta for each mode
exactly cancel at each frequency along opposite  directions, momentum is locally conserved in all interactions.

This approach introduces quantum-mechanical nonlocality   ``one direction at a time'':
it describes  quantum states that are coherent at arbitrary separation along each direction.  The amplitudes and phases  of the final classical perturbation modes in orthogonal directions are independent random variables, determined by the initial vacuum state.
A quantized local scalar field does  not allow for coherent, causal, nonlocal entanglement of $\Delta$ at each point among all directions on its past null cone.

The alternative hypothesis considered here invokes a different hypothesis about the   localization of coherent quantum states: the structures that define $\Delta$ at each point are invariant causal diamonds, composed of  null surfaces converging on the point from all directions, instead of plane waves. The  space-time structures of  coherent quantum states in
the two scenarios are illustrated in Fig. (\ref{conformals}). 

The causally-coherent hypothesis is not made up just to explain CMB anomalies. As discussed in the Appendix, it is based on general principles motivated by a wide range of studies. In essence, it extends to geometry a view of causal nonlocality widely adopted for quantum states of matter that are prepared and measured on a timelike world line:
the  entangled quantum states of geometry and matter must be consistent  with  causal, nonlocal relationships demanded by both general relativity and quantum mechanics. This correspondence principle is illustrated in the Appendix by thought experiments with noncosmological  systems such as  EPR-type particle decays (as in Fig. \ref{particles}), black holes, and interferometers.  Because  information about nonlocal  space-time relationships is encoded on 2D null surfaces, a causally-coherent  theory of emergent quantum locality can be  called  ``holographic''. 

 A quantum theory of gravity that incorporates causal nonlocality would lead to radical physical differences from the standard cosmological scenario during inflation, some of which are are  discussed  in the Appendix.  {\it In spite of these differences, the principal observable consequence at the end of inflation,   classical perturbations $\Delta$ with a nearly-scale-invariant 3D power spectrum, is not changed}.\cite{PhysRevD.99.063531,Hogan_2020}

This paper primarily focuses on  phenomenological signatures that could differ conspicuously from the standard picture, associated with universal symmetries of correlations on  spherical surfaces at large angular separations. 
The goal is to determine if there is real-world evidence of causal constraints that may result from
 a  causally-nonlocal   quantum theory of gravity.
}
 
\subsection{Horizons and formation of perturbations}

  As in the standard model, there is a local invariant curvature perturbation $\Delta(r,\vec\Omega)$ on every comoving world line at the end of inflation\cite{PhysRevD.22.1882}.  
In the causally-coherent scenario, the  value of $\Delta$ at each point  is determined by the state of its coherent    inflationary horizon $ {\cal H}$,  the inbound null surface that
arrives  at the end of inflation,  $t_I$.
 The horizon $ {\cal H}$
forms the future boundary of a series of causal diamonds of nearly constant area $ 4\pi (c/H)^2$ during the slow-roll phase,  when the bulk of inflation occurs  and the observed perturbations form, with expansion rate $H$.

As shown in Fig. (\ref{conformals}), and elaborated in the Appendix, surfaces of constant $t$ intersect $\cal H$ in 2-spheres. 
For  observer $\cal O$, its horizon $\cal H_{\cal O}$ defines a series of  spheres of nearly constant physical radius $c/H$ during the slow-roll phase.
The comoving  radius is
${\cal R}(t)= R_0 a_0/a(t)$,  where  $R_0$ denotes the physical radius at time $t_0$ and scale factor $a_0$, so the comoving radius is inversely proportional to the scale factor when a sphere matches the horizon radius.
We refer to these 2-spheres as ``frozen horizons'' for $\cal O$.

 In standard  inflation, perturbations form from the gravitational effect of zero point vacuum fluctuations of the inflaton field $\phi$. The amplitude and phase of a mode of $\phi$ freeze out coherently when its comoving wavelength approximately matches  ${\cal R}(t)$. The process of freezing occurs as a natural cooling process, as the oscillation rate of each mode falls below $H$.
 As shown  in Fig. (\ref{conformals}),  standard inflation does not prepare or collapse these quantum states causally\cite{penrose}.

 Causally-coherent quantum states differ radically from the standard effective quantum field model\cite{Banks2011,Banks:2011qf,Banks:2015iya,Banks:2018ypk,Banks:2020dus,Banks:2021jwj}: 
causal diamonds, not wave modes, now form  the coherent  quantum objects.
As discussed in the Appendix, there is no broad  consensus about the nature or  amplitude of quantum fluctuations in this system, which are associated with the emergence of locality. 
We will assume that their relic perturbations obey causality, but not locality.
Causal symmetries are defined by the classical background, so we do not  need  to know how the fundamental
quantum degrees of freedom of a horizon work in detail.

\begin{figure}
\begin{centering}
\includegraphics[width=\linewidth]{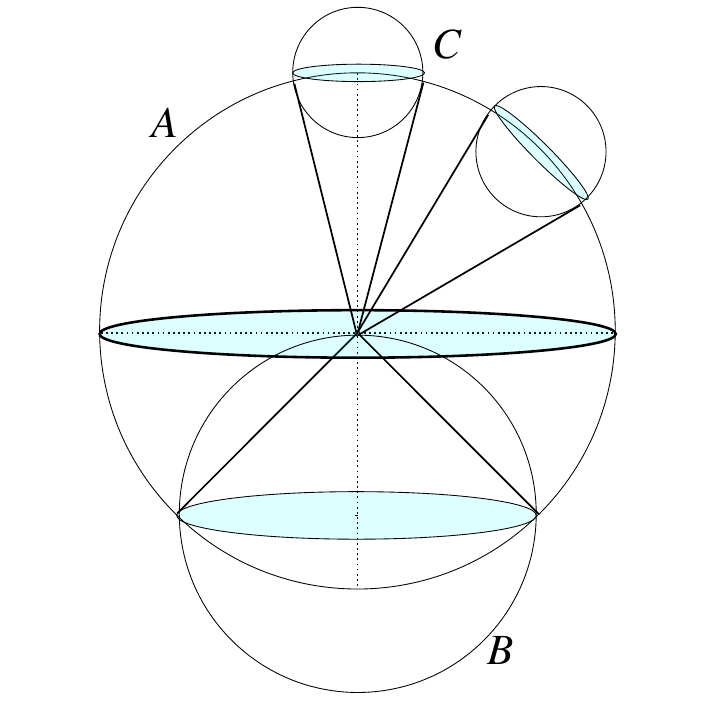}
\par\end{centering}
\protect\caption{3D view of  2D spheres that intersect comoving  horizons of  $A$,  $B$ and $C$. The smaller $B$ and $C$ spheres represent single-time slices of their horizons from a later time during inflation than that of $A$. Correlations on each horizon are controlled by quantum information in their interior.  Small angle relationships are the last to freeze; the final local, classical values of $\Delta$  do not freeze until inflation ends. 
The spheres shown here have particular  exact causal relationships; the  $C$ spheres are centered on the $A$ horizon, and $A$ has its center on  the $B$ horizon.  The  latter is shown with a  $45^\circ$ opening angle on $A$, which has particular causal significance for antihemispherical symmetries.  The shaded ellipses represent great circles, or intersections  with infinitely large spheres.
A universal holographic correlation function  is  constrained by causal  relationships and symmetries of  overlapping spherical horizons.
\label{frozen}}
\end{figure}

Fluctuations from slow-roll inflation create a variance $\langle\Delta^2\rangle$ that depends only on the slowly-varying physical radius of causal diamond boundaries, so the 3D power spectrum is nearly scale-invariant, as in the standard picture. 
However, the  formation of perturbations is not localized on surfaces of constant $t$, but is correlated  on  null surfaces $\cal H$ that   extend throughout the history of inflation. 
The horizons of the different observers are entangled where their interiors  overlap. 
On  each  comoving sphere, the  largest angle relationships  form from the earliest entanglements: they ``freeze  first'', as shown in Fig. (\ref{frozen}) and described in more detail below.  The detailed  pattern of $\Delta$ on each sphere varies with comoving radius and location, but as discussed below,  the angular power spectrum and correlation function vary much less than in the standard picture.



%

\begin{figure}[hbt]
\begin{centering}
\includegraphics[width=\linewidth]{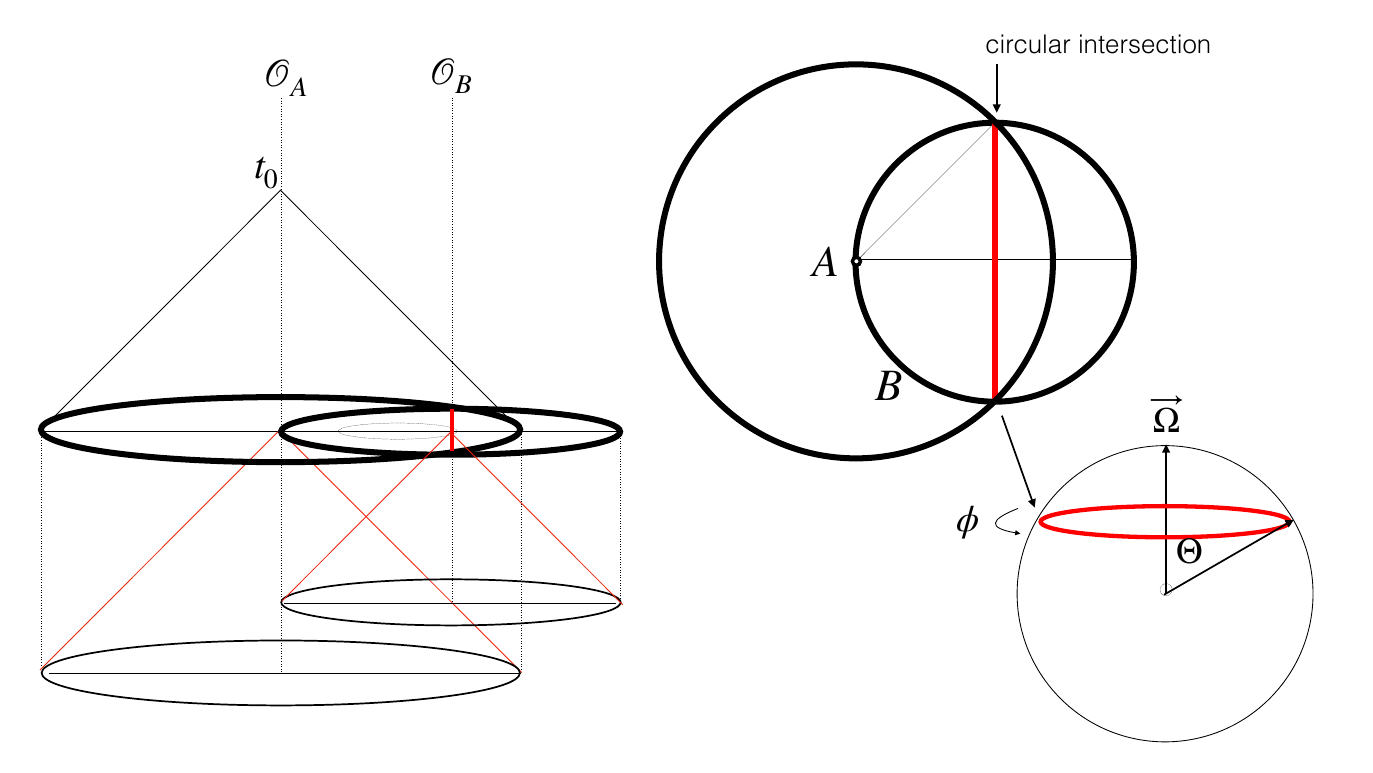}
\par\end{centering}
\protect\caption{4D and 3D views of  intersections of  comoving spherical horizons. 
The center of $A$ lies on the surface of the $B$ horizon, and their projected 2-spheres at the end of inflation represent footprints of null horizons from a particular time.  The relationship with the correlation function is shown at the right:  the mean on the common circle intersected by the horizons  contributes  to the polar angular average in the correlation function, Eq. (\ref{Ctheta}).
The particular $B$ sphere shown is chosen to have a special causal relationship with $A$: their intersection lies on a great circle of $B$. It defines a boundary of an exact causal symmetry, as described below. 
\label{spheres}}
\end{figure}

  \subsection{Correlations on  spherical horizons}\label{definitions}
  
The primordial pattern of invariant scalar curvature $\Delta(r,\Omega)$ at the end of inflation is still intact on the largest scales  today, and dominates the observed pattern of anisotropy in CMB temperature.
In the current analysis, we assume  the Sachs-Wolfe approximation\cite{1967ApJ...147...73S} where correlations of temperature and curvature are equivalent, as discussed further below.

The pattern  of  $\Delta(r,\Omega)$ for a sphere of radius $r$ 
can be described equivalently as a correlation function in the angular domain or as  a power spectrum of spherical harmonics (Fig. \ref{data}).
  The  two-point correlation function is defined as an all-sky average  over all pairs of points $a,b$  at angular separation $\angle ab = \Theta$:
\begin{equation}\label{corrdef}
C(\Theta) = \langle{\Delta}_a {\Delta}_b \rangle_{\angle ab = \Theta}.
\end{equation}
It can also be written as an angular average,
 \begin{equation}\label{Ctheta}
C(\Theta) = \langle {\Delta}(\vec\Omega)   \langle {\Delta} \rangle_{\Theta,\vec\Omega}  \rangle_{\vec\Omega},
\end{equation}
where $ \langle \rangle_{\Theta,\vec\Omega}$  denotes an azimuthal mean on a circle at a  polar angle $\Theta$ about direction $\vec\Omega$.
This form is helpful for showing how frozen spherical horizons should lead to symmetries of $C_{\Delta}(\Theta)$ (as in Fig.
\ref{spheres}).

 In the the spectral domain, the distribution is decomposed into spherical harmonics
$Y_{\ell m}(\theta,\phi)$:
\begin{equation}\label{decompose}
\Delta(\theta,\phi,r)=
\sum_\ell \sum_m Y_{\ell m}(\theta,\phi)  a_{\ell m} (r).
\end{equation}
The angular power spectrum, determined by the spherical harmonic coefficients $a_{\ell m}$,
\begin{equation}\label{powerpiece}
C_\ell= \frac{1}{2\ell+1}
\sum_{m= -\ell}^{m=+\ell} | a_{\ell m}|^2,
\end{equation} 
is related to the two-point angular correlation function $C(\Theta)$ by  the  standard formula  (e.g., \cite{WMAPanomalies,Schwarz:2015cma,Hinshaw_1996})
 \begin{equation}\label{harmonicsum}
 C(\Theta) = \frac{1}{4\pi}\sum_\ell (2\ell +1) { C}_\ell P_\ell (\cos \Theta),
\end{equation}
where $P_\ell $  are the Legendre polynomials. 
Anisotropy has  odd and even parity components, with $P_\ell(0) = 0  $  for odd  $\ell$, $P_\ell(0) \ne  0 $ for even $\ell$.

 \subsection{ Universal holographic angular correlation}

 Arguments  reviewed  in the Appendix suggest that  consistency of  general relativity and quantum mechanics requires  quantum states of a whole system  to  be coherent on null surfaces. 
 In a covariant formulation of  holographic quantum space-time\cite{Banks:2020dus,Banks:2021jwj}, the coherent quantum elements are causal diamonds, each defined by the unique  light cone pair associated with a world line interval.
 For a black hole, the coherent null surface is the event horizon.
 For correlations with  a point,  the coherent null surfaces are light cones centered on the point.
 
 A quantum theory of space-time  predicts a definite correlation function for operationally defined measurements.
  A simple example  discussed in the Appendix is  the decay of a particle into two null particles. In the linear, weak-field limit, it can be proven that 
 the  gravitational  wave function, a superposition of metric distortions measured by clocks on the spherical boundary of a causal diamond, has a universal angular power spectrum\cite{mackewicz2021gravity}.
 
 
 In  causally-coherent inflation, the null surfaces are  inflationary horizons.
 The ``measurements'' in the cosmological case are comparisons of scalar curvature $\Delta$ among different world lines at the end of inflation.
  All information about  relationships of $\Delta(A)$ with other world lines is contained within its inflationary horizon, ${\cal H}_A$.  
  
In the (unphysical) limit of a maximally symmetric, scale-invariant cosmology--- the infinitely slow-roll,  quasi-de Sitter limit of inflation---   coherent causal diamonds are all identical 
quantum systems,   prepared and measured in the same way.
All locations and directions are equivalent, so the two-point correlation function of fluctuations on the surface of a causal diamond should not depend on the location of its world line.

These considerations motivate  a concrete universality  conjecture about  scalar curvature perturbations associated with an idealized,  maximally symmetric, scale-free  system in holographic quantum gravity:
{\it   All spheres  have the same   angular power spectrum ${ C}_{\ell}$.} 
This universality adds  a holographic constraint consistent with  cosmological homogeneity and isotropy, the statistical equivalence of all locations and directions. 
As in standard cosmology, statistical homogeneity and isotropy imply a universal 3D power spectrum $\Delta^2_k$ on infinite spacelike hypersurfaces that depends only on the magnitude of $k$. A maximally symmetric holographic system also has a universal   
2D power spectrum $C_\ell$  on the compact surfaces of causal diamonds that  depends only on the magnitude of $\ell$.


The use of maximal symmetry as a foil is akin to the use of (unphysical) eternal black holes to analyze  symmetries and correlations created by coherent horizons\cite{Hooft:2016cpw,Hooft:2016itl,Hooft2018}. 
In this paper, we consider a universal angular correlation function  as an  approximation, to analyze the differences between  the holographic picture and standard quantum inflation, especially on large angular scales.

The  standard picture predicts that  projected onto any horizon sphere, there is a substantial cosmic variance in $C(\Theta)$  at any large angle, from independent, previously-frozen 3D modes larger than the horizon.
By contrast, exact causal constraints on $C(\Theta)$  impose correlated constraints on ${ C}_\ell$'s at all $\ell$.   In general,  a  universal causal constraint that extends over a range of angles
  requires significantly reduced variance in ${ C}_\ell$, especially at low $\ell$.
For actual measurements of CMB anisotropy, the two pictures lead to simlar predictions   at 
  small angular scales
($\ell\gtrsim 30$),  where the constrained angular phase information is scrambled by the effects of
 3D  baryon motion on anisotropy.


 Even for exact universality,  the detailed pattern of $\Delta$ is not identical on all horizons. In the harmonic expansion  (Eq. \ref{decompose}), the $a_{\ell m}(r)$ are not the same for all $r$, or for all choices of origin.     Up to resolution $\ell$, the power spectrum is specified by $\ell$ numbers (the $C_\ell$'s), but the details of the  pattern require specification of $2\ell+1$  numbers for   each $\ell$, the phase information represented by  the $a_{\ell m}$ harmonic coefficients.
This variation occurs because while all causal diamonds are the same kind of quantum system, they have different  information on their 2D boundary, which defines the relationship of each diamond with the rest of the universe. The differences between  spheres (that is,  skies around different observers) arise because of  different preparation of  states on the surface of each diamond.
 
If we assume that this picture applies to all causal diamonds down to the   Planck length, 
 the information in any causal diamond is consistent with holographic gravity.  The  number of degrees of freedom is  the number of modes up to some resolution scale $\ell_{max}$, which is about  $\ell_{max}^2$.  
 The  number of degrees of freedom of an inflationary horizon of radius $c/H$ agrees  with the Bekenstein-Hawking entropy $S_{BH}= \pi (R/ct_P)^2 $ of a black hole horizon of the same radius  $R=c/H$, if the maximum resolution $\ell_{max}\sim  R/ct_P$  is determined by a cut off at 
 the Planck length $ct_P= \sqrt{\hbar G/c^3}$. 
In other words,  {\it the entropy of an inflationary horizon with universal angular correlation  matches the thermodynamic entropy of a black hole event horizon}, if both are regarded as  coherent quantum systems, with a minimum duration for coherent causal diamond states  equal to the  Planck time. They have the same number of different possible configurations, and the same finite number of states in a discrete Hilbert space. 

 For comparison, the information in standard, independent   field states in the  volume contained by the horizon with a UV cutoff at the Planck scale is much larger, of order $(R/ct_P)^3 $.  In black holes, the extra  information exceeds the thermodynamic entropy, which leads to well-known paradoxes.  In standard cosmology, the extra information is the source of the predicted cosmic variance in the statistical properties of different horizons. 
If  horizons  are indeed causally coherent, this apparent extra information  is unphysical, because it is derived from an approximation that does not correctly account for causal constraints on coherent quantum states.


Holographic universality implies new and surprising symmetries of angular correlations.
Because  $C_\ell$ is the same for all spheres around a given point, the value of the universal monopole   $C_{\ell=0}$, the difference of the mean from the center, is the same for a sphere of any size.  As in the standard picture, this difference  approaches zero as the radius ${\cal R}$ of the   sphere shrinks to a point, ${\cal R}\rightarrow 0$; but unlike the standard picture, the value does not change with ${\cal R}$, so it must vanish for all ${\cal R}$.
 For a scale-invariant universal function, this requires that the  monopole harmonic  $C_{\ell=0}$ vanishes for all spheres, that is,  {\it the average perturbation $\langle\Delta\rangle$ on any sphere is the same as $\Delta$ at its center}. 
 
 This property  exemplifies  one apparent conspiracy required  for holographic universality. For a coherent horizon, such conspiracies are consistent with causality by construction, since the spheres all intersect the same null surface, the inflationary horizon of the center.

 



\begin{figure*}[t]
\begin{centering}
\includegraphics[width=.45\linewidth]{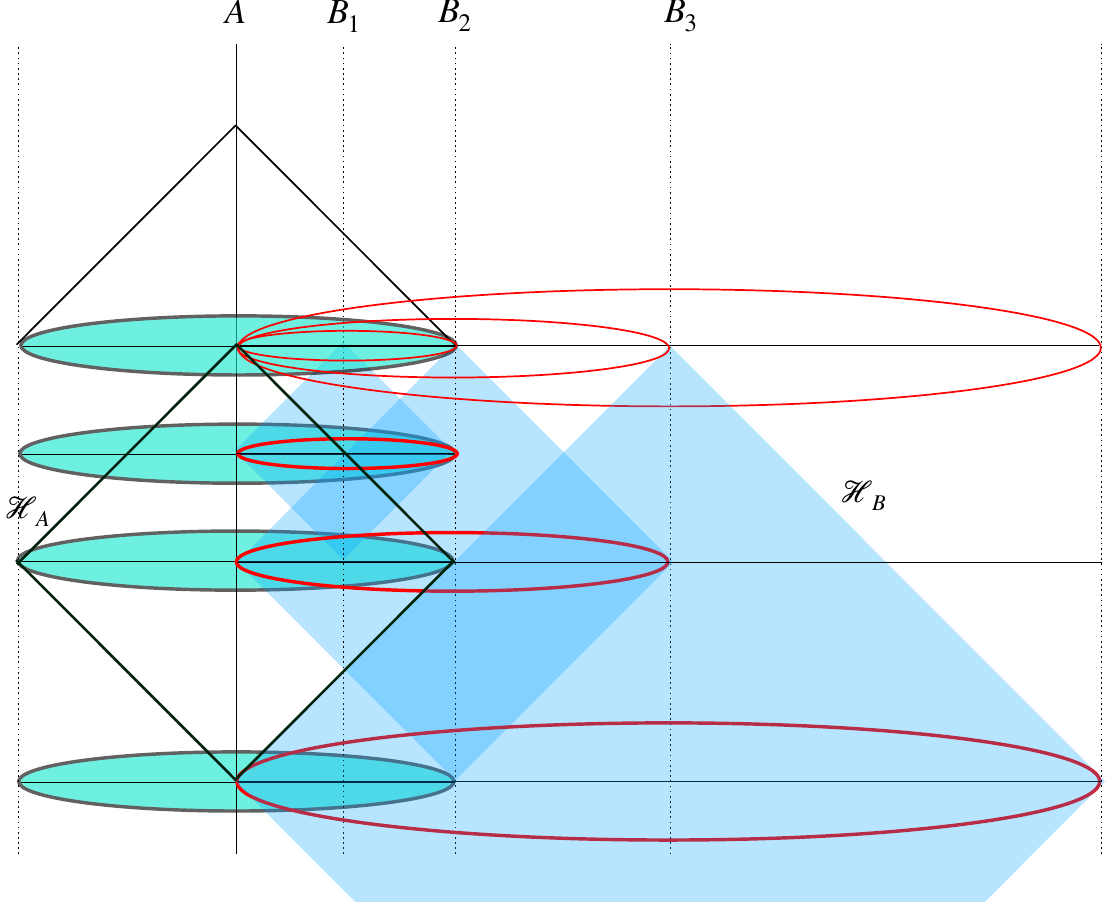}
\includegraphics[width=.49\linewidth]{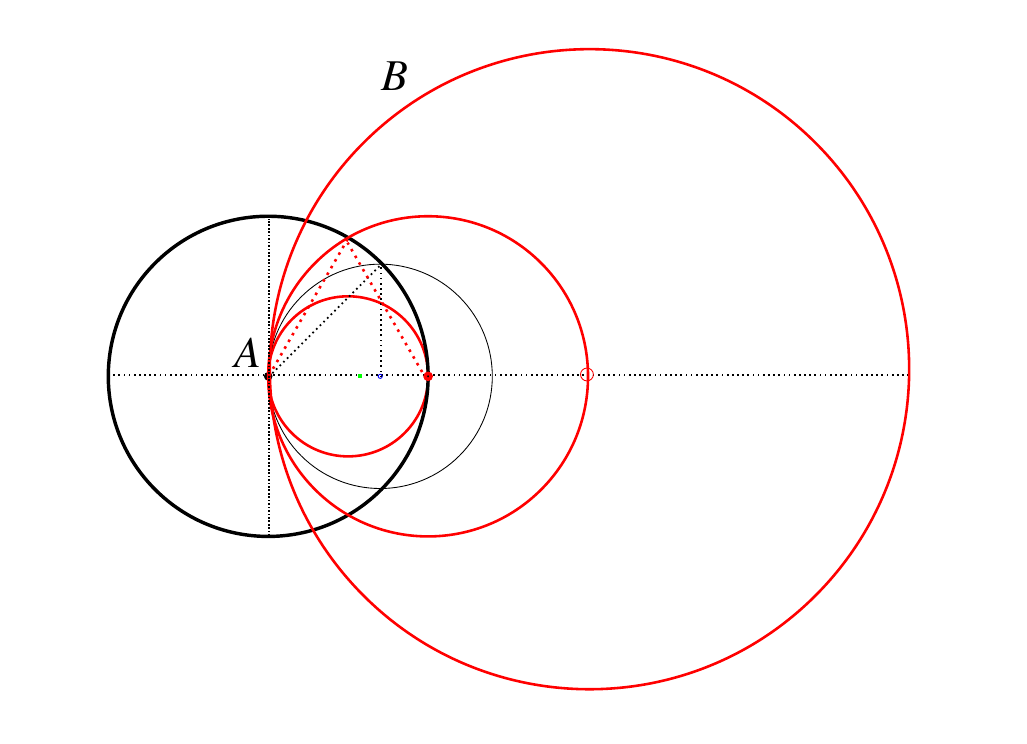}
\par\end{centering}
\protect\caption{At left, a causal diagram shows the cylinder swept in time by a comoving sphere of  world lines that bound a particular causal diamond of $A$ that ends at the end of inflation. Causal diamonds are shown for several other   world lines $B_1, B_2, B_3$ along a single axis, for which $A$ lies on their bounding   surfaces; for these examples,  ${\cal R}_B/{\cal R}_A= .5, 1$ and $2$. The surfaces bound  regions of space-time causally connected with  $A$  along the $AB$ direction, and domains of angular correlations associated with the polar angle of the intersection.   At right,  slices of their  comoving 2D spherical bounding surfaces are shown at the end of inflation.  Dotted lines show  angles of their circular intersections with the $A$ sphere  for special angles from the $AB$ direction,   $\Theta=\pi/3$
and $\Theta=\pi/4$.  
  \label{ABcausal}}
\end{figure*}

\begin{figure*}[t]
\begin{centering}
\includegraphics[width=.45\linewidth]{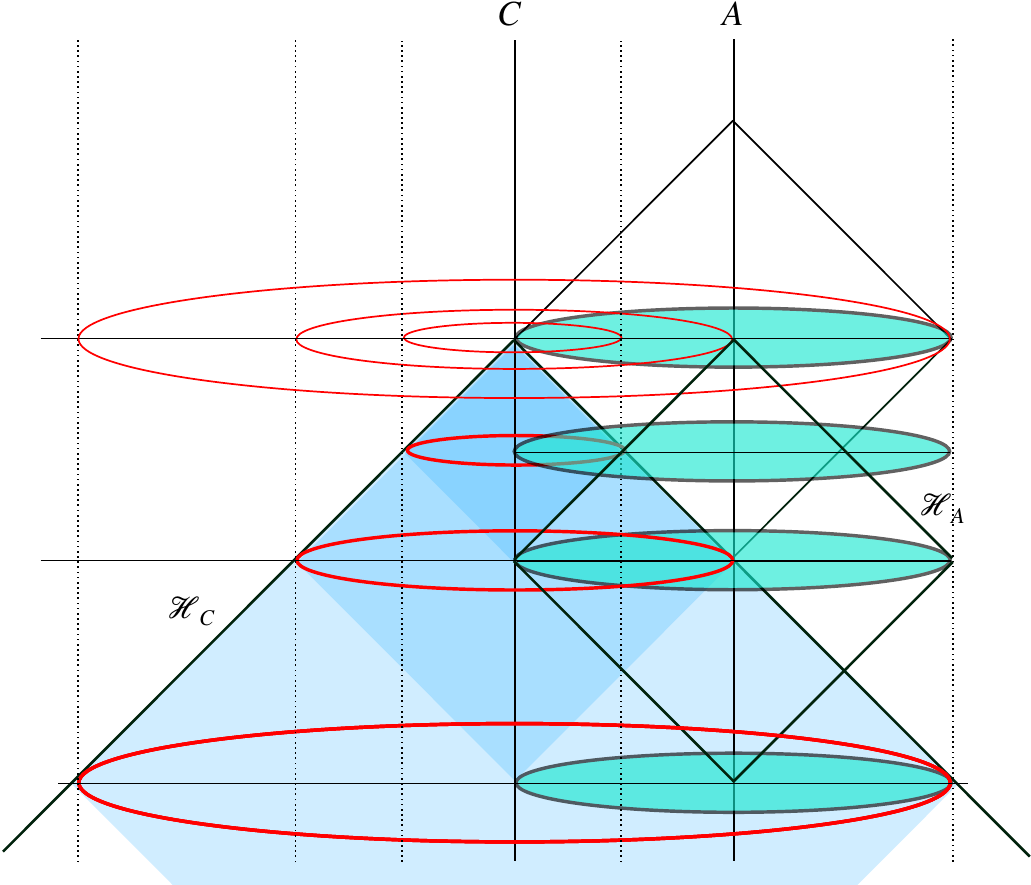}
\includegraphics[width=.49\linewidth]{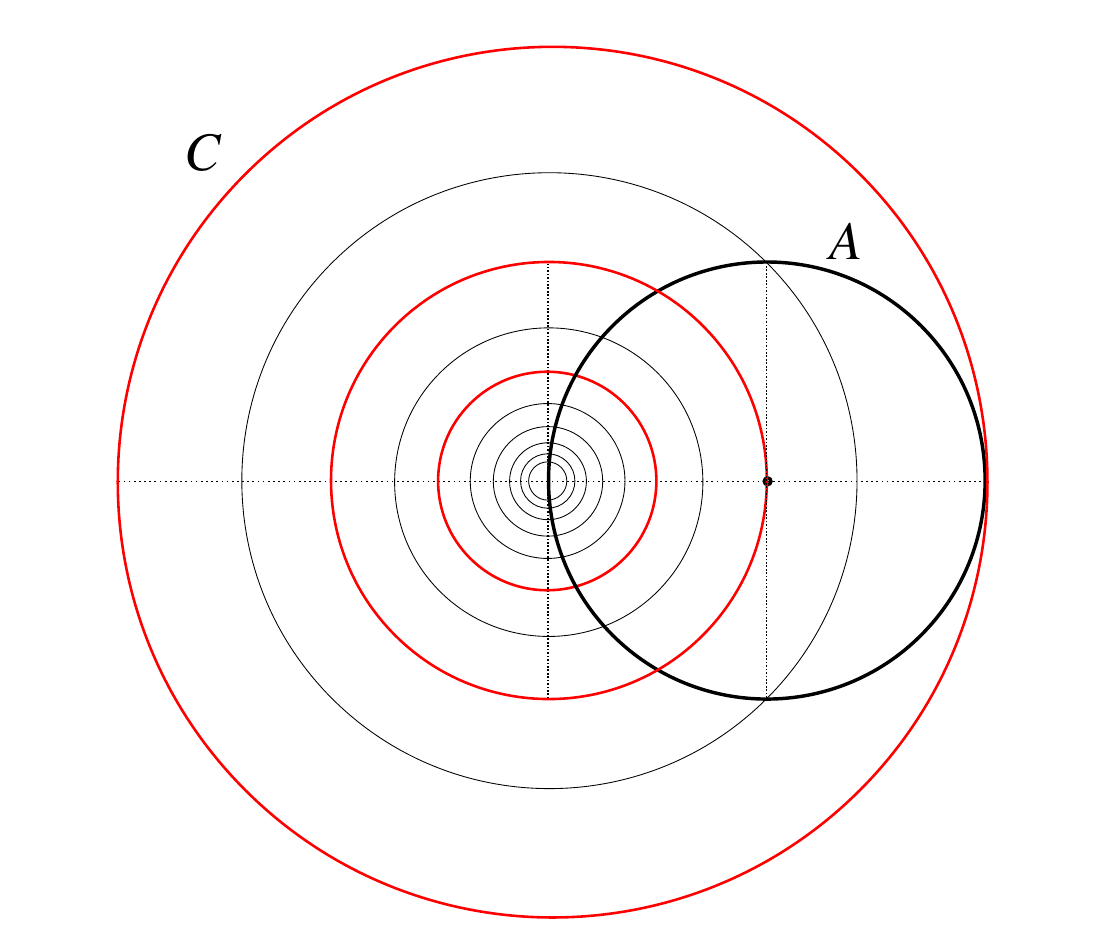}
\par\end{centering}
\protect\caption{At left, a causal diagram shows  the inflationary horizon of a world line $C$ that lies on a spherical surface centered on $A$, and  bounding surfaces of some of its embedded causal diamonds.  At right,  slices of these comoving bounding surfaces are shown at the end of inflation. 
The value  of $\Delta$   on the $A$ surface  in the $C$ direction is causally connected with other points in the shared interior of causal diamonds. 
The  series of $C$ surfaces is shown on  a linear scale,  with constant logarithmic spacing in factors of $\sqrt{2}$. 
The largest shown here, with ${\cal R}_C= 2{\cal R}_A,$ is also the earliest to freeze out, and the largest to causally entangle with  angular correlations on the $A$ surface. A similar sequence continues to smaller scales over the last $\sim 60$ $e$-foldings of inflation. 
  \label{ACcausal}}
\end{figure*}



 \begin{figure}
\begin{centering}
\includegraphics[width=\linewidth]{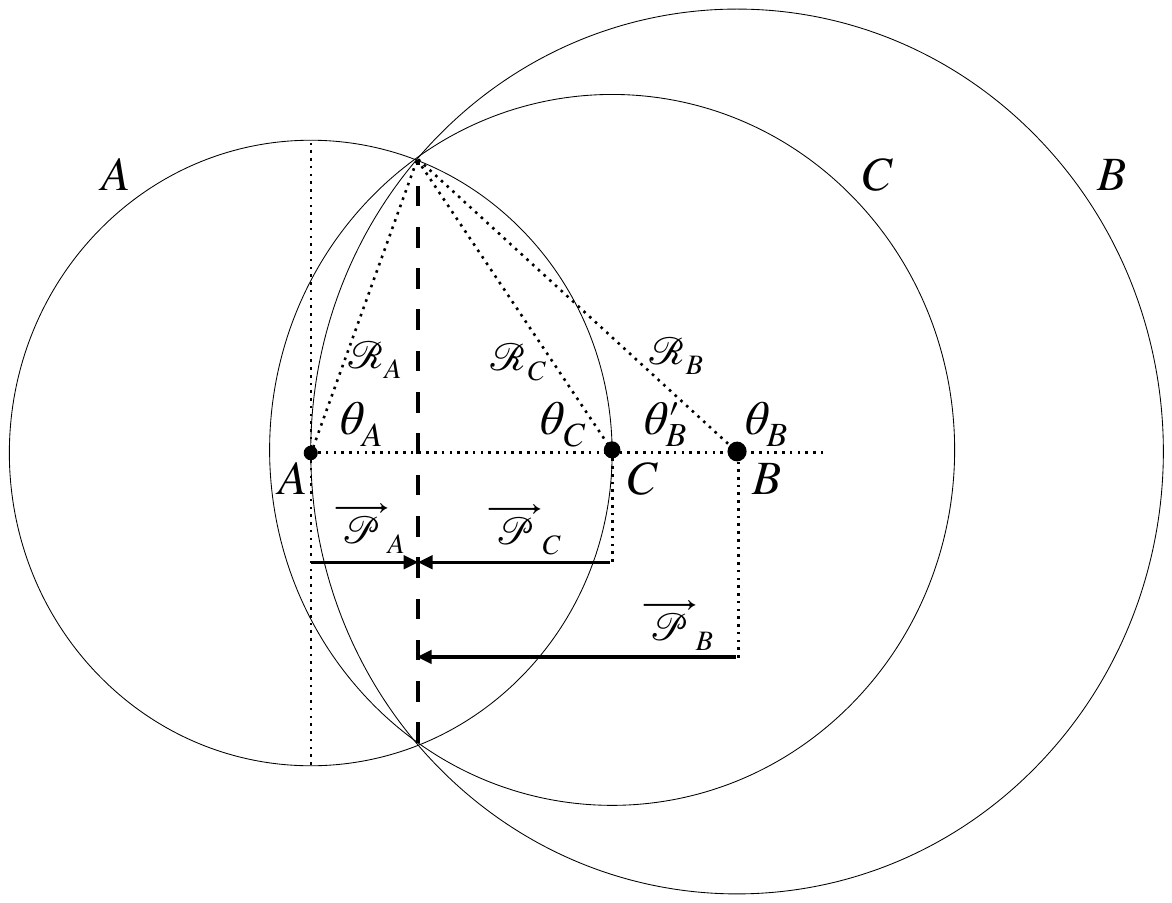}
\par\end{centering}
\protect\caption{ A section of three comoving spheres,   labeled by their centers $A,B,$ and $C$ along a single axis. 
Definitions of angles $\theta_{A,B,C}$,  comoving radii ${\cal R}_{A,B,C}$, and  polar axis projections $\vec{\cal P}_{A,B,C}$ are shown for  their common circular intersection, shown as a dashed line.
Each circular mean contributes to the correlation function on $A$ and $B$  at angles $\theta_A$ and $\theta_B= 2\theta_A$ respectively with respect to the polar axis defined by $AC$. The complementary angle $\theta'_B= \pi-\theta_B$  applies for an antihemispherical, parity-reversed twin of $B$.
  \label{trig}}
\end{figure}

 \begin{figure}
\begin{centering}
\includegraphics[width=\linewidth]{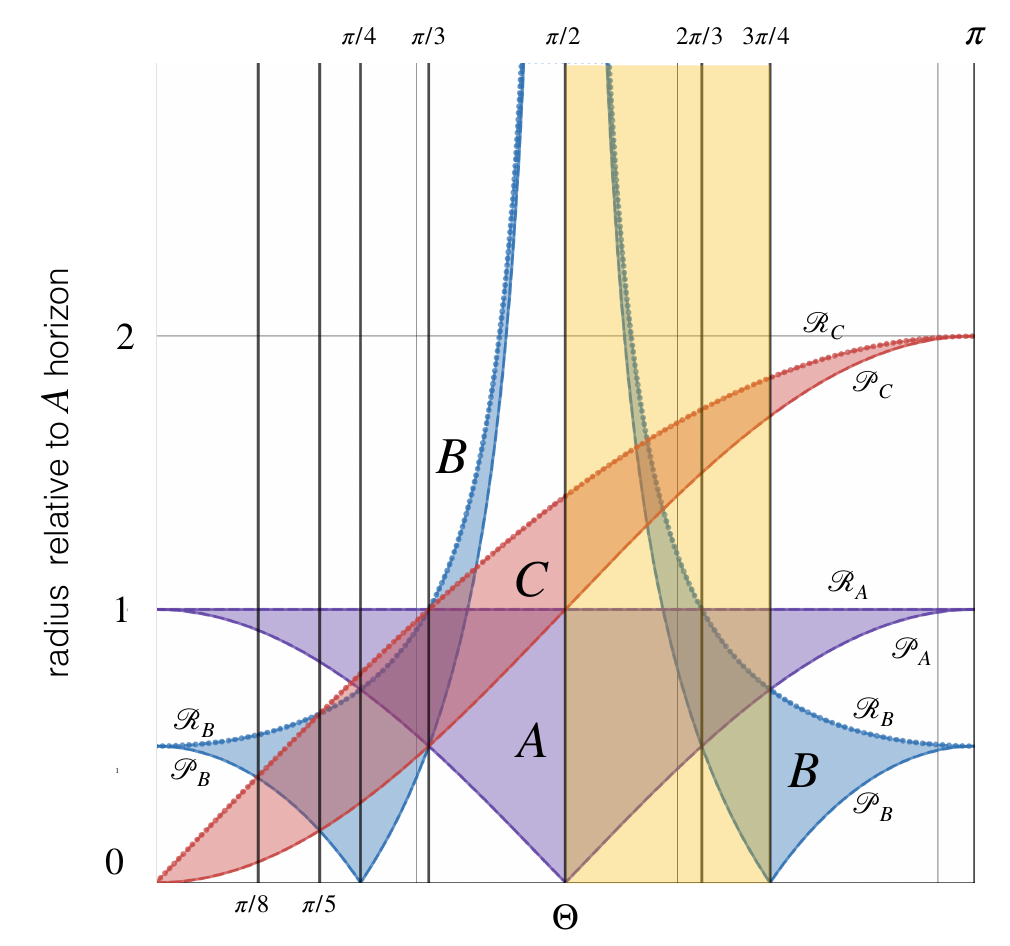}
\par\end{centering}
\protect\caption{ Comoving sizes of horizon radii ${\cal R}_{ABC}$ and polar-axis projections ${\cal P}_{ABC}$, as a function of polar angle $\Theta$. Units are linear,  relative to the freeze-out of the $A$ horizon.  On this plot, the history of inflation maps onto the range of polar angles $0<\Theta<\pi$ for observer $A$, and the curves represent distances to the quantum-classical boundary where horizons intersect. A sphere that freezes out at the  end of inflation  has a radius $\sim e^{-60}$ on this scale if $A$ matches the size of our  horizon.
Shaded regions bound causally entangled information that  affects  angular correlations. The two branches of  $B$ also map onto  complementary, twin $B$ spheres in  opposite hemispheres, with opposite parities.
An animated scan showing a cross section of the spheres and their intersections over time can be found at \href{https://www.desmos.com/calculator/qhqkzntxsg}{\underline{this hyperlink}}.
\label{ABCradii}}
\end{figure}

\subsection{Polar angles of intersecting horizons}



To translate a bulk  causal constraint into constraints on universal angular correlations, 
we introduce a geometrical construction to characterize polar angles of intersections of  causal diamonds of different world lines. 

 The nested comoving causal diamond surfaces around every  world-line resemble a layered ``onion'' of     concentric 2-spheres that represent the locations of its frozen horizon from different times during inflation. 
 Consider relationships of   points $A,B,C$, each one  associated with its own layered onion.
As shown in Figs.  (\ref{ABcausal}, \ref{ACcausal}), we focus on  particular relationships for which
the  center of $A$ lies  on the $B$ horizon, and the
center of $C$ lies on  the $A$ horizon. 
These configurations are chosen because
  $B$ spheres bound information related to the center of $A$, while  $C$ spheres bound  information related to the boundary of $A$. 
Causal boundaries of angular correlations are defined by  intersections of $ABC$ causal diamonds, as a function of  polar angle from a common $ABC$ axial direction.
 
 Thus, causal constraints on  $C(\Theta)$ are governed by   families of $A,B,C$ spheres along a single axis, with a common intersectional circle
 (Fig. \ref{trig}).
Denote the angles subtended by the common circle with respect to the common axis from each  of the centers  by $\theta_A,\theta_B,\theta_C$.  
The comoving radii as a function of angle are related by
\begin{equation}\label{RB}
{\cal R}_B={\cal R}_A/2\cos(\theta_A)
\end{equation}
and
\begin{equation}\label{RC}
{\cal R}_C=2{\cal R}_A\cos( \theta_C)= 2{\cal R}_A\sin(\theta_A/2).
\end{equation}
The projected axial separations of centers from the circle plane  are   
\begin{equation}
{\cal P}_A={\cal R}_A \cos(\theta_A),
\end{equation}
\begin{equation}
{\cal P}_B={\cal R}_B\cos(\theta_B)={\cal R}_A \cos(2\theta_A)/2\cos(\theta_A),
\end{equation}
and
\begin{equation}\label{Ccirclecos}
{\cal P}_C={\cal R}_C\cos(\theta_C)=  2{\cal R}_A \sin^2(\theta_A/2).
\end{equation}

Scaling to a universal correlation on the $A$ horizon, we  adopt the notation $\Theta$ instead of $\theta_A$ to refer to the angular separation.
Each value of $\Theta$ maps  onto comoving freeze-out radii in units of ${\cal R}_A$,  as shown in Fig. (\ref{ABCradii}).
The inverses of these  give the corresponding scale factor where they match the $A$ horizon.   
 Scaled projections of intersecting circles onto  $B$ and $C$ spheres along the polar axis are: 
\begin{equation}\label{Aratio}
\frac{\cal P_A}{\cal R_A}= \cos(\Theta),
\end{equation}
\begin{equation}\label{Bratio}
\frac{\cal P_B}{\cal R_B} = \cos(\theta_B)=\cos(2\Theta),
\end{equation}
and
\begin{equation}\label{Cratio}
\frac{\cal P_C}{\cal R_C}=\cos(\theta_C)=\sin(\Theta/2).
\end{equation}
 The  projection  ${\cal P}$  controls the relationship between polar values  and azimuthal circular averages, and  hence the  angular correlation function. These angular  relationships form the basis of causal constraints on symmetries of  ${ C}(\Theta)$.

 \section{Antihemispherical causal constraints}

\subsection{Vanishing correlation in the information shadow}

We now show that causality requires the angular correlation of curvature on spheres at the end of inflation  to vanish  over an antihemispherical interval of angles:
\begin{equation}\label{exact}
 { C}_\Delta(\pi/2<\Theta<3\pi/4) = 0.  
\end{equation}
The  rationale for this exact symmetry
is based on causal constraints on  information entangled among  causal diamonds.

The classical  scalar  $\Delta$ refers to a scalar perturbation on a classical background.   For causal correlations on the null surface of a horizon, quantum  information about  $\Delta$ is  directional in relation to  the apex of  a null cone.   It does not become a localized classical scalar until  the end of inflation, after which it becomes measurable. 


Information, bounded by causality, produces correlations  inside causal diamonds.   A radial null trajectory that starts on the surface of a sphere at the moment its radius matches the  horizon arrives at the center only at the end of inflation. The same is true for points on the great circle normal to any polar direction:
 there is no opportunity for information from a plane tangent at a pole  to causally affect correlations of $\Delta$  on its great circles.
Since  values of $\Delta$ on any great circle are independent  of  polar values, it follows\cite{PhysRevD.99.063531,Hogan_2020,Hagimoto_2020} that  ${ C}(\pi/2)=0$.

More generally, causal constraints on directional information flow at $\Theta>\pi/2$ are bounded by  colinear $A$, $B$, and $C$ horizon spheres (Fig. \ref{trig}).
Recall that {\it  $B$ spheres bound the flow of information to and from the center of $A$, while  $C$ spheres bound the flow of information to  points on the boundary of $A$.}
The $C$ spheres along the $ABC$ axis bound   regions of entangled information for an apex $C$ on the 
horizon of $A$, while the $B$ spheres  bound regions of  information in relation to the  $A$ center entangled with the   $C$ axis. Comoving radii ${\cal R}_{ABC}/{\cal R}_A$ and radial projections ${\cal P}_{ABC}/{\cal P}_A$ of their intersections with $A$,
 as a function of its angular radius  (Fig. \ref{ABCradii}),
 define causal boundaries of entangled information in the angular domain  observed by $A$.

\begin{figure}
\begin{centering}
\includegraphics[width=\linewidth]{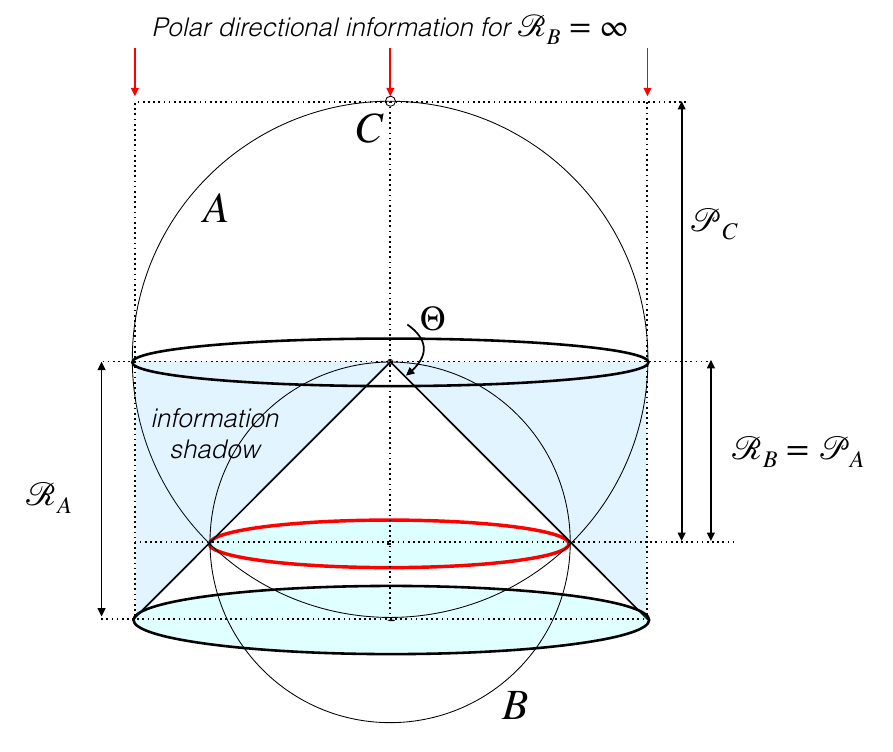}
\par\end{centering}
\protect\caption{A 3D view of frozen 2D horizon spheres  at the end of inflation, as in Fig. (\ref{frozen}). 
The spheres shown are frozen horizons that  intersect at  $ \Theta=3\pi/4$, with a causal history shown in Fig. (\ref{causal135}). The geometrical significance of  antihemispherical correlations   is illustrated  by  gnomonic projection of circles of  the 2D celestial (horizon) sphere onto an antipolar tangent plane in the emerged 3D space.  
  The  $C$  center lies on the polar axis at the top in this figure, with $\Theta$ defined as indicated. 
The lower  hemisphere at $\Theta>\pi/2$ lies in an  ``information shadow'' where all of the  information from the polar $\Theta=0$ direction (normal to an infinitely distant $B$-sphere horizon) has been ``absorbed'' into the mean potential $\langle\Delta\rangle$ of the $A$ sphere when it freezes out to  its central value, as measured by an infinitely distant observer in that hemisphere. 
\label{gnomic}}
\end{figure}

It is helpful to visualize constraints on information arriving  from any polar direction (say, from the top as in Fig. \ref{gnomic}) as an ``information shadow'' in the opposite hemisphere, at $\Theta>\pi/2$.  In the emerged 3D space, the potential  ``in the polar direction'' refers to  an infinitely  distant $B$ sphere whose horizon is an equatorial plane. Relative to a  very distant world line,
correlations at $\Theta>\pi/2$   are ``absorbed'' as coherent perturbations  of the whole  $A$ sphere.

\bigskip
\begin{figure}
\begin{centering}
\includegraphics[width=.8\linewidth]{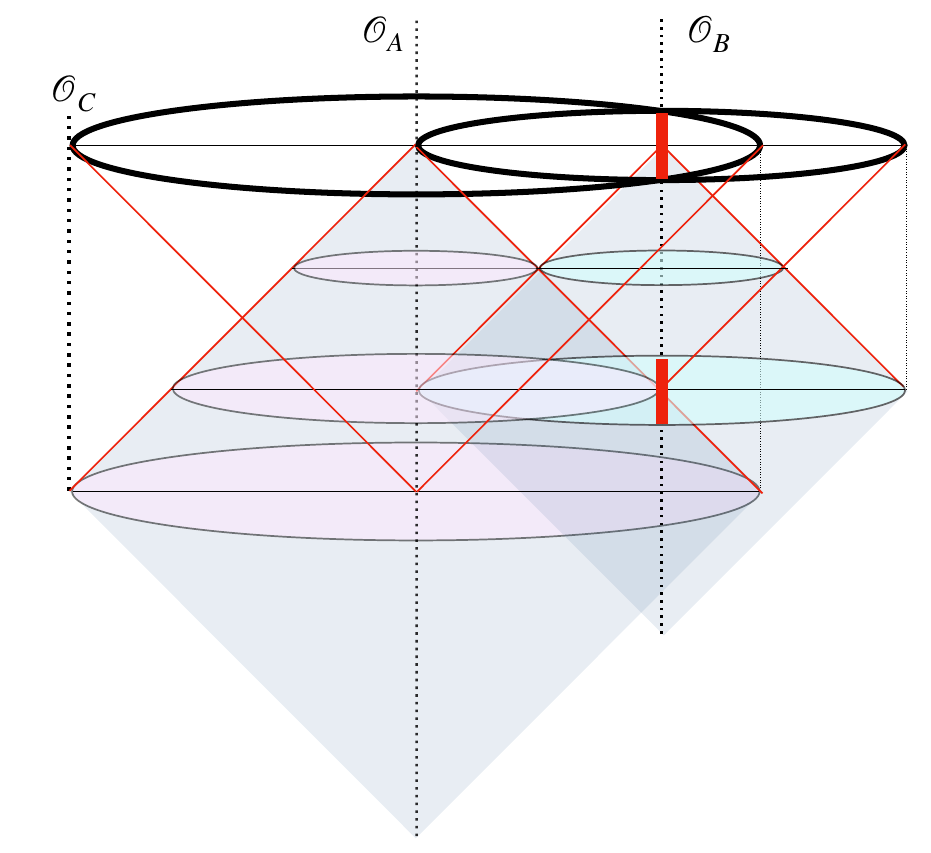}
\par\end{centering}
\protect\caption{ 4D view of causal structure of horizons for   $\Theta= 3\pi/4$, corresponding to the 3D arrangement in Fig. (\ref{gnomic}). The upper plane is the end of inflation, and 
the fiducial polar axis direction  is at the left. The bold bars indicate the $AB$ intersection,  a great circle on $B$ with ${\cal R}_B={\cal P}_A.$
The range $\pi/2<\Theta< 3\pi/4$,  where ${\cal R}_B>{\cal P}_A$, lies in the antihemisphere for both $A$ and $B$, so no axially correlated information reaches 
their intersectional circle.
 At $\Theta>3\pi/4$, 
axial information from the antipodal direction  leads to negative correlation.
  \label{causal135}}
\end{figure}

Figure \ref{ABCradii} includes curves for the two families of $B$ spheres, those from $\Theta<\pi/2$ and those from $\Theta>\pi/2$, that lie in opposite hemispheres, along an axis defined by $C$,  with opposite signs of 
$\vec {\cal P}_C \cdot \vec {\cal P}_B$ at each ${\cal R}_B$.
For a range of angles that is antihemispheric for both the $A$ and $B$ spheres,
\begin{equation}
\pi/2\ <\Theta<3\pi/4 \ \ \leftrightarrow\  \pi >  \theta_B > \pi/2,
\end{equation}
the $ABC$ intersection circle  lies between  $A$ and $B$:
\begin{equation}\label{shadowcondition}
    {\cal R}_B> {\cal P}_A.
\end{equation}

This condition defines the information shadow for angular correlations.
The correlation function is controlled by  information propagating in the axis direction  that entangles the $AB$  centers with the $ABC$ intersection circle: 
{ \it Causal entanglement does not occur for antihemispherical angles where the
 $AB$ separation  is larger than the
 axial distance of $A$ to the $ABC$ circle plane (Eq. \ref{shadowcondition}), 
 so  correlations exactly vanish for  $\pi/2<\Theta<3\pi/4$  (Eq. \ref{exact}).}
 A  correlation function with this symmetry is compatible with  any information  arriving from ${\cal R}_B=\infty$; correlations vanish for both $A$ and $B$.

\begin{figure}
\begin{centering}
\includegraphics[width=\linewidth]{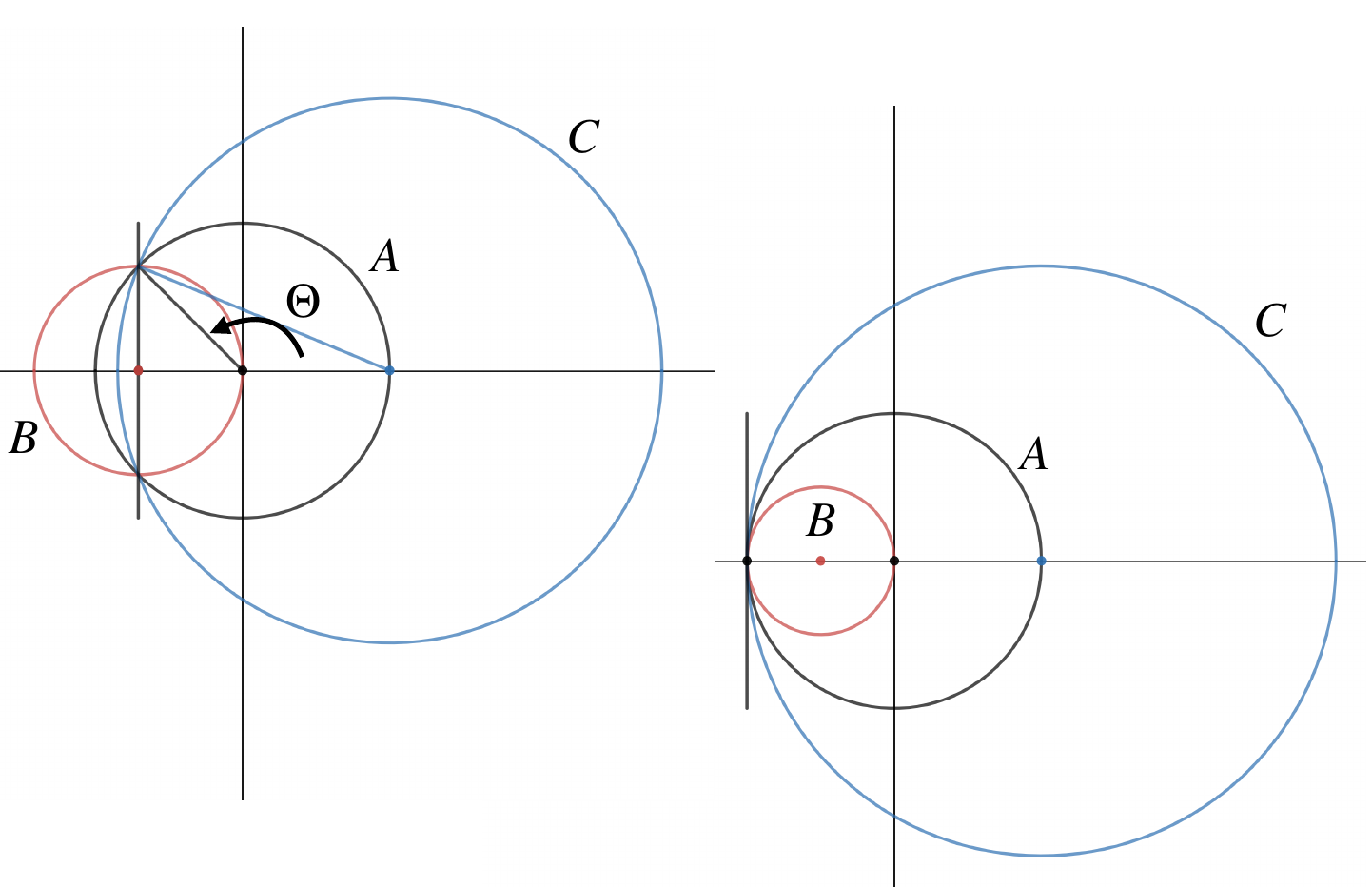}
\par\end{centering}
\protect\caption{Axial slice through the $ABC$ horizon spheres for $\Theta=3\pi/4$ and $\Theta=\pi$.  
      Antihemispherical anticorrelation  arises at $\Theta>3\pi/4$, where   ${\cal R}_B<{\cal P}_A.$
In this range of angles, anticorrelation arises from axially correlated information already inside the  $A$ horizon surface when it freezes out.  
  \label{135}}
\end{figure}

\subsection{Antihemispherical anticorrelation}

On the boundary of the shadow region at $\Theta = 3\pi/4$,  a null signal can just reach the $A$ center from the $B$  equator, which is also the $AB$ intersection, by the end of inflation.
 The  3D configuration 
is shown in Fig. (\ref{gnomic}), and the causal diagram in Fig. (\ref{causal135}).
Inside the shadow,  there is no causal opportunity for entanglement to create correlations.

Closer to the axis, 
\begin{equation}\label{largeABmap}
3\pi/4\ <\Theta<\pi \ \ \leftrightarrow\  \pi/2 \ >  \theta_B> 0,
\end{equation} 
causal bounds on  information change.
Here, in contrast to the shadow condition (Eq. \ref{shadowcondition}), we have
\begin{equation}
    {\cal R}_B<{\cal P}_A.
\end{equation}
The 
 $A$ and $B$ centers now lie on the same side of their intersectional plane (Fig. \ref{135}),
 and $\theta_B$ has wrapped around so that it again has a value less than $\pi/2$, so it no longer has an antihemispherical relationship to the intersection circle.
The $AB$ separation is now less than ${\cal P}_A$, so
 {\it nonzero causal correlations  in the antipolar region can arise from axially correlated information  that is  already inside the $A$ horizon when its mean value freezes to match the $A$ center}. 
In this range of angles,  overlap of $A$ and  $B$ horizons  allows directional information to causally entangle  among  circum-antipolar directions.

\begin{figure}
\begin{centering}
\includegraphics[width=.7\linewidth]{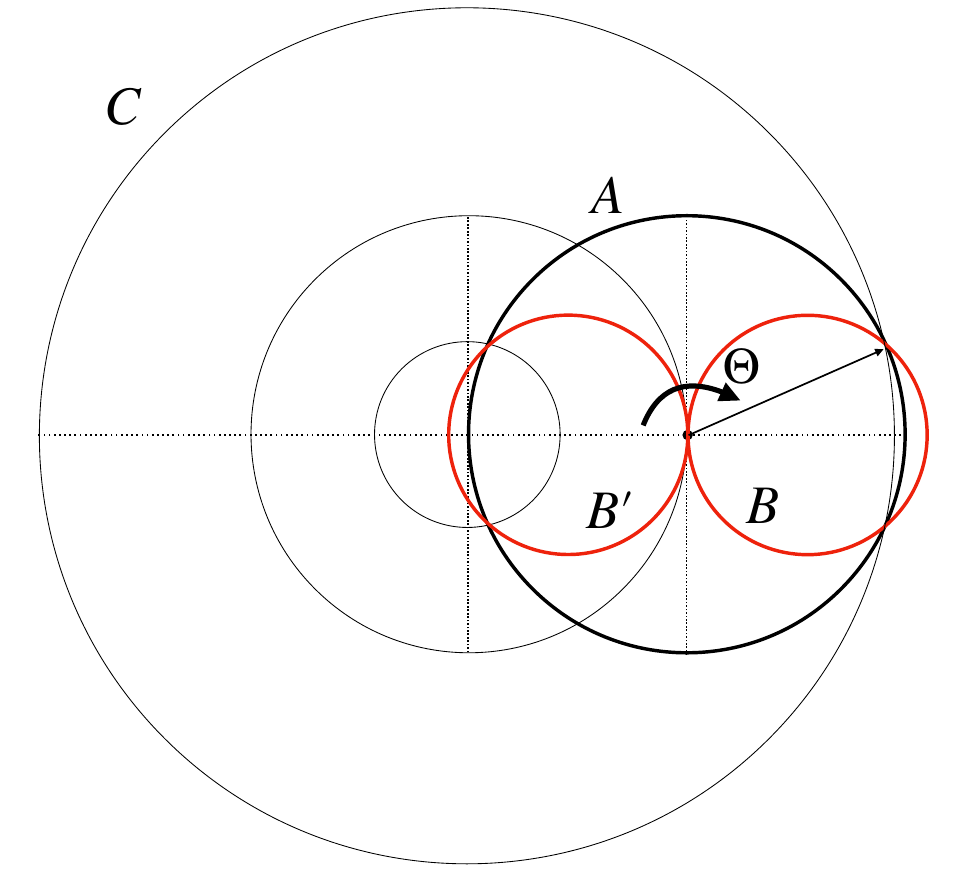}
\par\end{centering}
\protect\caption{Twin  $B$ horizons in opposite hemispheres of $A$ associated with   angle $\Theta.$ They share the same axis and the same value of $\Delta_C$, but have opposite parity with repect to $A$.
  \label{bpair}}
\end{figure}

The antihemispherical information is entirely internal, so it gives rise to correlations with incident information from  the $C$  direction only via coherent displacements of $A$. It has purely odd parity, so it adds perturbation power only in odd harmonic components.   Moreover, this power is added only on large angular scales, dominated by low-$\ell$ harmonics.

These properties can  be understood by examining in more detail the  $ABC$ causal entanglement in the antihemisphere.  
For a given polar ($C$) direction  and polar angle $\Theta$, consider a 
  pair of axially aligned $B$ spheres, one in each hemisphere (Fig. \ref{bpair}), with  opposite signs of
$\vec{\cal P}_C \cdot \vec{\cal P}_B$. They have matched circular intersections with $A$, one with polar angle  $\Theta$ and the other with polar angle $\pi-\Theta$. Their internal polar angles are $\theta_B= 2\Theta$ and $\theta_{B'}= -2\Theta$,
 so they map onto the same circles around the $B$ pole, but with opposite parity.

The twin $B$ spheres have a  mirror-image relationship: they  face the $A$ center from  opposite directions along the same polar axis, with the same $\Delta_C$. Since
 their spatial relationships with $A$ and $C$ are  parity reversed, 
the incoming $C$-axis information   contributes equal values with opposite  signs to the correlation function  (Eq. \ref{Ctheta}) at $\Theta$  and $\pi-\Theta$.
This adds purely odd parity perturbation power to
antihemispherical correlations induced on $B$.
Thus,
{\it the large-angle antihemispherical correlation function is  negative definite},
 \begin{equation}\label{antipodalstronger}
{ C}(3\pi/4< \Theta< \pi) < 0.
\end{equation}


In the  angular spectrum, anticorrelation introduces more overall perturbation power in negative  parity (odd) than positive parity (even)  harmonics. Moreover, the antihemispherical anticorrelations  arise  only from  a limited range of scale factors during inflation, close to the time a horizon freezes out. Put another way, it arises from $ABC$ entanglement at a time when all three horizons have comparable size.
In this range of angles, there is no entanglement of $A$ with much smaller or larger $C$ or  $B$ horizons: for $\pi>\Theta>3\pi/4,$
\begin{equation}
2>{\cal R}_C/{\cal R}_A= 2\sin(\Theta/2)>1.85
\end{equation}
and 
\begin{equation}
  1/2<   {\cal R}_B/{\cal R}_A= |1/{2\cos(\Theta)}|< .71.
\end{equation}
Since there is no antihemispherical entanglement with small-scale (${\cal R}/{\cal R}_A\ll 1$) horizons, 
 small-angle correlations  closely approximate the independent modes assumed in the standard picture. 
The lack of  fine-scale structure in antihemispherical  anticorrelation leads to an odd/even spectral parity asymmetry  dominated by  low values of $\ell$.
As discussed below, this qualitative behavior  corresponds to one of the well-known anomalies found on the real sky.

A  more specific  formulation of  antihemispherical anticorrelation  is promoted  below  to  a quantitative estimate in the context of an approximate model function.
We also propose below a direct model-independent statistical test that can include both  of the angular-domain symmetries, Eqs. (\ref{exact}) and (\ref{antipodalstronger}).

 \section{Holographic Model of the Angular Correlation Function}

The exact causal symmetries of the angular correlation function are  hidden from direct tests by measurement limitations, in particular the challenge of high-fidelity subtraction of foreground emission at large angular scales.
For quantitative tests, we 
adopt a hybrid approach to study
hidden symmetries,  by constructing an approximate holographic model
${\cal C}_{} (\Theta)$ that incorporates both causal relationships and symmetry constraints.
 This approach allows us  to separate some important physical signatures from measurement artifacts, and to compare predictions with measurements for both the correlation function and the power spectrum.
 The construction of the model also provides physical insights into how 3D relationships emerge holographically, particularly the relationships between spherical surfaces around different world  lines. We will refer to this specific approximate model in the following as the ``Holographic Model'' (HM), to distinguish it from the more general (and exact) causally-coherent symmetries just discussed.
 
\subsubsection{ Dipole}
Our local motion generates a temperature  dipole anisotropy, so  
the  primordial dipole component cannot be measured in practice. 
 We will invoke symmetries to choose a model from a  family of functions that differ from each other by a dipolar profile.

A  pure dipole corresponds to  an  $\ell=1$ harmonic perturbation.  
From Eq. (\ref{decompose}), a dipole aligned with the $z$ axis has the form
\begin{equation}\label{rawdipole}
\Delta(\theta)= a_{10}  \cos(\theta),
\end{equation}
where $a_{10}$ is the harmonic coefficient and $\theta$ is the polar angle.
From Eq. (\ref{powerpiece}), 
 it  has a power spectrum component
\begin{equation}
 C_{1}=a_{10}^2/3,
\end{equation}
and from Eq. (\ref{harmonicsum}), it produces 
 an angular correlation function 
\begin{equation}\label{puredipolecorrelation}
{ C}_D(\Theta)= \frac{3}{4\pi} C_1 \cos(\Theta)= \frac{a_{10}^2}{4\pi} \cos(\Theta).
\end{equation}

The correlation function is directly measured only up to this  degeneracy.
It can be written as a  sum of the dipole harmonic part,  and  all the higher harmonics:
\begin{equation}
{\cal C}_{}(\Theta)={\cal C}_{\ell=1}(\Theta) +  {\cal C}_{\ell>1} (\Theta).
\end{equation}
The unmeasured-dipole contribution has a pure cosine dependence, which we will parametrize using an overall normalization
$\Delta^2_\theta$ and an intrinsic dipole parameter $a_{1,\ell=1}$:
\begin{equation}\label{hiddendipolecorrelation}
{\cal C}_{\ell=1}(\Theta)=  a_{1,\ell=1} \Delta^2_\theta \cos(\Theta).
\end{equation}


The  value of the intrinsic dipole parameter $a_{1,\ell=1}$  is determined physically from the relationship of perturbations on a whole horizon with  external directional information; in this sense, it resembles, but is not the same as, a classical peculiar velocity. 
It is related to the other  contributions ${\cal C}_{\ell>1}$    by causal  symmetries, which we exploit below to constrain the model.

 \subsubsection{Match to standard perturbations at small angles}
 
On scales small compared to any horizon, the emergent 3D power spectrum is the same as in standard inflationary cosmology: 
the perturbations  in three dimensions have
 a nearly constant scale-invariant variance per $e$-folding of 3D wavenumber $k$:
\begin{equation}
 d \langle \Delta^2\rangle/ d \ln k= -\epsilon,
 \end{equation}
 where $|\epsilon|\ll 1$.
The 2D angular power spectrum of the  holographic model must  agree with the expected spectrum in  standard cosmology  at  $\Theta\ll 1$ and   $\ell\gg 1$,  where the curvature of the horizon is negligible.
 That is,  perturbations approximate  standard  plane-wave coherence in the limit $k\gg 1/R$ where a horizon of radius $R$ is  nearly flat. 

 In the $ABC$ sphere construction, perturbations at small angles correspond to ${\cal R}_C\ll {\cal R}_A$,  so that the $A$ horizon curvature is negligible.
The spheres represented by the smallest layers  of a  $C$ onion in a given direction all have the same mean offset from $A$, so in the limit of small angular separation, the  curvature perturbation is coherent over  circular patches which share the polar offsets of the $C$ centers  from $A$.

As in standard inflation, each e-folding of  ${\cal R}_C$   contributes about the same to the total variance, so in the small-angle limit  the logarithmic derivative of ${\cal C}_{}(\Theta)$  should be approximately constant at small angles. The form of the function at small angles must approximate
\begin{equation}\label{smalltheta}
(d {\cal C}_{}/d \ln \Theta )_{\Theta\rightarrow 0} \approx  \Delta_\theta^2 
 d \ln {\cal R}_C/d \ln \Theta,
 \end{equation}
where $ \Delta_\theta^2$ is fixed by matching the standard normalization. 
  As discussed below, it is straightforward to include a  linear correction for a slightly tilted spectrum ($\epsilon\ne 0$), as is observed in the real universe.

\subsubsection{Integral of  dipole fluctuations}

As seen above, correlations are governed by the  trigonometric projections  (Eqs. \ref{Aratio}, \ref{Cratio} and \ref{Bratio}), which connect  polar projections and radii of  horizons to measured  angles. 

As a first approximation, consider a correlation function with angular derivative of the form
\begin{equation}\label{candidatekernel}
\frac{ d{\cal C}_{ABC}(\Theta)}{d \Theta} 
= 
 \frac{\cos(2\Theta) }{ \sin(\Theta/2)}. \  \  \ 
 \end{equation}
This expression can be interpreted physically as the effect of  $B$-sphere dipoles projected onto $A$, fluctuating  by random values of  $C$-sphere noise, $\Delta_C$.
The $\cos(2\Theta)$ factor comes  from the coherent angular pattern of correlated distortions on  $A$ and $B$
(Eq. \ref{Bratio}), the $\sin(\Theta/2)$ factor from the scale of $C$ (Eq. \ref{Cratio}).
The   small-angle limit of Eq.(\ref{candidatekernel}) agrees with Eq. (\ref{smalltheta}); the small-scale behavior, dominated by the denominator, arises  from  coherent displacements  of  nearly-independent   $C$ horizons on the boundary of $A$.

\begin{figure}
\begin{centering}
\includegraphics[width=\linewidth]{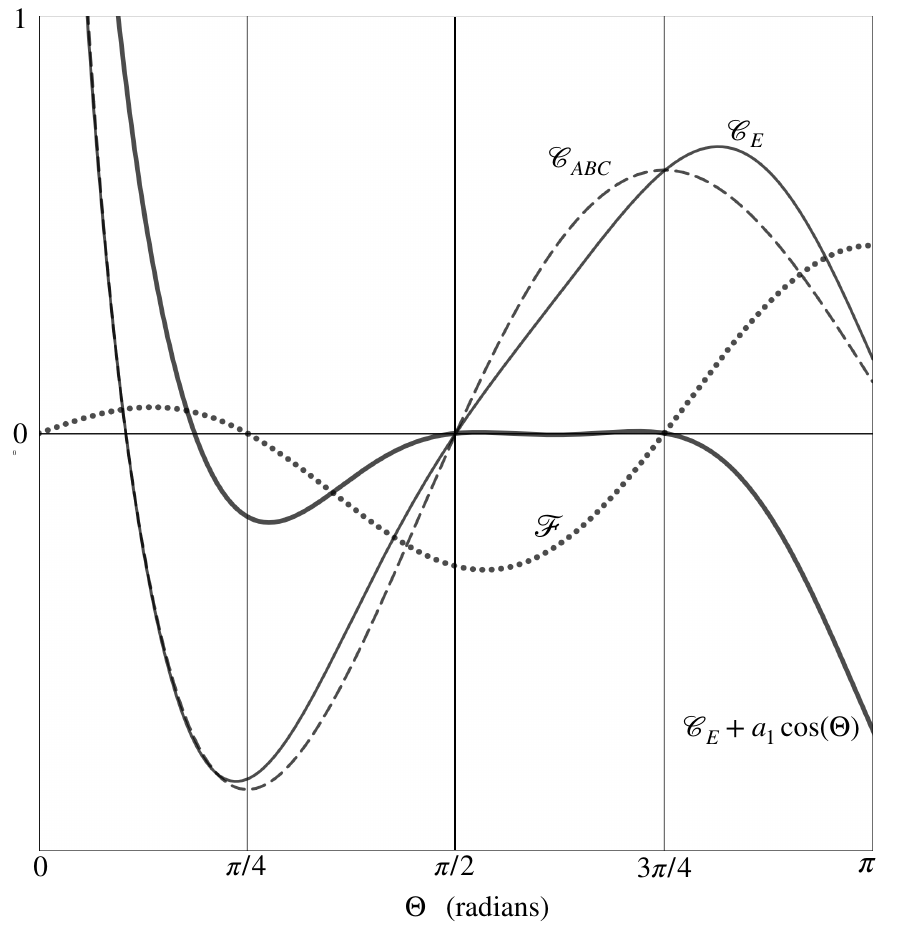}
\par\end{centering}
\protect\caption{Functions used to construct a model  correlation function  ${\cal C}(\Theta)$ (Eq. \ref{model}), in natural units with $\Delta_\theta^2=1$.  The parameters  $a_1$ and ${\cal F}_0$ used for  ${\cal C}_E$ in this plot are fixed to approximate the exact causal null symmetry  (Eq. \ref{exact}) at $\pi/2<\Theta<3\pi/4$.  
\label{functions}}
\end{figure}

Integrating the accumulation of fluctuations over conformal time during inflation (Fig. \ref{ABCradii})  is the same as
integration  of Eq. (\ref{candidatekernel}) over polar angle:
\begin{equation}\label{simplefunction}
{\cal C}_{ABC}(\Theta)=  \int_\Theta^{\pi/2} 
d \Theta \ \  \frac{\cos(2\Theta) }{ \sin(\Theta/2)}.  \  \  \ 
\end{equation}
This function is shown  in
Fig.  (\ref{functions}).

\bigskip

\subsubsection{Entangled-dipole shape correction}  

With a dipole term $\propto \cos(\Theta)$ added, Eq. (\ref{simplefunction}) roughly resembles the correlation function of the real sky.  The agreement is remarkably close at 
 angles  $\Theta\lesssim \pi/4$, but it is not exact at larger angles. 
Indeed we expect that the simple  expression in Eq. (\ref{simplefunction}) cannot be the whole story, since it assumes that the  quantum interiors of the $B$ and $C$ spheres are independent of each other. A holographic correlation  needs to include  coherent entanglement  of causal diamonds on all scales, so
overall consistency with a universal function shape requires  an angle-dependent  modification of the  dipolar approximation.

To approximate this effect, we add an amplitude modulation to Eq. (\ref{simplefunction}) that varies slowly with angle, to allow for an extra consistency constraint on the even-parity modes of $B$ and $C$ spheres: 
\begin{equation}\label{corrected}
 {\cal C}_{ABC}\rightarrow {\cal C}_E=  {\cal C}_{ABC}(1+{\cal T}_\epsilon+ {\cal F}) ,
\end{equation}
which includes a small tilt correction ${\cal T}_\epsilon(\Theta)$ discussed below, and   a parameterized entangled-dipole correction of the form
\begin{equation}\label{correction}
{\cal F}(\Theta) 
= {\cal F}_0\cos(2\Theta) \sin(\Theta/2),
\end{equation}
 which is based on   the $B$ and $C$ dipolar projection factors  (Eqs. \ref{Bratio} and \ref{Cratio}).
As discussed below (see Eq. \ref{a2range}), we will fix the value of the shape  parameter ${\cal F}_0= 0.45$ to  approximate  vanishing correlation at 
$\pi/2<\Theta<3\pi/4.$

This modification is shown in Fig. (\ref{functions}).  The  modification is small below $\Theta\sim \pi/4$, but makes a measurable difference on large angular scales.
Notice that the zeros of ${\cal F}$ are extrema of ${\cal C}_{ABC}$.
In particular, the modification vanishes at  $\Theta=  \pi/4$ and $ 3\pi/4$,  the angles where  circular intersections with $A$ are great circles on $B$.

\subsubsection{Correction for  tilted power spectrum}

A truly universal ${ C}(\Theta)$ assumes exact scale-invariance, that is, all of the spherical horizons are assumed to have ${ C}(\Theta)$ of the same form, determined entirely by trigonometric projections.  However, in the slow-roll inflationary background, the expansion rate and horizon radius change slowly with time. This leads to  a small logarithmic tilt correction that has  a measurable effect at small angles.

  A simple first-order correction  can allow for the fact that the 3D power spectrum $\Delta^2(k)$ has a spectral index $n_S$ that differs slightly from unity. The deviation is characterized by a tilt parameter $\epsilon$,
\begin{equation}
n_S-1 = d \ln \Delta^2/d\ln k  = -\epsilon,
\end{equation}
which has a  value $\epsilon = 0.035\pm 0.004$ 
 measured by 
  {\sl Planck}  \cite{Aghanim:2018eyx,Akrami:2018vks}, from  a fit to the spectral index  at  $\ell>30$. 
We model the effect of tilt as a slow variation of    normalization $\Delta_\theta^2$   with    ${\cal P}_C$:
 \begin{equation}
\frac{\delta \Delta_\theta^2}{\Delta_\theta^2} \approx \frac{d \ln \Delta^2}{d \ln {\cal P}_C} {\delta \ln {\cal P}_C}\approx  - \epsilon \ln [{\cal P}_C(\Theta)/{\cal P}_C(\pi)].
\end{equation}
Eq. (\ref{Ccirclecos})
then leads to a   linear correction, 
\begin{equation}\label{tiltcorrection}
 {\cal T}_\epsilon(\Theta)\equiv -\frac{\delta \Delta_\theta^2}{\Delta_\theta^2}  \simeq 2\epsilon \ln[\sin(\Theta/2)].
\end{equation}

\bigskip
\begin{figure}
\begin{centering}
\includegraphics[width=\linewidth]{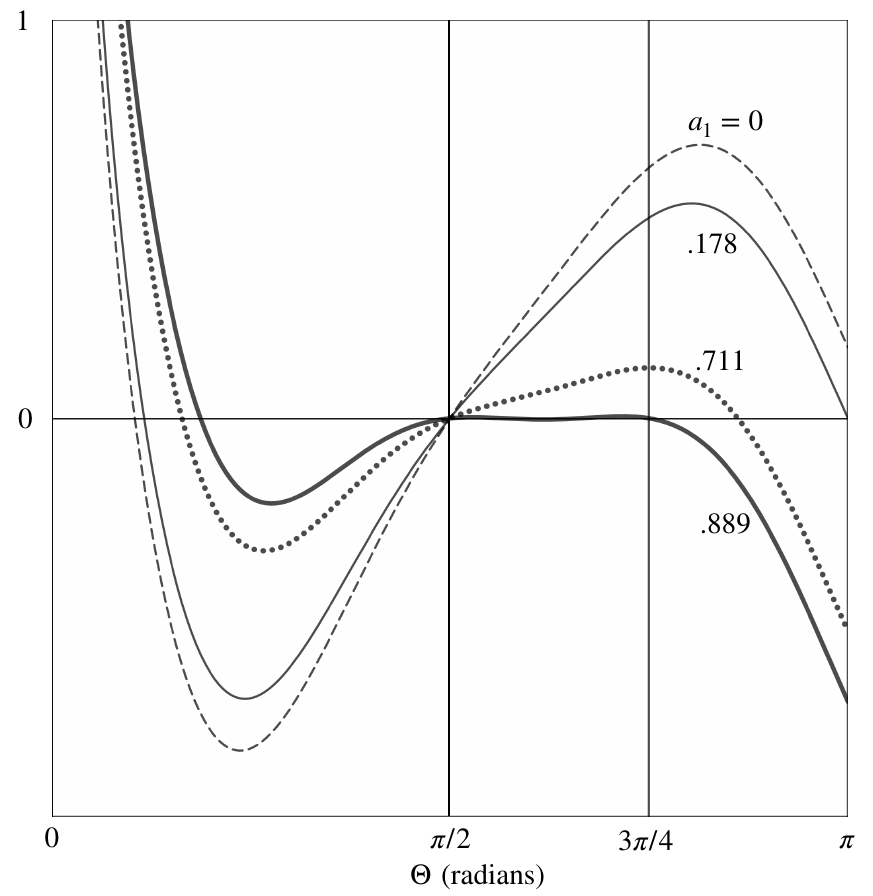}
\par\end{centering}
\protect\caption{The family of model functions (Eq. \ref{model}) for $\Delta_\theta^2=1$, ${\cal F}_0=.45$,  $\epsilon= 0.035$, and a range of values for the cosine coefficient,  $a_1$. The heavy curve reveals the approximate null symmetry at $\pi< \Theta<3\pi/4$  used to fix $a_1$ and ${\cal F}_0$ for the true function.  The value $a_1= a_{1U}= .889$   is taken to approximate  the true  function.  For the observed correlation function,    antipodal parity symmetry at $\Theta=\pi$ fixes the value of the unmeasured dipole component at  $a_{1U,\ell=1}=  .178$, and therefore the value  $a_1=a_{1U}-a_{1U,\ell=1}= .711$ is predicted to fit sky maps. 
\label{functionfamily}}
\end{figure}

 \subsubsection{Model function family with parameters}
 
The holographic model for the measured correlation is  obtained by including all of the above elements.
The  complete family of  functions can  be written
\begin{equation}\label{model}
{\cal C}(\Theta)=  \Delta_\theta^2   \ {\cal C}_{E}+  \Delta_\theta^2 a_1 \cos(\Theta)
\end{equation}
where
\begin{equation}\label{CE}
{\cal C}_E(\Theta)=  (1+{\cal T}_\epsilon+{\cal F}) \ {\cal C}_{ABC}
\end{equation}
A function from this family is fixed by four parameters, $a_1$, $ {\cal F}_0$,  $\Delta_\theta^2$, and ${\cal T}_\epsilon$.
Some examples from this family of  functions  are plotted in Fig. (\ref{functionfamily}), for different values of   $a_1$.

The two shape  parameters $a_1$ and $ {\cal F}_0$ are fixed below,  by constraining the model to agree with antihemispherical causal symmetry.
The normalization parameter $ \Delta_\theta^2$ sets the overall scale of the function in physical units, and must match the standard cosmological normalization at small angles.   Similarly, in a slow roll inflationary background,  the tilt $\epsilon$ must agree with the value  measured by {\sl Planck} on small angular scales.
(As in standard  inflation, these parameters   are determined  physically from the scale and rate of change of the  inflationary horizon in Planck units, as discussed in the Appendix.)
The function thereby provides a unique approximate model of a universal large angle correlation function that can be compared with data.

\subsubsection{Model  fixed by causal  symmetry}
The antihemispherical null symmetry   (Eq. \ref{exact}) for the true universal correlation function cannot be exact for the approximate  function we have written down (Eqs. \ref{model}, \ref{CE}).  Even so, it can  be used to  fix the two shape parameters, $a_{1U}$ and ${\cal F}_0$.

The   value  $a_{1}$ for the model  universal function ${\cal C}_{}$
 can be fixed  by requiring Eq. (\ref{exact}) to hold exactly at one angle,
\begin{equation}
{\cal C}(3\pi/4)=0,
\end{equation}
which leads to a value $a_{1}= 0.889$.  At this angle,    ${\cal F}(3\pi/4) =0$ so this value     is   independent of ${\cal F}_0$.

The  parameter ${\cal F}_0$  controls the  shape of  ${\cal C}_{E}$. It describes the effect of entangled $B$ and $C$ dipoles, and  serves to flatten the  shape of the model function so that  it can nearly vanish over $ \pi/2<\Theta<3\pi/4$. 
It  is not precisely fixed,  because the null is not exact over for this approximate function.  However, it is closely constrained by minimizing 
$|{\cal C}(\Theta)|$ over the range   $ \pi/2<\Theta<3\pi/4$:   for  ${\cal F}_0$ in the range
 $.448<{\cal F}_0<.454$, we find
 \begin{equation}\label{a2range}
|{\cal C}(\pi/2<\Theta<3\pi/4)|<.005.
\end{equation}
The corresponding fractional variations in ${\cal C}$ over this range are less than one percent, much smaller than the differences between different foregrounds subtracted maps of the sky.
For the fits below, we adopt ${\cal F}_0=.45$.  (A  better approximation would include higher order terms in powers of the cosine to agree better with Eq. (\ref{exact}), but this is not needed to address current data.) 
To this precision,  antihemispherical causal symmetry leads to a unique model for the universal function ${\cal C}_{}$, up to a physical normalization  and tilt.

\subsubsection{Dipole fixed by exact antipodal  parity symmetry}

The coefficient $a_{1}$ includes the  sum of the  dipole term  $a_{1,\ell=1}$ and  an integrated contribution from dipoles over the history of  inflation
$a_{1,\ell>1}$,  which arises from  ${\cal C}_{E}$. 
Thus, a fit to actual data  still has an unknown parameter, the part of the  cosine coefficient $a_{1,\ell=1}$ that comes from the unmeasured dipole harmonic.

If we posit an exact antipodal causal   symmetry, we can predict the invisible dipole correlation coefficient $a_{1U,\ell=1}$, even without a fit to the data. 
The model and its dipole  are both then uniquely determined  by  symmetries, so that we can fit the large-angle sky with only two parameters, the normalization and tilt, that are already constrained at small scales.

The  function ${\cal C}_E$ (Eq. \ref{CE}), which represents the projected sum of $BC$ dipole fluctuations on $A$,  has  an antipodal  value ${\cal C}_E(\pi)=0.178$, a net positive perturbation power.
That means that for  a particular  choice   $a_1=0.178$,   $\cal C$   vanishes at the antipode, $\Theta=\pi$ (see Fig. \ref{functionfamily}). 
We posit that this value corresponds exactly to the amplitude of the true dipole component of ${\cal C}$:
\begin{equation}\label{antiexact}
{\cal C}_E (\pi)={\cal C}_{\ell=1}(\pi).
\end{equation}
This antipodal correlation reflects an exact  parity relationship of total perturbation power:   {\it  the net positive parity perturbation from  ${\cal C}_E$ is exactly  canceled by the  $\ell=1$ dipole  of ${\cal C}_{}$}.  

This symmetry can be interpreted in terms of   our previous causal argument based on the parity of internal, unabsorbed polar information. The  dipole   of the universal function is determined  by the external, ``shadowed'' relationships of a horizon.   The antipodal parity symmetry (Eq. \ref{antiexact})  relates the $A$ dipole  to  all the other harmonics on $A$ that entangle with large $C$ spheres in the antihemisphere. The   
  correlation  is then  zero or negative everywhere in the antihemisphere  (Eqs. \ref{exact}, \ref{antipodalstronger}), where the $C$ dipole is entangled only with odd-parity perturbations on $A$.

Antipodal parity symmetry in this model requires an
 unmeasured dipole with a cosine coefficient
\begin{equation}\label{invisiblecoefficient}
a_{1,\ell=1}= a_{1} - a_{1,\rm fit} = .178.
\end{equation}
 The  $a_1$ for a  fit to the sky data is thus predicted to be $a_{1,\rm fit}=.711$.

\subsubsection{Two-parameter holographic model to fit data}

After imposing these exact causal symmetries, a  fit of the holographic model to a dipole-subtracted temperature anisotropy map has two   parameters,   the physical normalization $\Delta_\theta^2$ and tilt parameter $\epsilon$:
\begin{equation}\label{finalmodel}
{\cal C}_{}(\Theta)=  \Delta_\theta^2   \ {\cal C}_{E}+   0.711 \Delta_\theta^2\cos(\Theta),
\end{equation}
\begin{equation}
{\cal C}_E(\Theta)=  (1+{\cal T}_\epsilon+{\cal F}) \ {\cal C}_{ABC},
\end{equation}
\begin{equation}
 {\cal T}_\epsilon(\Theta)= 2\epsilon \ln[\sin(\Theta/2)],
\end{equation}
\begin{equation}
{\cal F}=
.45 \cos(2\Theta)\sin(\Theta/2),
\end{equation}
\begin{equation}
{\cal C}_{ABC}= \int_\Theta^{90^\circ } 
d \Theta \  \frac{\cos(2\Theta) }{ \sin(\Theta/2)} .
\end{equation}
The two parameters are not arbitrary: they must match the amplitude and tilt of the standard inflationary power spectrum, which are already tightly constrained  by  {\sl Planck} data
at smaller angular scales $\ell>30$. 
In this sense, the large angle correlation in this model is fixed:  once the small angle part is fitted,  no independent free parameters are available to adjust the shape of $C(\Theta)$ at angles larger than a few degrees.

\section{Comparison with data}

\begin{figure}
  \centering
\includegraphics[width=0.49\textwidth]{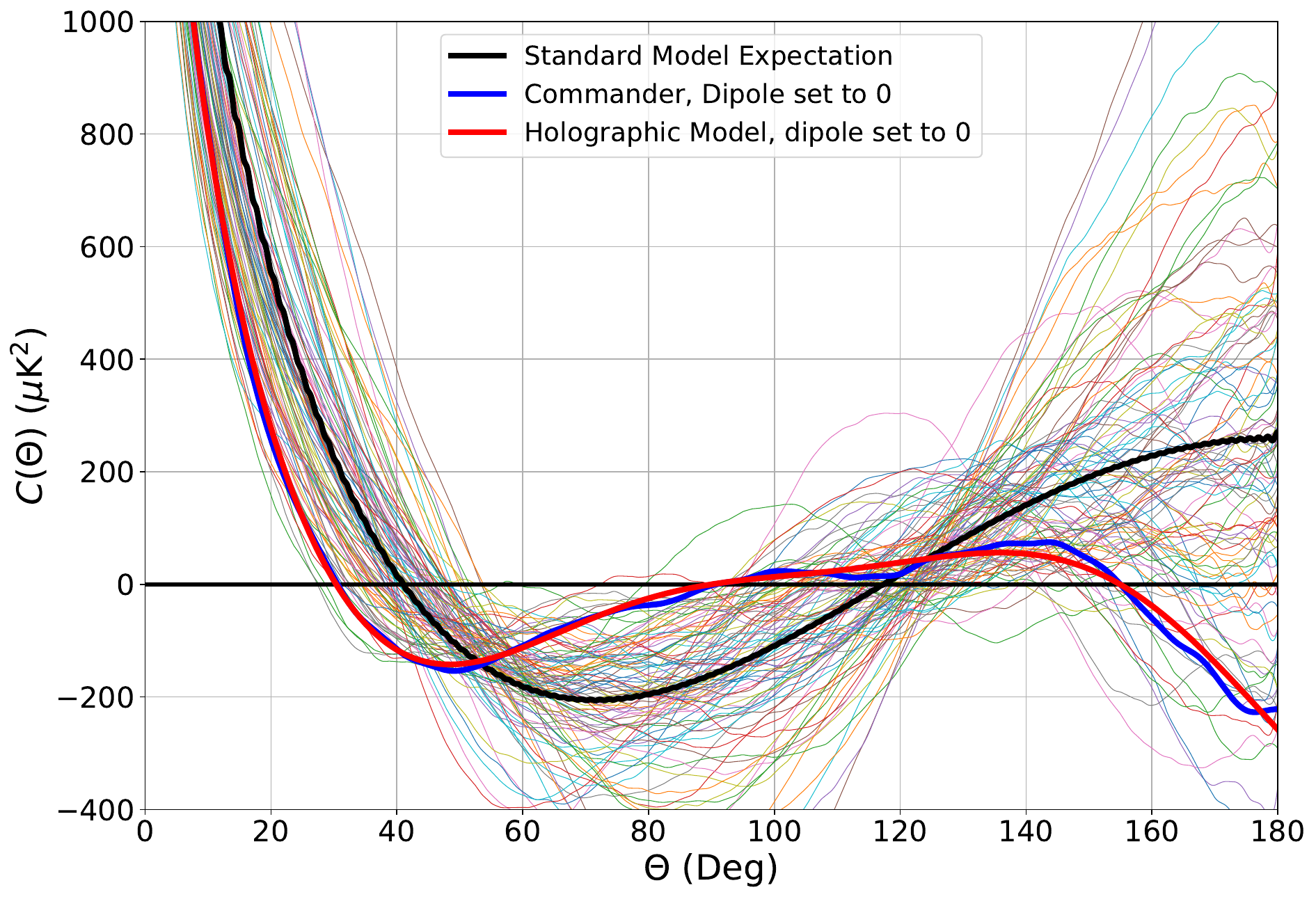}
  \caption{The holograpic model function (Eq. \ref{finalmodel}), along with the measured  Commander correlation function.
   In the holographic picture, all  of the predicted  positive correlation at $\Theta>90^\circ$ is due to the absence of the unmeasured dipole.  
   100 random realizations of the standard model are shown for comparison.
Unlike the standard picture, the  holographic correlation model fits well  in the angular domain over the whole range of angles from  $\Theta\sim 5^\circ$ to $\Theta\sim180^\circ$, with  no need (or opportunity) for cosmic variance to match  the model. 
The deviation from that model is attributed mostly to  imperfect foreground subtraction.
  \label{modeldata}}
\end{figure}

\subsection{Curvature and temperature perturbations}
 
 The pattern of curvature $\Delta(\theta,\phi)$ on the sky does not exactly follow the pattern of temperature anisotropy $\delta T(\theta,\phi)$:
 on small angular scales,  the Sachs-Wolfe approximation  breaks down.
 On scales smaller than the horizon at recombination,   the pressure of the radiation-baryon plasma creates   acoustic waves whose velocity  creates Doppler anisotropy, and whose compression and rarefaction modify the phase relationship of the scalar contributions from local gravitational redshift and temperature perturbations\cite{Hu:2001bc,Wright:2003ig}.

 For this reason,
 we compare models and symmetries with data only on  angular scales  well outside the horizon at recombination, $\gg 1^\circ$, where these effects lead to deviations of less than a few percent.  To this accuracy, the Sachs-Wolfe approximation\cite{1967ApJ...147...73S,PhysRevD.22.1882} works well on the scales we fit to the data,  $\Theta> 5^\circ$. By the same token,  the model  power spectrum should be a good approximation up to   $\ell \sim 20$.
 For the current analysis,  smaller  scales are accommodated  in the fits where needed by smoothly fitting to actual data at $\Theta<5^\circ$.  The model is a good fit to the data (and to the standard model) at this scale, so the matching works well.
Since the new, exotic angular correlations are only  significant at even larger scales, $\Theta\gtrsim 20^\circ$ and  $\ell\lesssim 9$, the interpolation between our model and the standard approach is handled consistently, with a significant region of common overlap.

  
Small  departures from the Sachs-Wolfe approximation due to 
 the Integrated Sachs-Wolfe (ISW) effect\cite{Hu:2001bc,Wright:2003ig} and cosmic reionization
 \cite{Aghanim:2018eyx} are expected to
 modify $C(\Theta)$ at   large angles by less than the other measurement uncertainties, and will be ignored here.



With these caveats and appropriate normalization, a  smoothed sky temperature map, measured at $\Theta\gtrsim 5^\circ$,   should have approximately the same angular correlation function and power spectrum as a spherical slice of the primordial potential, so they can be directly compared (Fig. \ref{modeldata}).

\begin{figure*}[hbt]
  \centering
  \includegraphics[width=.9\textwidth]{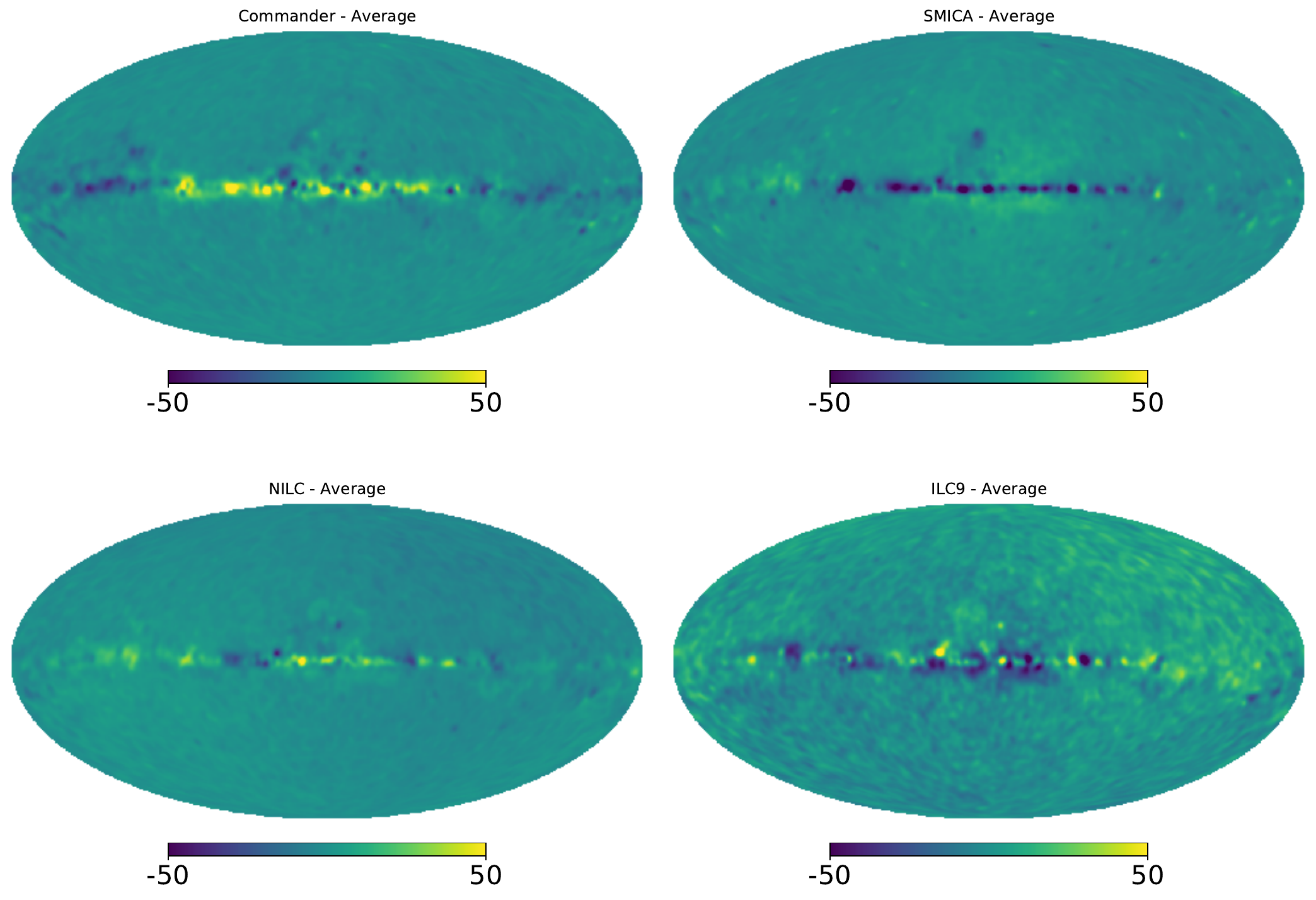}
  \caption{Images of the difference between each sky map and the average of the maps, with higher contrast than Fig. (\ref{Averagemap}), shows the actual pattern of Galaxy-foreground model differences at  $\Theta\gtrsim 5^\circ$ and $\ell\lesssim 30$.  The differences clearly show the  imprint of an imperfectly subtracted Galaxy model over a wide range of angular scales.   In the angular domain,  the fine-grained differences in the large-scale pattern, organized mainly along the Galactic plane,  create prominent differences between  correlation functions at   $C(\Theta> 120^\circ)$, as seen in Fig. (\ref{dataandmodelwithdipole}). Units are $\mu$K. \label{differencemap}}
\end{figure*}

\subsection{CMB Data at large angular scales}

Two satellite experiments, {\sl WMAP} and {\sl Planck}, have published the full-sky CMB maps and angular power spectra at $\ell < 20$ with galactic foregrounds removed. These can be used to test the prediction of the holographic model and evaluate the relative likelihood of the of the data given the Standard Model (SM) and the Holographic Model (HM) outlined above.  We use the 6 parameter $\Lambda$CDM model with parameters values taken from the Planck 2018 publication \cite{Akrami:2018odb} as the SM because additional parameters come into play only at smaller angular scales than we are considering. This also avoids concern about tension between experiments and cosmological models.

As already noted, the unmeasured dipole and errors in foreground subtraction limit the strength of the comparison of the relative probabilities. For the specific holographic model described above, we can use the exact symmetry of the vanishing 2-point function in the range $90^\circ < \Theta < 135^\circ$ to predict what the dipole should be. Alternatively, we can allow for the addition of an arbitrary dipole amplitude and evaluate the residual symmetry directly in a model-independent way. We use both approaches here.

The residual galactic foregrounds are the overriding limitation of the comparison of  models and symmetries to the data. For most cosmological studies, the information constraining models is concentrated at $\ell > 20$ where the foreground subtraction is a  much smaller problem. The 6 parameter SM fit is almost unaffected if $\ell < 20$ data is not included. For the HM, the large angular scales are key. We therefore are constrained to work with the foreground subtracted maps published by the satellite experiments. Both satellite groups warn that the bias and uncertainty of the foreground subtraction is difficult to quantify, especially at the largest angular scales. Neither group has published an estimate of the foreground-subtracted map pixel-pixel covariance, thus proscribing direct quantitative likelihood comparisons with models.

The empirical approach considered here takes the foreground-subtracted maps from both satellites and uses the map and 2-point correlation function differences as a measure of the likely uncertainty. This procedure, while not statistically rigorous, allows  quantitative estimates of relative liklihood. We use the average of three different maps from the {\sl Planck} team  (Commander, SMICA and NILC) \cite{planckforegroundsub2018}, and the foreground subtracted map from the {\sl WMAP} team, ILC9
\cite{2013ApJS..208...20B}.
 Differences between these maps  shown in Fig. (\ref{differencemap})  clearly show systematic errors due to imperfect Galaxy subtraction. The fact that these tests are sub-optimal is motivation for better large angular scale foreground subtraction techniques and new data. 

\begin{figure*}[t]
\begin{centering}
\includegraphics[width=.49\linewidth]{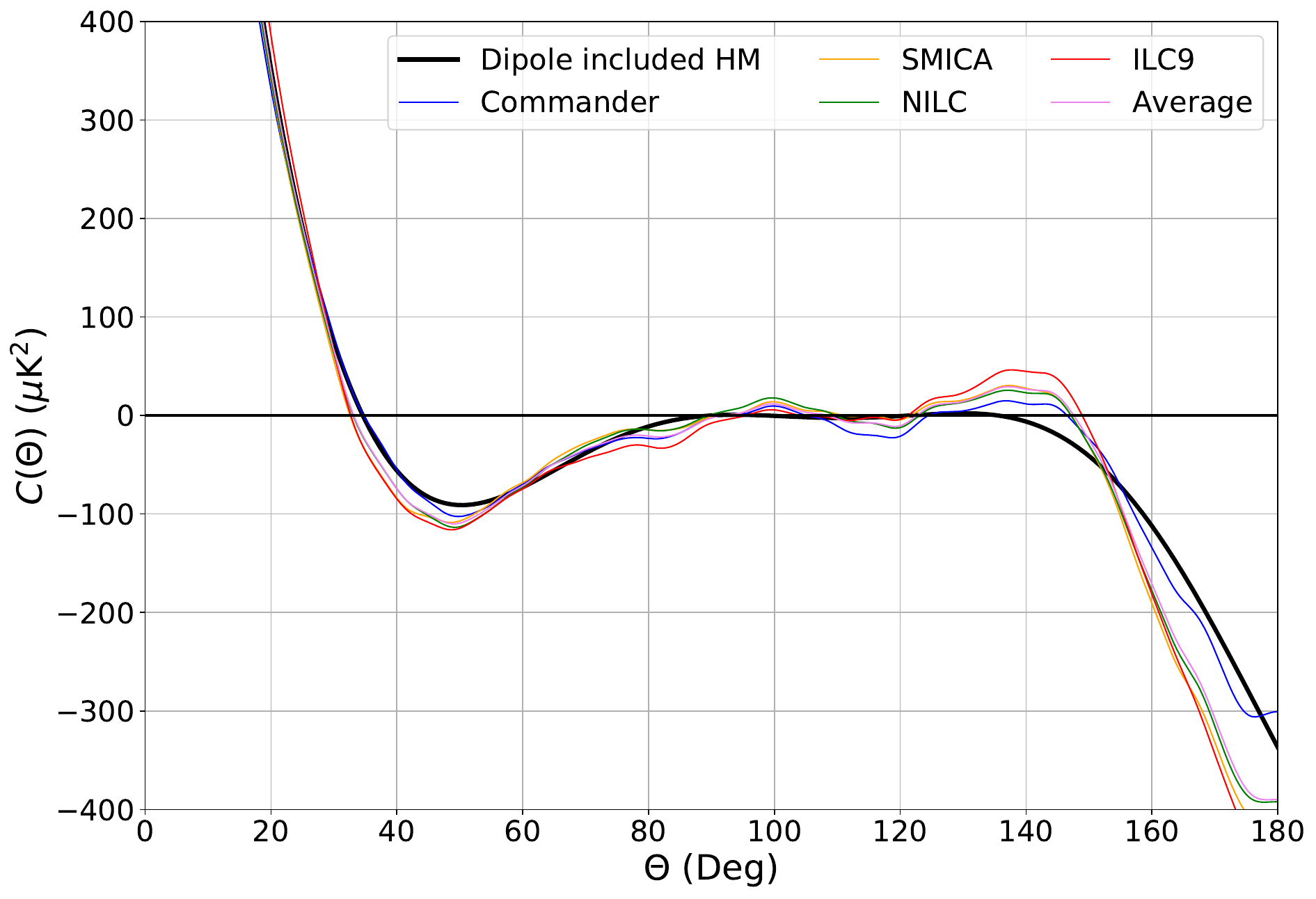}
\includegraphics[width=.49\linewidth]{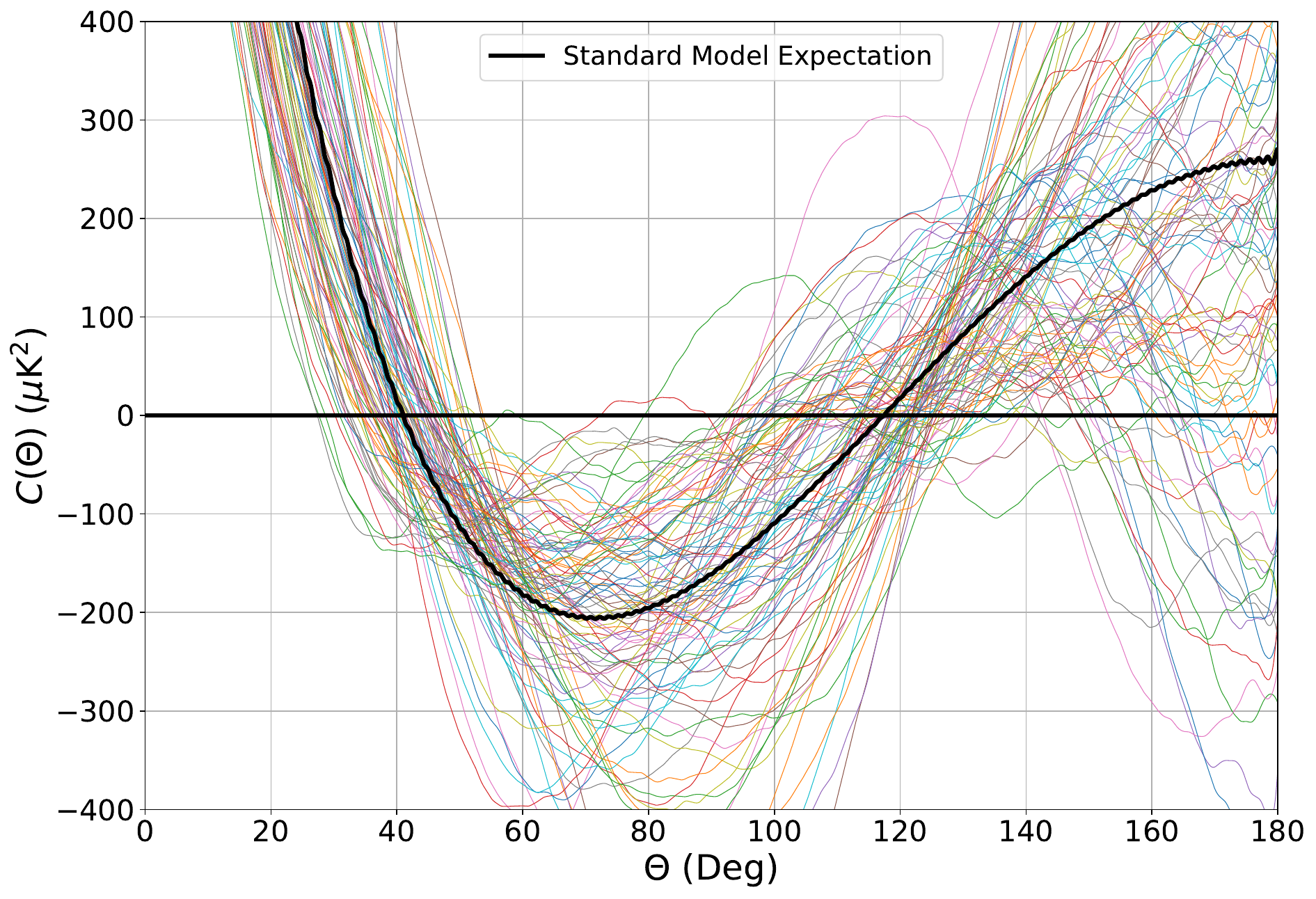}
\par\end{centering}
\protect\caption{On the left, the Holographic Model (HM, Eq. \ref{finalmodel}) is shown together with correlation functions of  maps,  on an expanded scale.  The data are shown with the cosine contribution of  an invisible dipole correlation (Eqs.  \ref{hiddendipolecorrelation}, \ref{invisiblecoefficient}) restored, to compare with  what  holographic symmetry would predict if the dipole  were measured. 
The differences between maps are 
mainly due to the Galactic contamination, visible in Fig. (\ref{differencemap}).
For comparison, on the right, 100  standard-model realizations are shown on the same scale to show the range of cosmic variance typically expected in the standard picture. 
  \label{dataandmodelwithdipole}}
\end{figure*}

\begin{figure}
\begin{centering}
\includegraphics[width=\linewidth]{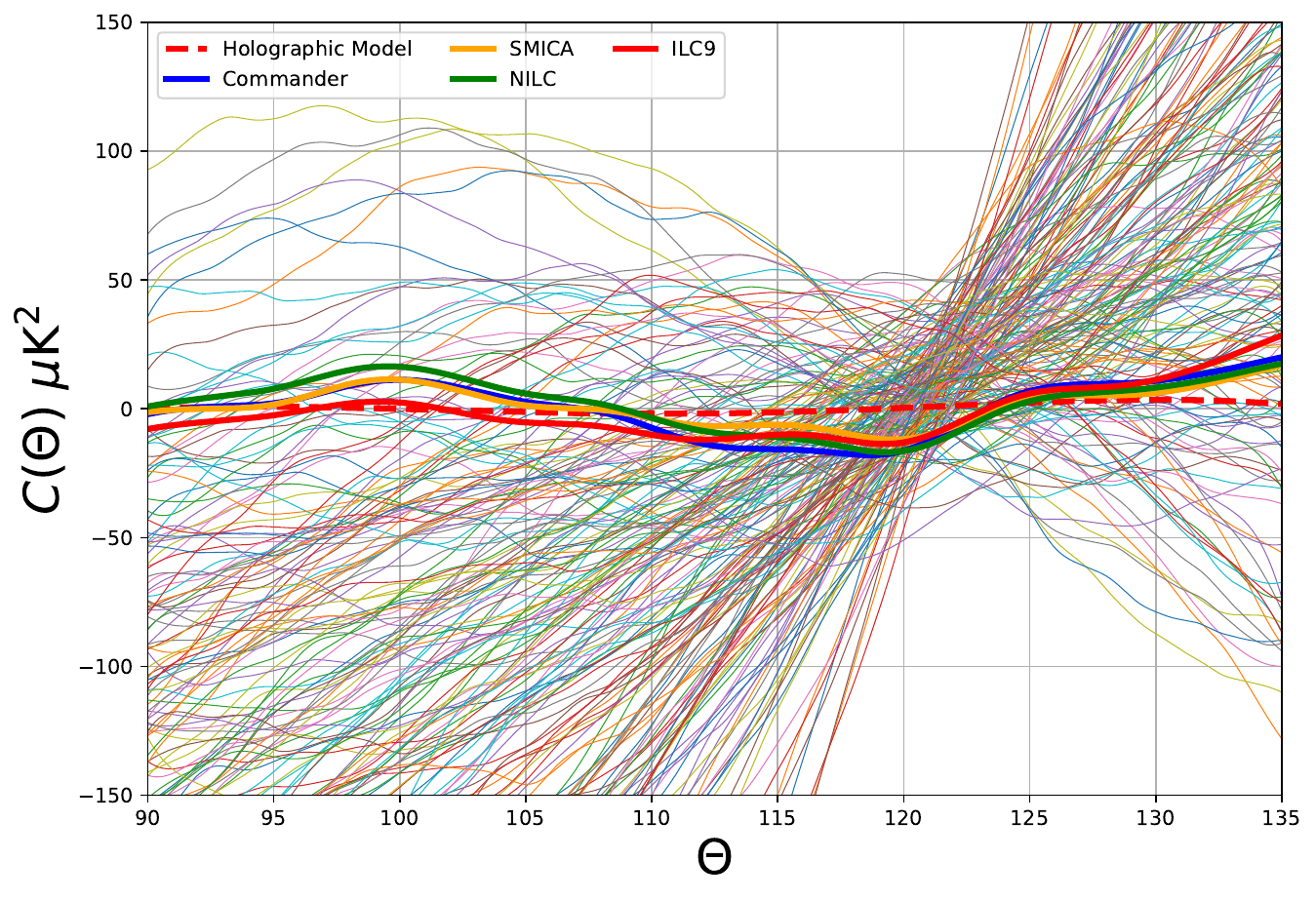}
\par\end{centering}
\protect\caption{Measured and simulated correlation functions, after addition of an unmeasured  dipole contribution that minimizes the  difference from zero for each dataset over this range of angles.  
Correlation functions of four maps are shown, as well as 200 random standard-model realizations. Very occasionally, a dipole-corrected realization happens to lie as close to zero as the actual sky across this range of angles.
\label{standardresiduals}}
\end{figure}

\subsection{Angular 2-Point Function}
\label{sec:2-point_function}

The holographic hypothesis predicts exact values for correlations at specific angles so the 2-point function is the natural basis for comparing it to data. 


Figure~\ref{modeldata} shows a fit of the HM to the average 2-point function of foreground-subtracted maps.The fit to the model,  (Eq. \ref{finalmodel}) is good overall but closer inspection (Fig. \ref{dataandmodelwithdipole}) shows important differences. Apart from the unmeasured dipole, the variation the 2-point functions of the four maps varies approximately as much as the difference between the model and the data, indicating that systematic errors in this subtraction, clearly seen in difference maps  (Fig. \ref{differencemap}), is an important limitation.

The HM fit has two parameters: an overall amplitude $\Delta^2_\theta$,  and the tilt parameter, $\epsilon$. The fit minimizes the residual of the HM to the 2-point function from the map evaluated every 1 degree in angle. These points have a large off-diagonal terms in the map noise covariance matrix. This is not being considered in this fit. The probability of the data given the model can therefore not be directly computed from this fit and this is the object of future analysis. However, with this fit technique, the parameters vary by $\lesssim$ 5\% depending on which foreground-subtracted map is used. Of the four maps we considered, Commander has the smallest residuals to the model. The 2-point function residuals of the four data sets differ from each other by more than the Commander residuals with the model. Therefore, further refinement of the test awaits a better understanding of the foregrounds.

 \begin{figure}[hbt]
\begin{centering}
\includegraphics[width=\linewidth]{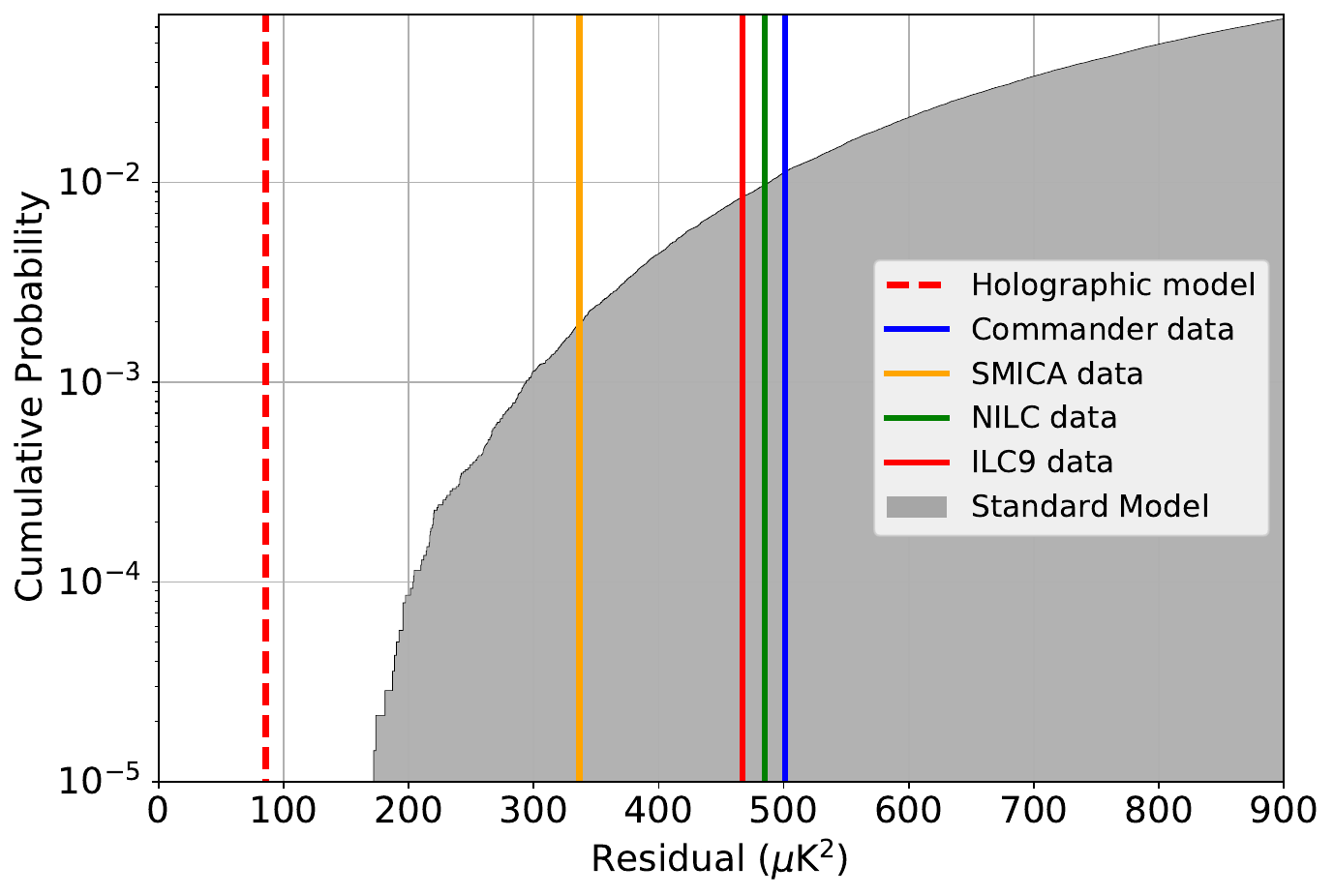}
\par\end{centering}
\protect\caption{ Comparison of cumulative residual correlation $\delta C_{90,135}$ (Eq. \ref{residue}) at $90^\circ<\Theta<135^\circ$.
Measured residuals are shown for the four maps.
For exact antihemispherical null symmetry, the  residual is predicted to vanish, which is approximated by the holographic model. The standard model has no angular symmetry, so a significant departure is  predicted from cosmic variance. The histogram shows the cumulative probability of values for standard model realizations. Less than one percent  of  standard model realizations come as close to zero as any of the galaxy-subtracted maps.
The difference between the maps, which is an artifact of Galactic subtraction errors, is comparable to their differences from zero, consistent with the null hypothesis for the true CMB sky. 
\label{histogram}}
\end{figure}
 
\subsection{Direct test of  antihemispherical null correlation}

With coherent  inflationary horizons, null  correlation at $90^\circ<\Theta<135^\circ$ could be an  exact causal symmetry.
According to this picture, if the intrinsic dipole were measured, the  symmetry would have been obvious in  measurements of the CMB, as shown in Fig. (\ref{dataandmodelwithdipole}).

The standard picture does not allow any such exact  symmetries, so an approximate  null symmetry is regarded as a statistical fluke. 
Yet especially after allowing for an unmeasured dipole, the correlation appears to be remarkably close to  zero over a significant range of angles, $90^\circ<\Theta<135^\circ$.

As seen in Fig. \ref{dataandmodelwithdipole}, SM realizations seldom come close to zero at $90^\circ$, and when they do, they tend to wander away quickly at other angles.
Fig. \ref{standardresiduals} shows the correlation for each dataset and one set of realizations, now adding the dipole that optimizes  match to the symmetry for each realization.

The procedure followed to evaluate the relative probability of the SM and the HM is to carry out a fit for the four measured 2-point functions, the HM and a large number of 6-parameter SM realizations. For each map or realization, the fit minimizes the squared residual $\delta C_{90,135}^2(A_i)$, which measures the departure from zero
 of the 2-point function allowing for an additive dipole component controlled by a single parameter $A_i$:
\begin{equation}\label{residue}
  \delta C_{90,135}^2 (A_i) = \sum_{\Theta_j = 90^\circ}^{135^\circ}  [C(\Theta_j)+ A_i\cos(\Theta_j)]^2.
\end{equation} 
Figure~\ref{histogram} shows the value of the residual for the four maps and the HM as well as the  distribution of residuals for a large number of SM realizations.

The coherent-horizon hypothesis predicts that the null symmetry is exact in the Sachs-Wolfe approximation.   In this scenario, any measured nonzero value  $\delta C_{90,135}>0$ for a  real-sky dataset above that of the expected measurement noise must be attributed to foreground emission and systematic measurement error.

The standard picture does not obey any exact angular symmetry.
It predicts a cosmic variance in the angular correlation, so a very close agreement with the symmetry has to  be a statistical fluke. The probability of that fluke is shown by Fig. (\ref{histogram}) to be less than 1\% for the standard model for all of the maps, and significantly less than that for the SMICA map.  As noted above, this limit is set by the sky subtraction uncertainties demonstrated by the variation in the four maps.

The differences between the foreground subtracted maps is comparable to the residual from the holographic model. Therefore, it is plausible that the true sky is compatible with holographic symmetry. The strength of the comparison between holographic and standard models is directly related to the quality of the foreground subtraction at these scales because of the exact nature of the holographic prediction. The holographic hypothesis can be falsified to higher accuracy with better control of the large angular scale measurements.

Given the variations of the current data, it would be premature to rule out either the standard picture or the holographic picture based on this test. 
A more precise, direct, model-independent test of antihemispheric null symmetry  should be possible with better reconstruction of the  primordial potential, especially better Galaxy subtraction.

\begin{figure*}
\begin{centering}
\includegraphics[width=.49\linewidth]{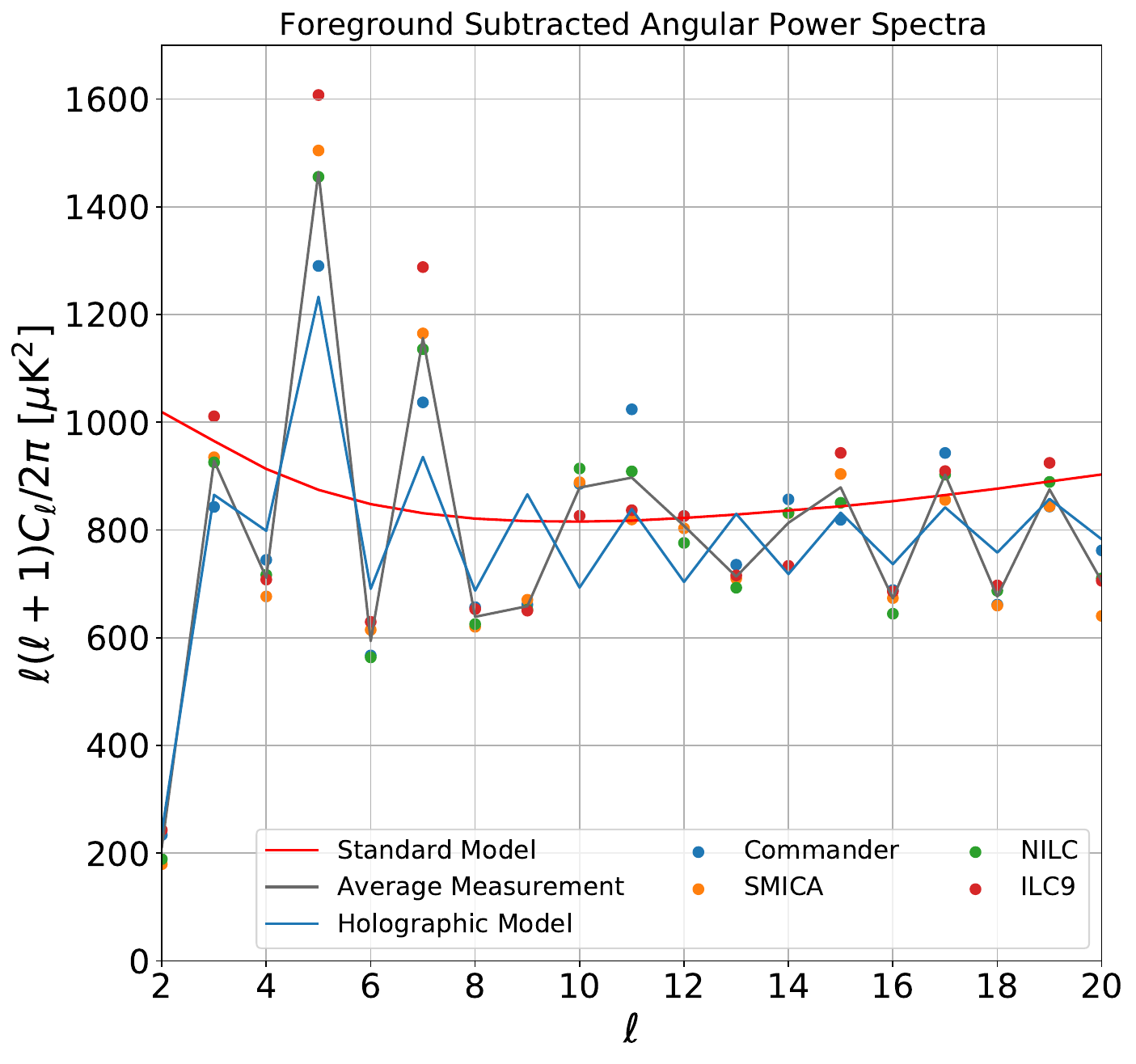}
\includegraphics[width=.49\linewidth]{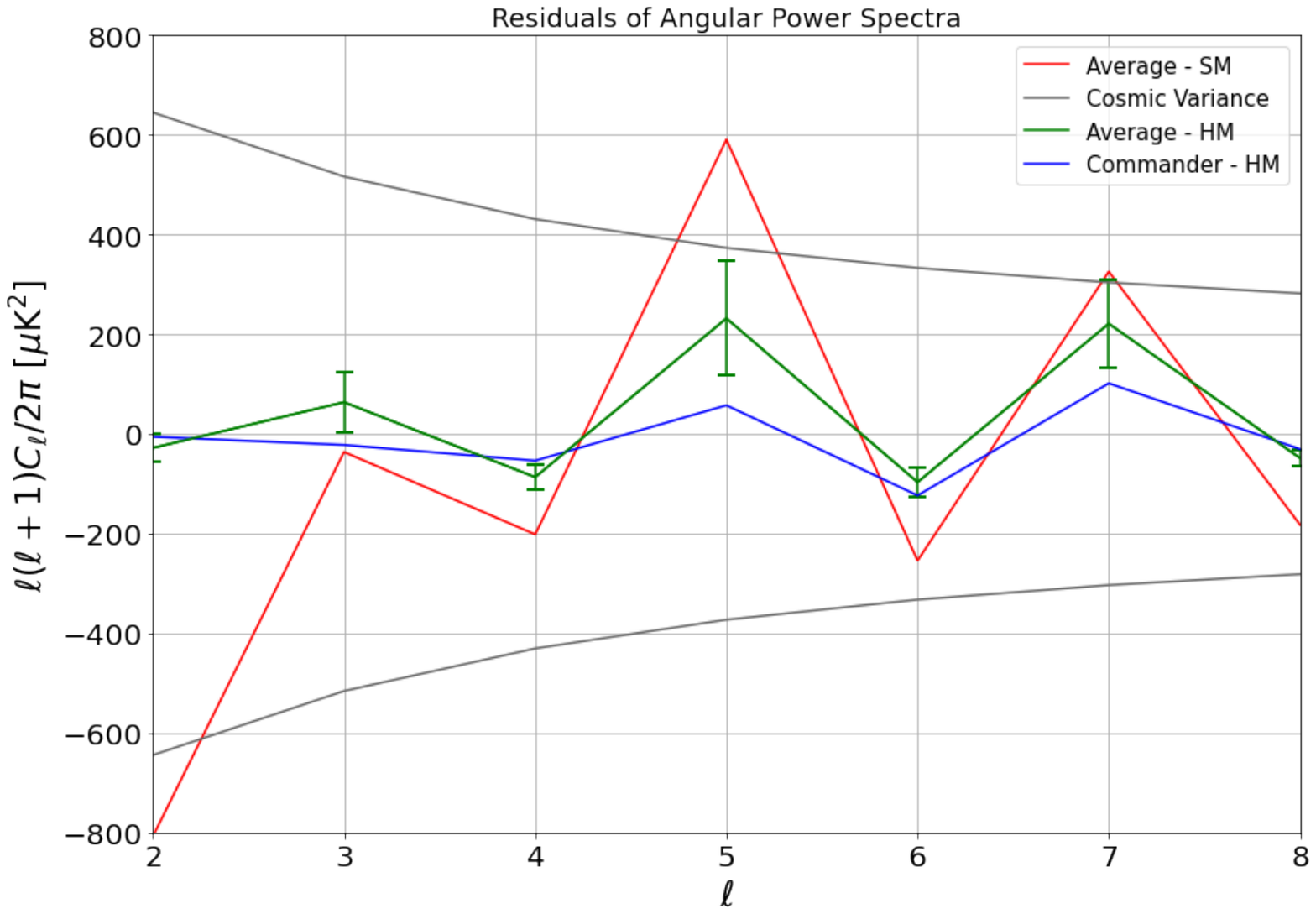}
\par\end{centering}
\protect\caption{Comparison in the spectral domain of holographic and standard models with  different maps.  Data and model maps were first smoothed with a $3^\circ$ beam to reduce aliasing from higher harmonics, which suppresses the high-$\ell$ power slightly compared to Fig. (\ref{data}).  The holographic model  approximately agrees  with the standard expectation at $\ell\ge 9$,
with the notable new feature that it reproduces the well-known excess of odd over even parity harmonics.
As shown at the right, it is a much better match to the data, especially the Commander  map,  at $\ell\le 8$. 
Differences between maps in the low-$\ell$ harmonics, especially at $\ell=3,5,7$, lead to the conspicuous differences at large angles apparent in the angular domain  (Figs. \ref{dataandmodelwithdipole}  and \ref{differencemap}), and are comparable to the differences from the holographic model. The even-parity harmonics, where the maps agree best, also have the smallest residuals.
  \label{spectralcomparison}}
\end{figure*}

\subsection{Angular Power Spectrum at low $\ell$}
\label{sec:power_spectrum}
Comparing the HM with data in $l$-space has the advantage that the data points have smaller covariance. However, the foreground subtraction model parameters and the spatially non-uniform uncertainty in the maps correlates the spectral points as well. In this initial look at the relationship of the model with the foreground subtract data, we have ignored the off-diagonal elements in the spectrum covariance. On the left of Figure~\ref{spectralcomparison} is a low-$l$ spectrum showing the data from the four maps (points) and the average of the data (black line) as well as the SM prediction (red line) and the HM prediction (blue line). On the right are plotted the residuals for the average measured spectrum  from the SM and the HM. The error bars on the HM residuals with the average spectrum show the sample standard deviation of the spectra of the four maps. Because of the covariance between points, this can not be used to calculate a probability of the model. Also plotted are the 1$\sigma$ cosmic variance predicted for the SM and the residual of the HM with the Commander data.

 As expected,  the holographic model spectrum approximately matches the standard expectation at $ \ell\gtrsim 8$.
One notable difference that survives at higher $\ell$ is the saw tooth pattern, the excess of odd- over even-parity power. 
We do not expect agreement at  $\ell\gtrsim 30$, due to Doppler anisotropy and other physical effects that have been ignored here.

 The largest differences between the models occur at $2\le\ell\le 8$. Again, this was expected from the symmetries imposed in the angular domain: the exotic correlations make a significant difference only at  large angular scales. Differences include  the small value of the quadrupole, as well as the specific apparently-conspiratorial pattern of alternating odd and even harmonics that leads to the exact angular symmetries.
 In this range, the holographic model matches the data better than the standard scenario. These low-$\ell$ differences are mainly responsible for the cosmic variance between the $C(\Theta)$ realizations seen in the previous plots, even for for angular separations as small as $\Theta \sim 5^\circ$.

 As expected from angular domain fit, 
 the holographic spectrum agrees best with the Commander map. 
 In  the standard scenario, the  $C_\ell$'s are independent of each other, so each one can be regarded as a separate measurement for a comparison of models. 
For $2\le\ell\le 8$, all of  the holographic residuals  are much  smaller  than the standard scenario  (Fig.  \ref{spectralcomparison}), which can be interpreted as evidence of hidden holographic symmetry.

\subsection{Interpretation of Previously Known Anomalies}

As noted in the introduction, several well-known anomalies  in the pattern of anisotropy have received detailed study, both by the 
satellite collaborations and by other authors
\cite{deOliveira-Costa:2003utu,WMAPanomalies,Ade:2015hxq,Akrami:2019bkn,2017MNRAS.472.2410A,2012MNRAS.419.3378A,2015MNRAS.449.3458C,Schwarz:2015cma}.
The main contribution here is to interpret some of these anomalies as signatures of particular new fundamental physical symmetries.
The true holographic symmetries are hidden, but still leave conspicuous signatures on large angular scales. As shown above, this physical interpretation leads to precise predictions that can be tested.

The most conspicuous symmetry, the near-vanishing of correlation at $\Theta>90^\circ$, has  been conspicuous since the first few years of {\sl COBE DMR} data\cite{1992ApJ...396L..13W,1994ApJ...436..423B,Hinshaw_1996}.
Higher-resolution, less noisy maps by {\sl WMAP} and {\sl Planck} confirmed this property: the anomalous character of the large angle correlations in relation to standard expectations became more sharply revealed as the precision of the measurement and Galactic foreground subtraction improved
\cite{WMAPanomalies,Ade:2015hxq,Akrami:2019bkn}.
In our scenario, the suppression of large-angle correlation   
 arises directly from the lack of  antihemispherical causality. The anomaly is even more striking  after allowing for the addition of an unmeasured dipole. 
 
The spectral domain also shows  anomalous patterns.
 The spectrum approximately matches the standard picture when averaged over a broad band of $\ell$, 
but has very 
 low quadrupole amplitude, and   a systematic preference for odd parity harmonics up to  $\ell\simeq 30$, which maps onto a negative antipodal correlation,  $C(\Theta \rightarrow 180^\circ)<0.$
This pattern was found in WMAP data \cite{2010ApJ...714L.265K,2011ApJ...739...79K,2012AdAst2012E..34K}, and confirmed by Planck \cite{Ade:2015hxq,Akrami:2019bkn}.
Again, this  feature can be explained as a direct result of the antihemispherical antisymmetry of causally-coherent correlations.



The  correlation function on its own cannot  explain apparent  anomalies in the shapes and alignments of harmonics\cite{deOliveira-Costa:2003utu,2017MNRAS.472.2410A,2012MNRAS.419.3378A,2015MNRAS.449.3458C,Schwarz:2015cma}, or in
 hemispherical or dipolar asymmetry of high-$\ell$ power.
These  additional hidden correlations involve not just   spectral power (the $C_\ell$'s) but also spectral phases (the $a_{\ell m}$'s).
In principle, some of these patterns might arise from higher order causally-coherent correlations, but they are not studied here.

\section{Conclusion}

The present study  shows that  the measured correlation function of CMB temperature anisotropy has properties that can be interpreted as exact symmetries imposed by the causal structure of intersecting horizons.  A generalization of causal arguments that predict an exact null correlation  at $\Theta=90^\circ$  accounts for specific anomalous  correlations over a wide range of angles, including a near vanishing value over a large range of angles and anticorrelation at the largest angles.  A simple geometrical model, constrained by  precise null symmetries,  agrees quantitatively with data.  These results  motivate further investigation of the possibility that the 2D angular correlation function of primordial curvature may be governed by causally-coherent quantum gravity.

Demonstration of an  exact  symmetry of the  two-point correlation function on large angular scales would have a profound significance.   Instead of being an uninteresting,  anomalous fluke of random cosmic variance, 
the large-scale pattern of anisotropy could carry unique, precise  signatures of  basic principles underlying holographic quantum gravity.  
More precise  tests  of this possibility, and of specific candidate holographic  models and symmetries  using CMB data, will require better control of Galactic foregrounds.  
Because ground-based surveys suffer from directional sampling bias due to limited spectral coverage, incomplete sky coverage, and constrained scan patterns,  
precision probes of holographic symmetries proposed here add new motivation to the science case for a next-generation CMB satellite.
\cite{kogut2014}\footnote{More detail about the proposed satellite LiteBIRD can be found at  \href{https://rdcu.be/ceUJT}{ { \underline {this link}} to an article on its current design}.}




\bigskip

\censor{
 \begin{acknowledgments}
  This work was supported by the Fermi National Accelerator Laboratory (Fermilab), a U.S. Department of Energy, Office of Science, HEP User Facility, managed by Fermi Research Alliance, LLC (FRA), acting under Contract No. DE-AC02-07CH11359. We acknowledge the use of HEALPix/healpy and the Legacy Archive for Microwave Background Data Analysis (LAMBDA), part of the High Energy Astrophysics Science Archive Center (HEASARC), and   the NASA/ IPAC Infrared Science Archive, which is operated by the Jet Propulsion Laboratory, California Institute of Technology, under contract with the National Aeronautics and Space Administration. The paper is based on observations obtained with {\sl Planck} (http://www.esa.int/Planck), an ESA science mission with instruments and contributions directly funded by ESA Member States, NASA, and Canada. We are grateful for useful discussions with T. Banks, T. Crawford, O. Kwon, N. Selub, and F. Wehlen.
\end{acknowledgments}

}
\bibliography{universal}

\section{Appendix}

The model of  causally-coherent  correlations presented here does not address the deep problems of quantum gravity and cosmology, such as the emergence of space-time and locality, the extrapolation to the infinite timelike past, and specification of the initial quantum state. We have derived some constraints of coherent causal symmetries on late-time correlations  without appeal to a deeper theory, based on causal  and directional relationships between world lines in the emergent classical system.
Indeed, our treatment does not even use quantum mechanics: it is based entirely on the trigonometric properties of intersecting 2-spheres in 3-space,  the comoving spherical ``footprints'' of causal diamonds  at the end of inflation.  

However, the constraints derived here do require 
a radical modification of the standard quantum inflation model in some regimes.
Given the remarkable success of the standard scenario in matching observations on angular scales smaller than a few degrees, this Appendix is included to  summarize some physical and mathematical reasons to consider such foundational modifications.

\subsection{Gravitational quantum coherence and locality}

The basic framework of standard quantum inflation 
 is based on effective field theory.  Quantum coherence is assigned to comoving momentum eigenstates of a scalar field coupled to a scalar gravitational potential.   As the cosmic  expansion cools and eventually freezes the mode oscillations, the  standing waves convert into modes of a classical metric perturbation.

This framework has been used for almost all quantum models of inflation.  
However, it may not be the correct model of physical quantum gravity. 
In addition to  widely-studied UV difficulties of effective field theory beyond the Planck scale, often addressed using string theory,  
 gravity creates infrared constraints\cite{CohenKaplanNelson1999,PhysRevD.101.126010,2021arXiv210304509C} on field states,  a sign that quantum-gravitational correlations on macroscopic scales  are not included in the standard framework.
Effective field theory uses approximations that  do not fully account for  the effects of  nonlocal quantum coherence  on correlations of  measurements that compare the physical effects of gravity  at spacelike separations in  different directions.
This coherence has a negligible effect on current laboratory measurements, but makes a significant difference for gravitational effects of vacuum fluctuations.

The  framework for quantum inflation developed here is based on  principles derived from the requirement that quantum states are  compatible with classical relativistic invariance  and  invariant causal structure. First,
 {\it   quantum states underlying the matter-geometry system (and the origin of locality) are coherent on  null surfaces}. 
And, in particular,   {\it  reduction of the quantum states of the coupled matter-geometry system  occurs on invariant null surfaces, so that quantum 
collapse correlates a light  cone state with a classical scalar localized at  its apex.}
Geometrical properties of quantum coherence then guarantee that collapse of a particle state to a definite location in space and time always  produces a  classical geometry consistent with  its mass-energy in that location. 
 A quantum matter state creates coherent gravitational correlations everywhere in a causal diamond emanating from where a state is prepared. 

Considerations of such a general nature have  motivated previous holographic theories of quantum gravity based on 
coherent quantum states of  covariant causal diamonds
\cite{Banks2011,Banks:2011qf,Banks:2015iya,Banks:2018ypk,Banks:2020dus,Banks:2021jwj}. Causality and holographic scaling\cite{tHooft:1993dmi,Susskind1995,Bousso:2002ju} are also built into semiclassical thermodynamic or entropic theories\cite{Jacobson1995,Jacobson:2015hqa,Verlinde2011,Padmanabhan:2013nxa}  where classical space-time emerges statistically, in a continuous fluid approximation, from an 
 ensemble of covariant null surfaces.
The inflation model developed here assumes
the same kind of coherent null holographic boundary adopted in these proposals:  the scalar potential perturbation at a point is assumed to depend on the quantum state of its past null cone.

These proposals  address a basic unsolved problem in quantum gravity,  
the physical origin of locality.  It is not known how  a classical  localized scalar  gravitational potential emerges from
a nonlocal quantum system.  
 In standard quantum mechanics, preparation and measurement of  states is causal,  even in  spooky ``acausal'' systems such as EPR \cite{RevModPhys.71.S288,2009FoPh...39..677B}.
 Time ordering of spacelike-separated events depends on observer motion, so collapse of a quantum state onto spacelike surfaces  implies an unacceptable observer dependence of physical effects\cite{Zych_2019}.
Quantum coherence of geometrical causal structures, such as horizons and causal diamonds,
is needed to avoid apparent paradoxes of causality associated with active gravity of nonlocal quanta.
{\it The quantum state of an apparatus that makes a macroscopic  measurement of geometry---
 such as a comparison of clocks at spacelike separations in different directions,  or a comparison of potential differences on an inflationary horizon---  must be  a  coherent, Schr\"odinger-cat-like superposition of possible outcomes, each one entangled with the coherent geometrical state of the space-time it inhabits. The nonlocal state of the geometry encompasses  the whole causal diamond of the measurement.}

Coherent,  holographic or emergent quantum  gravity has been studied in various forms over many decades\cite{wheeler1946,Rovelli_1991,Pikovski_2017}.   
 Holography and coherent horizons  have been studied  for flat space-time\cite{Verlinde:2019xfb,Zurek:2020ukz,Banks:2021jwj}, where it might have experimental consequences\cite{holoshear,Richardson:2020snt};  for black holes \cite{Hooft:2016cpw,Hooft:2016itl,Hooft2018,Giddings:2018koz,Giddings:2019vvj,Haco:2018ske,Almheiri:2020cfm}, where nonlocal entropy and entanglement can address so-called information paradoxes;  for early-universe cosmology\cite{Banks2011,Banks:2011qf,Banks:2015iya,Banks:2018ypk}, where coherence might resolve  paradoxes associated with initial conditions;  for the gravitational effect of field vacuum states, where it might explain famously wrong predictions for the value of the cosmological constant\cite{Weinberg:1988cp,Hogan:2020aow};  and 
extensively in the context of  anti-de Sitter space\cite{Ryu:2006bv,Verlinde:2019ade}, where a rigid classical infrared boundary condition provides precise control for studies of holographic quantum states.  

In spite of much formal progress in theory, there is no broad community consensus
about  physical consequences of holographic causal coherence in quantum gravity. The phenomenology of  coherent geometrical fluctuations on horizons and causal diamonds is still an active area of study, with widely varying predictions.


\begin{figure*}
\begin{centering}
\includegraphics[width=.6\linewidth]{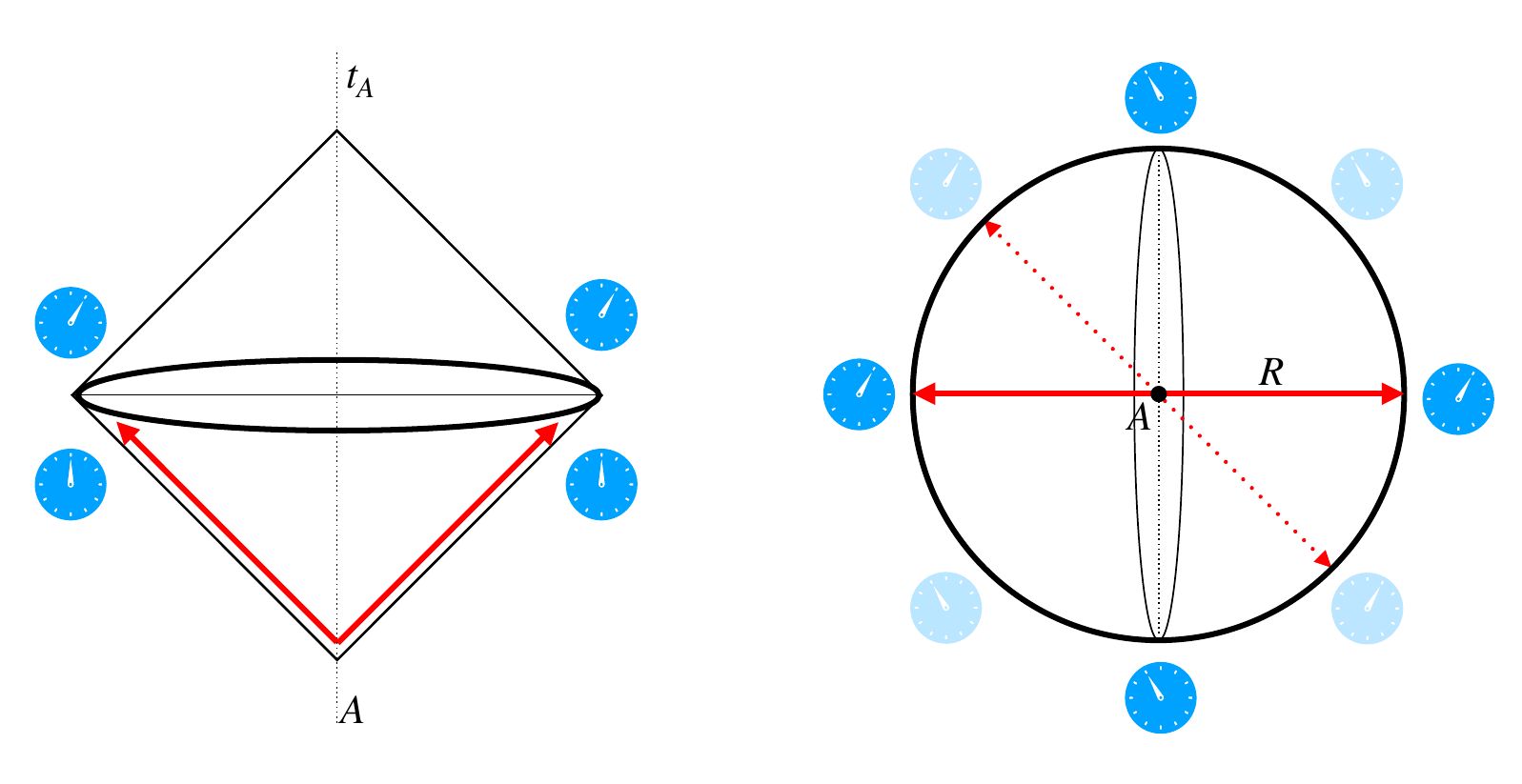}
\par\end{centering}
\protect\caption{The space-time structure of the  gravitational shock wave   from decay of a massive particle into two null point particles, and its effects on measurements of anisotropic time distortions, adapted from ref. \cite{mackewicz2021gravity}.  
The effect of the particle's gravity is to ``drag'' the space-time along with it on the null trajectory, causing an instantaneous displacement in time and velocity, with a decay   axis shown by the solid trajectory in space-time (left) and in space (right).
The causal diamond at the left shows the outgoing particle, and inward-propagating null paths from clocks at radius $R=ct$ observed from the origin at  time $t=2R/c$ after the decay.   The right panel shows a spatial slice at one time as the particle passes radius $R$.  The displacement  $\delta \tau$ on the clock faces shows the gravitational distortion of time on the causal diamond surface. The  spatial structure of the distortion varies coherently on the macroscopic scale $R$  of the causal diamond. The quantum state of the geometry has the same spatial coherence as this classical solution.  For an isotropic, $S$-wave decay, the  geometry  remains in an isotropic,  delocalized coherent superposition of all possible orientation states up to the duration $\tau$ of the measured causal diamond; one such state in the superposition, a Schr\"odinger-cat-like alternative macroscopic geometrical history, is shown by the dotted trajectory and the ghostly clock faces.  Gravity   nonlocally entangles quantum states that describe the distribution of matter and geometry in all directions within the radius $R$ of the causal diamond. 
  \label{particles}}
\end{figure*}

\subsection{Coherent gravitational fluctuations of horizons}

\subsubsection{Universal 2D spectrum of coherent gravitational shock waves from  particle decays}

In the standard scenario,  quantization of modes in orthogonal directions occurs independently.
In the causally-coherent scenario, gravitational effects on space-time in all  directions remain entangled as part of a single coherent, quantum system until the state of a causal diamond collapses.
 The  physical effect of this directional quantum-geometrical coherence can be  calculated in 
 a simple  system\cite{mackewicz2021gravity}:    
 a point particle   of mass $M$ at rest  decays into two null particles,  and their 
 gravitational  effect 
far from the decay axis is measured by clocks, observed from the world line of the body where the decay happens (see Fig. \ref{particles}).

The linear  solution for the classical metric of this system is described in ref. \cite{mackewicz2021gravity}.
The classical geometry  a time $t$ after the decay is a discontinuous spherical null shock wave that separates an exterior Schwarzchild metric at $R>ct$ from an interior flat space-time at 
$R<ct$.
 The anisotropic gravitational field on the shock wave varies coherently on the scale of the light cone,  $R=ct$.
 
The geometrical distortion of the shock wave leads to a coherent large-angle anisotropic distortion in the geometry. It can be measured by an observer at the origin who compares the differences $\delta \tau$ in
elapsed time intervals $\tau$ on clocks  at radius $R<ct$ in different directions. 
Far from the particle decay axis, 
the angular spectrum  is dominated by   $\ell=2$ quadrupolar harmonic components, whose typical
values $a_{2m}$ are
\begin{equation}\label{clockresponse}
    \Delta_{\ell=2}= (\delta \tau/\tau)_{\ell=2} \simeq  GM/\tau c^3.
\end{equation}

 Consider a
 superposition of  gravitational distortions from many  decays.  Each decay has a different orientation, so that  spherical harmonic distortion coefficients $a_{\ell m}$ from each decay pattern
$\Delta (\theta, \phi)_i$ differ.
However, the pattern from each decay is the same, and it is the same at all separations apart from a normalization. Thus, {\it an arbitrary sequence of decays  produces
 a universal 2D power spectrum of distortions}. 
 The  variation from $N$  decays add in quadrature; for example, the typical quadrupole variation is of order
\begin{equation}\label{clockdelta}
   \Delta_{\ell=2} \sim N^{1/2} GM/\tau c^3.
\end{equation}
 Gravitational distortions with this amplitude  
on a spacelike sphere of radius $R$ are generated by randomly oriented decays,
up to
a  maximum number 
\begin{equation}\label{Nmax}
N=N_{BH}\equiv c^2 R/2GM 
\end{equation}
where a black hole forms.  

The  macroscopic directional coherence of geometrical states makes an enormous difference to both  the nature and the amplitude of purely geometrical quantum fluctuations \cite{PhysRevD.99.063531,Hogan_2020}.
Gravitational time distortions  of virtual states are dominated by the most massive, shortest wavelength particle states.  
 For a UV cut off at the Planck length $ct_P$, the quadrupolar quantum-geometrical distortions
   on a timescale $\tau$ are given by Eqs. (\ref{clockdelta}, \ref{Nmax}) with
   $M=M_{Planck}=\sqrt{\hbar c/G}$:
 \begin{equation}\label{Planckdistortions}
 \Delta_{\ell=2}\sim \delta R/R\sim \delta \tau/\tau \sim \sqrt{t_P/\tau},
 \end{equation}
similar to  estimates derived from holographic gravity in flat space
   \cite{Verlinde:2019xfb,Verlinde:2019ade,Zurek:2020ukz,Banks:2021jwj}, but much larger than estimates from effective field theory.
  
 
 \subsubsection{Extrapolation to horizon fluctuations}
 
The decaying-particle  system is based on perturbation of a flat space-time, but it can be continuously extrapolated to estimate coherent  fluctuations of solutions with physical horizons, such as black holes and inflationary cosmologies.
Suppose that the outgoing null  trajectories of the decaying particles are time-reversed, so photons arrive from infinity at a point in the center.  For $N=N_{BH}$,
they form a black hole of radius $R$.  
By continuity, the large-scale angular distortions of the spherical-shock solution should  also apply to the black hole.
The fractional distortion (Eq. \ref{Planckdistortions}) for $N=N_{BH}$ then serves as an estimate of quantum-gravitational distortions of black hole horizons.   

For black holes, these distortions are associated with the radiation of the particles by Hawking evaporation: a typical evaporating particle is associated with a macroscopic distortion  
$\Delta \sim Gm_P/\tau c^3$, where $m_P=\sqrt{\hbar c/G}$ is the Planck mass.
For gravitational radiation, the same  magnitude  for the horizon distortions is also estimated  from the correspondence principle: if the quantum state of a black-hole geometry  corresponds to a coherent superposition of consistent classical solutions,  gravitational radiation from a black hole of radius $R$ is emitted at the Hawking rate, one quantum of energy $\sim \hbar c/R$ in a time $\sim R/c.$
  To radiate gravitational radiation with this frequency and amplitude, the classical quadrupole radiation formula requires distortions  of the horizon of the same magnitude,
  \begin{equation}\label{blackholes}
      (\delta R/R)_{\ell=2}\sim \sqrt{ct_P/R}.
  \end{equation}
 
 As viewed from outside the horizon,  particle states in a coherent model interact with a whole horizon as a single quantum object\cite{Hooft:2016cpw,Hooft:2016itl,Hooft2018,Giddings:2018koz,Giddings:2019vvj}. 
 The classical trajectories of all incoming particles are  focused onto the singularity after they enter the horizon, but in the quantum system they are nonlocally entangled with the geometry everywhere within the horizon.   
 A classical black hole horizon  scrambles orbits of incoming photons with a wide range of orbital parameters, including impact parameters up to $\sim R$, over a wide range of arrival times; an incoming null orbit at any point can have arrived from far away from many different directions and times.  Similarly, a quantum horizon nonlocally fast-scrambles and entangles quantum states arriving from or radiating into  all directions, and over a wide range of elapsed times  $\tau\sim t_P(R/ct_P)^3\gg R/c$ (the evaporation or Page time) in the distant frame.  
 
Quanta of any kind, not just gravitons,  generate similar gravitational distortions.
The correlations on the black hole horizon are encoded in the correlations of its evaporative decay products  far from the hole. 
As discussed previously, a universal 2D power spectrum of distortions  is consistent with holographic information content.

Several formal arguments\cite{Banks:2021jwj} suggest that large-angle, coherent distortions of similar  magnitude to Eqs.   (\ref{Planckdistortions}, \ref{blackholes})  occur on de Sitter  horizons. By extension, they should also occur on  inflationary horizons, where $R=c\tau$ is the horizon radius and $H=1/\tau$ is the expansion rate during slow roll.  The assumption in this paper that is that after inflation,  the nonlocal spacelike distortions of curvature on the quantum horizon  become directly measurable as cosmic perturbations.

  In contrast to these examples, the framework of linearized quantum gravity and  effective  field theory\cite{STAROBINSKY1982175,Starobinsky:1983zz,Parikh:2020kfh} used in standard inflation   produces scalar curvature fluctuations that scale as
  \begin{equation}\label{standarddistortions}
   \Delta\sim ct_P/R,   
  \end{equation}
  much smaller than Eqs. (\ref{Planckdistortions}, \ref{blackholes}).
  In a black hole, this is the typical amplitude of metric distortions for the quanta of Hawking radiation, not the classical horizon distortion needed to radiate them. For scalar inflationary fluctuations, which describe the gravitational response to virtual inflaton fluctuations,  this value is  multiplied by an additional 
  factor that depends on the slope of the inflaton potential. By contrast,  coherent geometrical fluctuations (Eq. \ref{Planckdistortions}) just depend on the Planck-scale cutoff and the horizon size.

  From a formal point of view, it is not really a mystery why there is such a big difference between these models: a more coherent geometry results in more coherent perturbations. A particle wave packet in the standard picture is only coherent in one direction, so broad band, transverse components of many modes cancel. 
  While standard cosmology has long adopted this approach, it is not known which (if either) model of quantum-gravitational coherence is physically correct.   


\begin{figure*}
\begin{centering}
\includegraphics[width=.6\linewidth]{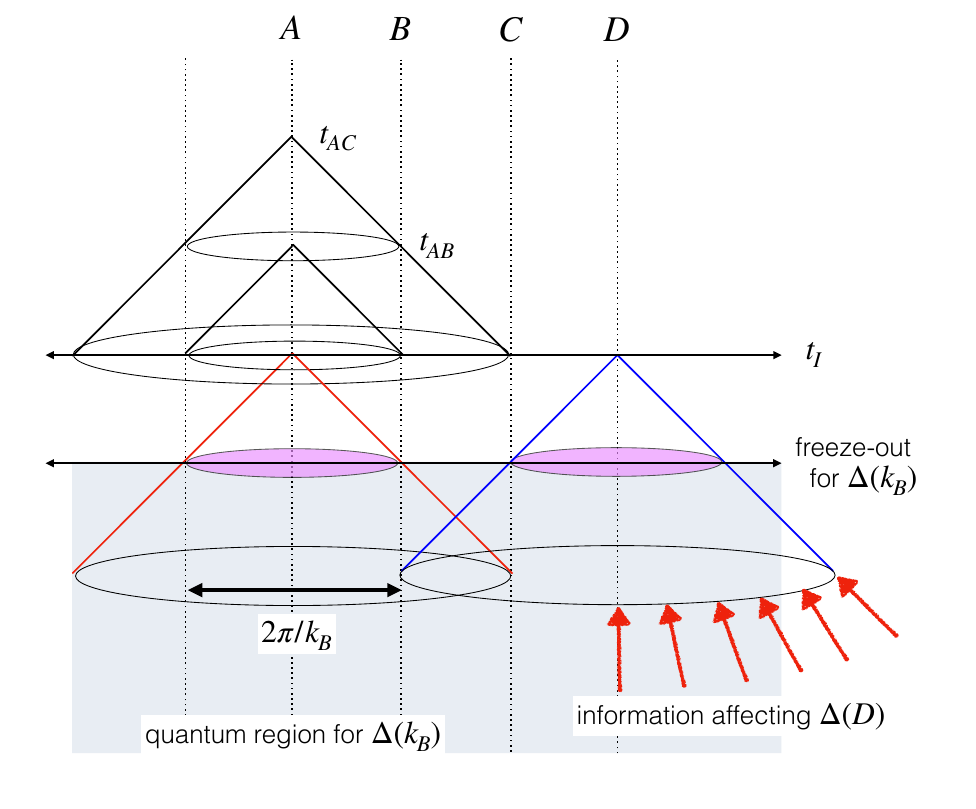}
\par\end{centering}
\protect\caption{Conformal causal diagram of standard inflation as in Fig. (\ref{conformals}), showing acausality of perturbations in standard quantum inflation. A mode of wavelength $2\pi/k_B$ freezes out with a  classical amplitude $\Delta(\vec k_B)$ determined by its initial phase. The classical amplitude  determines the value of $\Delta(\vec k_B)$ for  world lines  arbitrarily far away, even those such as   $D$ that are out of causal contact.
Moreover, modes of all $k$ with $\vec k$ normal to  $\vec {AD}$ contribute coherently to  $\Delta(A)$ and $\Delta(D)$ at arbitrarily large separations $AD$. 
 \label{standardcausal}}
\end{figure*}

\subsection{Coherence of inflationary perturbations}

 \subsubsection{Causally-coherent and   standard quantum models of perturbations}

 Standard inflation and causally-coherent inflation start with the same basic classical framework, a homogeneous slow-roll inflation solution of general relativity. They both consider perturbations from this framework formed by quantum fluctuations.  The difference between them lies in the model for the space-time coherence of the quantum states.  In the standard model, the quantum states are  coherent on spacelike comoving plane wave modes. In the causally-coherent model, the quantum states are coherent on null surfaces, the inflationary horizons of every comoving observer. 
 
 In both the causally-coherent and standard pictures, the local scalar curvature perturbation at any point is not determined classically until the end of inflation. Its value relative to other places freezes out at different times, depending on wavelength or location. In the standard picture, it freezes out as mode amplitudes on  coherent spacelike surfaces, with a coherent relationship that extends acausally out to spatial infinity for each mode when its wavelength matches the horizon size; for large wavelengths, the final classical mode amplitude and phase are fixed  early.  In the causally-coherent picture,  quantum states are coherent on  null surfaces, the light cones that uniquely map to the local curvature on each world line at the end of inflation; the quantities that fix early are the mean values of $\Delta$ on comoving spheres, frozen when they pass through a horizon.

  Coherent quantum gravity  creates correlations among all directions and on all  spatial scales,  as shown  in the decaying-particle system (Fig. \ref{particles}).
   During inflation, the phase of the quantum state that ultimately sets the final value of the scalar curvature at any point is determined by  information from all directions on the past light cone of that point.
  This directional coherence is not accounted for by the  independently-quantized coherent field modes used for inflation. 
  Indeed, it  appears to disagree with properties of local effective field theory, such as separation of scales and independence of orthogonal directions.

\subsubsection{Coherence in the standard inflation model}

As illustrated by the decaying-particle example,  the classical gravitational potential of  mass is not distributed like  the mass itself: the gravitational time dilation effect of a point mass extends coherently through all of space.
By contrast, for linear perturbations, the spatial distribution of a  mass perturbation matches its scalar potential perturbation.  Apparently, an effective field theory of scalar curvature only approximates  active gravity in the limit of  small angles or high wavenumbers, that is, in a causal diamond of size $R\gg k^{-1}$. This restriction is an example of the gravitational IR constraints on states discussed above.

The  critical assumption about coherence in the standard  scenario arises from the promotion of a classical scalar mode to a quantized scalar mode.
Fluctuations of the scalar potential $\Delta(\vec k)$ are connected to the fluctuations of the inflaton scalar $\phi(\vec k)$ by classical relations, with the same comoving spatial structure.
To quantize the system, a classical  mode amplitude $u_{\vec k}$ is identified with  a  quantum  operator $\hat u_{\vec k}$:
\begin{equation}\label{quantumstep}
    u_{\vec k}\rightarrow \hat u_{\vec k}= u_k(\eta)\hat a_{\vec k}
+ u_{-k}(\eta)\hat a_{-\vec k}^\dagger
\end{equation}
where standard raising and lowering operators obey the  noncommuting algebra of a simple harmonic oscillator,
\begin{equation}
    [\hat a_{\vec k},\hat a_{-\vec k'}^\dagger] =(2\pi)^2\delta(\vec k-\vec k')
\end{equation}
with an initial state defined by a field vacuum,
\begin{equation}
    \hat a_{\vec k} |0\rangle = 0.
\end{equation}

In its lowest-energy or ground state,  the wave function of  a harmonic oscillator amplitude has a finite width around zero.
Thus in its vacuum state, the field amplitude has vanishing mean but nonvanishing variance. For each oscillator,   zero-point fluctuations of field amplitude correspond to virtual particles of energy   $\delta E= \hbar \omega/2$ at frequency $\omega$ and physical wavenumber $k= \omega/c$.
Solving the dynamical equations for comoving modes of the field shows that the  quantum fluctuations of amplitude freeze to a definite value when the frequency goes below the expansion rate, the inverse of the horizon scale. The final classical amplitude of a  mode when it freezes out is fixed by  the  quantum phase, which  is coherent everywhere on an infinite surface of constant conformal time $\eta.$

This kind of quantum state does not collapse in a causally coherent way.    For example, it does not produce large-angle correlations in causal macroscopic superpositions of position states, like the system  shown in Fig. (\ref{particles}).
For that system, the quantum state of the geometrical time distortion $\delta \tau$ on a sphere is an entangled superposition of field modes in all directions. The state collapses coherently on a light cone from its center, not on infinite surfaces of constant time.


The spatial coherence assumed in Eq. (\ref{quantumstep}) 
creates acausal  correlations of $u(\vec x)$ at large separations, as shown in Fig. (\ref{standardcausal}).
The quantization procedure  quietly imports an unphysical, acausal coherence from the unperturbed background that extends to infinity, but affects local measurements. 
It preserves the spatial structure of  a comoving standing wave mode, so it  constrains coherent  phases  for the counterpropagating wave components everywhere on a spacelike hypersurface. It preserves the phase information introduced by the initial vacuum state, and does not allow for causally entangled quantum phase noise of transverse directions to affect the measurement  outcome.

\subsubsection{Observer dependence of 3D spectral decomposition}


A delocalized quantum system becomes ``classical''  only when it is measured at a particular place, while a perturbation mode of classical  scalar curvature  is defined only in relation to a  background defined by a spatially infinite system.
The two scalar perturbation fields
$\phi(\vec x)$ and $\Delta(\vec x)$ must have the same observable linearized classical spatial structure  at the end of inflation, but  the transform that uniquely connects $\Delta(\vec x)$ to $\Delta(\vec k)$ depends on an integral that extends to spatial infinity,  which is not part of the quantum system, and is not measurable.  {\it Different spectral decompositions $\Delta(\vec k)$ can describe  identical observable  differences of $\Delta(\vec x)$ in any finite region.} Thus, there can be an observer dependence to the 3D spectral decomposition, which is not a property of standard quantum inflation.

In the causally-coherent picture, different observers infer different decompositions into 3D modes to describe the same   distribution of $\Delta(\vec x)$ over the finite region of space where their horizons overlap. The quantum entanglement of their horizons occurs while the geometrical states are not yet localized scalars, so there is never a physically observable contradiction between actual observables, or  with causality. In the standard picture, a well-defined universal spectral decomposition or realization is achieved by inserting acausal correlations at the beginning, in the initial conditions of coherent infinite plane wave modes.




 \subsubsection{Interferometer analogy}

In real physical systems, nonlocal quantized states of  fields depend on their directional correlations over  the whole space-time volume of a measurement\cite{Caves1980,Caves1980a}.  The decomposition in  Eq. (\ref{quantumstep}) is well suited for analyzing particle collisions and tracks, but not for  coherent states delocalized over a macroscopic volume. For example, consider  the state of quantized light fields in a Michelson interferometer \cite{Caves1980,Caves1980a}: when the beamsplitting element is included to mix coherent states in different directions, photon-number  eigenstates are  macroscopically delocalized along two directions in a plane. 
``Squeezing'' of these states to reduce quantum noise in an interferometer signal,  now a staple of LIGO technology, depends on the injection of vacuum fluctuations with controlled phase into the dark port\cite{PhysRevLett.124.171102}; there is nonlocal spacelike macroscopic entanglement of field modes in different directions with an entire apparatus.
 Similarly,  the phase of zero-point fluctuations of a field vacuum at any point is determined by information from all directions on its past light cone: 
 the nonlocal quantum states of inflationary horizons  entangle information from all directions.
 
\subsubsection{Black hole analogy}
Causally-coherent inflation  resembles a coherent  quantum  black hole. 
The  analog of an inflationary perturbation is the back-reaction on a black hole horizon from Hawking radiation.    
 Viewed from outside, the coherent quantum horizon acts like an atom--- a single, delocalized quantum object\cite{Hooft:2016cpw,Hooft:2016itl,Hooft2018,Giddings:2018koz,Giddings:2019vvj}. 
 As noted above, this delocalization of  quantum information---  or more precisely, its localization to horizons and other null surfaces associated with a measurement of a state---  is likely the key to solving black hole information paradoxes.

 The same idea applies to horizons in de Sitter space\cite{Banks:2021jwj}, and by extension to the inflationary horizon.
A coherent  quantum horizon completely represents the relationship of a black hole to  an outside observer.  Similarly in our model of causally-coherent inflation, the interior of an observer's horizon is not part of the perturbation model: the classical concept of space-time locality melts away for quantum fluctuations of $\Delta$ inside the inflationary horizon.
Until the end of inflation,  the  curvature perturbation is nonlocal, observer-dependent and indeterminate.

\subsubsection{Extrapolation to past timelike infinity}

Standard  quantum inflation allows extrapolation of  quantum fields on a classical background indefinitely to small scales and into the far past on any world line, possibly even to a larger, eternal multiverse\cite{PhysRevD.27.2848,GUTH2000555}.   Quantum states of local field patches are  localized on trajectories of proper comoving time, rather than causal diamonds.
This model of  locality can lead to conceptual paradoxes\cite{Ijjas2015}. 

The proposed holographic resolution to both cosmological and 
 black hole information paradoxes is a causal entanglement of nonlocal quantum states.
In causally-coherent cosmology, a deeper quantum  theory of geometry is still needed to address how  space and time  emerge  in the  interiors of the causal diamonds\cite{Banks2011,Banks:2011qf,Banks:2015iya,Banks:2018ypk,Banks:2020dus,Banks:2021jwj},
and to specify the  initial quantum state\cite{penrose,Layzer2010,Aguirre2011}.


\subsubsection{Statistical isotropy}

An important symmetry of standard inflation, often called ``statistical isotropy,''  is the hypothesis
that  3D mode amplitudes  only  depend on $k$ and not on direction, $\vec k/k$.  
As a result, the  spectral coefficients $a_{\ell m}$ are statistically independent.  This property is not a symmetry of the sky, but of the model: the statistical independence refers to independent realizations.

The causally-coherent picture also does not define any physically preferred direction in space, since by construction, it has the same causal symmetries as relativity. In this  sense, it  is also statistically isotropic.
However, because it  has universal symmetries in the angular domain that apply to any realization, the $a_{\ell m}$'s  are no longer independent of each other, and
the $\Delta(\vec k)$'s are only independent of each other to a good approximation for $k\gg R^{-1}$, where $R$ is the size of any spatial volume. 
As in standard inflation, isotropy is spontaneously broken for a finite part of any  realization.  For example, perturbations on any sphere define a preferred axis, that of its intrinsic dipole.
In our scenario,  the large-angle uniformity of the sky and the near-flatness of the universe as a whole ultimately derive from an  underlying universal holographic symmetry of the quantum system\footnote{As Hermann Weyl wrote: `` As far as I can see, all a priori statements in physics have their origin in symmetry.''(ref. \cite{Weyl1952}, p. 126)}.

\subsection{Homogeneous classical inflation}

  A  standard inflation model\cite{Baumann:2009ds,Weinberg:2008zzc} is   assumed throughout this paper  for the classical background cosmology that defines the causal relationships between world lines.  The model of mass-energy is  a spatially uniform classical (that is, unquantized) inflaton field, with dimension of mass and vacuum expectation value $\phi(t)$, where $t$ is a standard FRW time coordinate.
  In standard notation where $\hbar= c =1$, the expansion rate $H$ and cosmic scale factor $a$  evolve according to classical general relativity and thermodynamics,
\begin{equation}\label{hubble}
H^2(t)\equiv (\dot a/a)^2= (8\pi G/3) (V(\phi)+\dot \phi^2/2),
\end{equation}
where the evolution of the  inflaton depends on the potential $V(\phi)$  via
\begin{equation}\label{roll}
\ddot \phi + 3H\dot \phi + V' = 0,
\end{equation}
and  $V'\equiv dV/d\phi$.
For  slow roll inflation,  the first term is negligible and the system evolves with slowly-varying  values of $H,\dot \phi,$ and $ V'$,  related by
\begin{equation}\label{slowroll}
3H\dot \phi \sim  - V',
\end{equation}
which produces  
a nearly-exponential expansion.
About 60 $e$-foldings of $a$ after the currently observable volume of the universe matches the scale $c/H$ of the inflationary horizon, inflation ends,  and the inflaton energy subsequently  ``reheats'' with the conversion of $V(\phi)$ to thermal matter.
This  background solution provides the  global
definition of surfaces of constant unperturbed cosmic time on comoving world lines.

  \subsection{Background parameters in a holographic model}
The slow-roll background solution assumed for inflation here  is identical to the standard picture.  
However, the values of  parameters of the background cosmology during inflation, which determine  the amplitude and spectral tilt of causally-coherent perturbations,  are not the same as in the standard picture. 

As an example, let us evaluate the constraints on $V(\phi)$  for  a model of  causally-coherent uncertainty\cite{PhysRevD.99.063531,Hogan_2020} where the perturbation power per $e$-folding is  given by 
the causal diamond radius in Planck units:
\begin{equation}\label{potentialvariance}
 \Delta^2  = d \langle \Delta^2 \rangle/ d\ln t = \alpha  H t_P,
\end{equation}
where $\alpha$ is a factor of order unity.  
This estimate applies to  causally-coherent states of a causal diamond that include a coherent superposition of all directional states. 
The linear dependence on $Ht_P$ is  different from (and generally larger than)  standard perturbations, which scale like $(Ht_P)^2$.  Unlike standard inflaton-driven perturbations, $\langle \Delta^2 \rangle$  does not depend on the slope of  $V$, only on its value.

Let  $\phi_0$ denote the value of $\phi$ when the measured comoving scales,  comparable to the current Hubble length,  match the inflationary horizon.  From  Eq.  (\ref{hubble}), 
\begin{equation}\label{scalar}
  \Delta^2 =     \alpha   ( 8\pi G t_P^2/3)^{1/2} \ \  V(\phi_0)^{1/2}
\end{equation}
The measured value\cite{Ade:2015lrj,Aghanim:2018eyx}  $\Delta^2= A_S=  2\times 10^{-9}$ implies an equivalent temperature or particle energy, 
 \begin{equation}\label{radiationscale}
V_0^{1/4} \sim \alpha^{-1/2} (3/8\pi)^{1/4}  \Delta_S \ m_Pc^2 \sim  3\times 10^{14}     {\rm GeV},
\end{equation}
where $m_P\equiv \sqrt{\hbar c/G}$.  This is  a somewhat lower value than in many standard inflation models, although still consistent with the idea of inflation at  grand unification.

As in standard inflation, the value of  $H$ is not constant during inflation, but  varies slowly, according to Eqs. (\ref{hubble}) and (\ref{roll}),  with a  spectrum described by a  spectral index $n_S$:  $\Delta_k^2 \propto k^{n_S-1} $. In the causally-coherent scenario, the tilt  in the spectrum,  
\begin{equation}\label{tilt}
 n_S-1 \equiv -\epsilon = \frac{d \ln \Delta_k^2}{d \ln k} = \frac{d \ln H}{d \ln k},
\end{equation}
is related to the potential    by
\begin{equation}\label{slowrollepsilon}
 \epsilon = (V'/V)^2 (16\pi G)^{-1} .
\end{equation}
Eqs.  (\ref{tilt}) and (\ref{slowrollepsilon}) imply that the measured tilt  depends only on  $V'/V$ at the epoch when a scale passes through the horizon. 
The measured value \cite{Aghanim:2018eyx,Akrami:2018vks}  $1-n_S = 0.035\pm0.004$ constrains   its logarithmic slope  to be close to the inverse  Planck mass: 
\begin{equation}\label{slope}
 \left(\frac{V'}{V}\right)_{\phi_0} = \frac{\sqrt{16\pi \epsilon}}{ m_P} = 1.32 m_P^{-1}  \left(\frac{1-n_S}{0.035}\right)^{1/2} .     
\end{equation}
 As usual, sufficient inflation to reach the current scale of the universe requires $N\sim 60$ e-foldings since $\phi=\phi_0$, depending on reheating and subsequent evolution. 
 In the slow roll approximation,   
\begin{equation}\label{Nfold}
|\dot \phi/\phi|_{\phi_0} \sim H(\phi_0)/N.
\end{equation}

Combination of  Eqs. (\ref{hubble}),  (\ref{slowroll}),  (\ref{slope})  and (\ref{Nfold}) leads to  a lower limit on the range in $\phi$ covered during inflation:
\begin{equation}\label{phi0}
\delta\phi >  \frac{N}{8\pi} \left(\frac{V'}{V}\right)_{\phi_0} m^2_P\sim  3.1 \ m_P \  \frac{N}{60}  \left(\frac{1-n_S}{0.035}\right)^{1/2}.
\end{equation}

It   is not trivial for a potential to satisfy  experimental constraints on both $N$ and $n_S$. 
Because the relation between $\Delta^2$ and $\epsilon$ differs  from  the standard picture,
slow-roll potentials preferred in the causally-coherent scenario are strongly excluded 
for standard models, and {\it vice versa}.
To choose one  example that does work for the causally-coherent scenario, a  potential of the simple form
\begin{equation}\label{phifour}
V= {\rm constant} \times\  \phi^4,
\end{equation}
fits current measurements with $N= 59\pm 7$.
 A  potential of  this  form   is now ruled out for standard inflation\cite{Akrami:2018odb}.

\end{document}